\documentclass[aps,prd,reprint,showpacs,floatfix,longbibliography,nofootinbib,superscriptaddress]{revtex4-1}

\usepackage{tikz,graphicx,booktabs,multirow,array,microtype,mathtools,float}

\usepackage{centernot}
\newcommand{\fsl}[1]{{\centernot{#1}}} 

\graphicspath{{figures/}}
\usepackage{enumerate}   
\extrafloats{200}
\usepackage[colorlinks=true,backref=false, linktocpage=true,
citecolor=blue,
urlcolor=blue,linkcolor=blue,
pdfpagemode=UseOutlines]{hyperref}
\newcolumntype{C}{>{$}c<{$}}
\usepackage{dcolumn}

\usepackage{anyfontsize}
\usepackage[shortlabels]{enumitem}

\usepackage{amsfonts}
\AtBeginDocument{
\heavyrulewidth=.08em
\lightrulewidth=.05em
\cmidrulewidth=.03em
\belowrulesep=.65ex
\belowbottomsep=0pt
\aboverulesep=.4ex
\abovetopsep=0pt
\cmidrulesep=\doublerulesep
\cmidrulekern=.5em
\defaultaddspace=.5em
}

\def\MeV {\mathop{\hbox{MeV}}}

\def\Tr {\mathop{\hbox{Tr}}}

\newcommand{\langl}{\begin{picture}(4.5,7)
\put(1.1,2.5){\rotatebox{60}{\line(1,0){5.5}}}
\put(1.1,2.5){\rotatebox{300}{\line(1,0){5.5}}}
\end{picture}}
\newcommand{\rangl}{\begin{picture}(4.5,7)
\put(.9,2.5){\rotatebox{120}{\line(1,0){5.5}}}
\put(.9,2.5){\rotatebox{240}{\line(1,0){5.5}}}
\end{picture}}

\usepackage{amsmath} 
\usepackage{dsfont}  
\usepackage{bm}      

\def\iden{\mathds{1}}

\def\beq{\begin{equation}}
\def\eeq{\end{equation}}
\def\beqs#1\eeqs{\beq\begin{split} #1 \end{split}\eeq}

\long\def\comment#1{}

\makeatletter 
    
\renewcommand\onecolumngrid{
\do@columngrid{one}{\@ne}%
\def\set@footnotewidth{\onecolumngrid}
\def\footnoterule{\kern-6pt\hrule width 1.5in\kern6pt}%
}

\renewcommand\twocolumngrid{
        \def\footnoterule{
        \dimen@\skip\footins\divide\dimen@\thr@@
        \kern-\dimen@\hrule width.5in\kern\dimen@}
        \do@columngrid{mlt}{\tw@}
}%

\makeatother    

\begin{document}

\title{
Charged pion electric polarizability from four-point functions in lattice QCD 
}
\author{Frank~X.~Lee}
\email{fxlee@gwu.edu}
\affiliation{Physics Department, The George Washington University, Washington, DC 20052, USA}
\author{Andrei~Alexandru}
\email{aalexan@gwu.edu}
\affiliation{Physics Department, The George Washington University, Washington, DC 20052, USA}
\affiliation{Department of Physics, University of Maryland, College Park, MD 20742, USA}
\author{Chris~Culver}
\email{C.Culver@liverpool.ac.uk}
\affiliation{Department of Mathematical Sciences, University of Liverpool, Liverpool L69 7ZL, United Kingdom}
\author{Walter Wilcox}
\email{walter\_wilcox@baylor.edu}
\affiliation{Department of Physics, Baylor University, Waco, Texas 76798, USA}

\begin{abstract}
Polarizabilities reveal valuable information on the internal structure of hadrons in terms of charge and current distributions. For neutral hadrons, the standard approach is the background field method. But for a charged hadron, its acceleration under the applied field complicates the isolation of the polarization energy. In this work, we explore an alternative method based on four-point functions in lattice QCD. The approach offers a transparent picture on how polarizabilities arise from photon, quark, and gluon interactions. We carry out a proof-of-concept simulation on the electric polarizability of a charged pion,
 using quenched Wilson action on a $24^3\times 48$ lattice at $\beta=6.0$ with pion mass from 1100 to 370 MeV.
We show in detail the evaluation and analysis of the four-point correlation functions and 
report results on charge radius and electric polarizability.
Our results from connected diagrams suggest that charged pion $\alpha_E$ is due to  a  cancellation between elastic and inelastic contributions. It would be interesting to see how the cancellation plays out at smaller pion masses in future simulations.

\end{abstract}
\maketitle

\twocolumngrid

\section{Introduction}
\label{sec:intro} 

Understanding electromagnetic polarizabilities has been a long-term goal of lattice QCD. The challenge in the effort lies in the need to apply both QCD and QED principles. The standard approach to compute polarizabilities is the background field method which has been widely used for dipole polarizabilities~\cite{Fiebig:1988en,Lujan:2016ffj, Lujan:2014kia, Freeman:2014kka, Freeman:2013eta, Tiburzi:2008ma, Detmold:2009fr, Alexandru:2008sj, Lee:2005dq, Lee:2005ds,Engelhardt:2007ub,Bignell:2020xkf,Deshmukh:2017ciw,Bali:2017ian,Bruckmann:2017pft,Parreno:2016fwu,Luschevskaya:2015cko,Chang:2015qxa,Detmold:2010ts}.  Methods to study higher-order polarizabilities have also been proposed~\cite{Davoudi:2015cba,Engelhardt:2011qq,Lee:2011gz,Detmold:2006vu} in this approach. 
Although such calculations are relatively straightforward, requiring only energy shifts from two-point functions, there are a number of unique challenges.
First, since weak fields are needed, the energy shift involved is very small relative to the mass of the hadron (on the order of one part in a million depending on the field strength). This challenge has been successfully overcome by relying on statistical correlations with or without the field. 
Second, there is the issue of discontinuities across the boundaries when applying a uniform field on a periodic lattice. This has been largely resolved by using quantized values for the fields, or Dirichlet boundary conditions.
Third and more importantly, a charged hadron accelerates in an electric field and exhibits Landau levels in a magnetic field. Such motions are unrelated to polarizability and must be disentangled from the deformation energy on which the polarizabilities are defined. For this reason, most calculations have focused on neutral hadrons. For charged hadrons, what happens is that the two-point correlator does not develop single exponential behavior at large times. 
In Ref.~\cite{Detmold_2009}, a relativistic propagator for a charged scalar is used to demonstrate how to fit such lattice data for charged pions and kaons.
This approach is improved recently in Ref.~\cite{niyazi2021charged} with an effective charged scalar propagator exactly matching the lattice being used to generate the lattice QCD data. A new fitting procedure is proposed where  a $\chi^2$-function utilizes information in both the real and imaginary parts of the correlator while remaining invariant under gauge transformations of the background field. 
For magnetic polarizability, a field-dependent quark-propagator eigenmode projector is used to filter out the effects of Landau levels~\cite{Bignell_2020,He:2021eha}.
These special techniques for charged particles involve fairly complicated analysis to treat the collective motion of the system in order to isolate the polarizabilities. 

In this work, we explore an alternative approach based on four-point functions in lattice QCD. 
Instead of background fields,  electromagnetic currents couple to quark fields to induce interactions to all orders. 
It is a general approach that treats neutral and charged particles on equal footing, but particularly suited for charged particles.
The trade-off is an increased computational demand of four-point functions.
Although four-point functions have been applied to study  various aspects of hadron structure~\cite{Liang:2019frk,Liang_2020a,Fu_2012,Alexandrou_2004,Bali_2018,bali2021double},  
not too much attention has been paid to its potential application for polarizabilities.
We know of two such studies from a long time ago~\cite{BURKARDT1995441,Wilcox:1996vx}, 
a recent calculation on the pion~\cite{Feng:2022rkr}, and a preliminary one on the proton~\cite{Wang:2021}.
A reexamination of the formalism in Ref.~\cite{Wilcox:1996vx} is recently carried out  in Ref.~\cite{Wilcox:2021rtt} for both electric and magnetic polarizabilities of a charged pion and a proton.
We also note that although Refs.~\cite{Engelhardt:2007ub,Engelhardt:2011qq} are based on  the  background field
method, they are in fact four-point function calculations. A perturbative expansion in the background field at the action level is performed in which two vector current insertions couple the background field to the hadron correlation function, leading to the same diagrammatic structures as in this work.

Experimentally, polarizabilities are primarily studied by low-energy Compton scattering. Theoretically,  a variety of methods have been employed to describe the physics involved, from quark confinement model~\cite{PhysRevD.45.1580}, to NJL model~\cite{Dorokhov:1997aa,PhysRevD.40.1615}, to linear sigma model~\cite{BERNARD198816}, to dispersion relations~\cite{Lvov:1993fp,Lvov_2001,PhysRevC.81.029802,Fil_kov_2017}, to chiral perturbation theory (ChPT)~\cite{Moinester:2019sew,Lensky:2009uv,Hagelstein:2020vog} and chiral effective field theory (EFT)~\cite{McGovern:2012ew,Griesshammer:2012we}.  Reviews on hadron polarizabilities  can be found in Refs.~\cite{Moinester:2022tba,Moinester:2019sew,Griesshammer:2012we}.

The presentation is organized as follows.
In Sec.~\ref{sec:method} we outline the methodology to extract polarizabilities, using the electric polarizability of a charged pion as an example. 
In Sec.~\ref{sec:cfun} we detail our notations and algorithms used to evaluate the four-point functions, including how the  Sequential-Source Technique (SST) can be applied in this context.
In Sec.~\ref{sec:results} we  show our analysis procedure and results from a proof-of-concept simulation.
In Sec.~\ref{sec:con} we give concluding remarks and an outlook.
Some technical details are put in the Appendices.

\section{Methodology}
\label{sec:method} 

\begin{figure}[ht]
\includegraphics[scale=0.55]{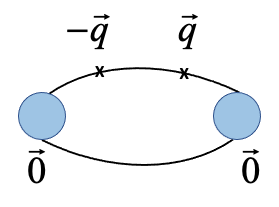}
\caption{Four-point function for charged pion polarizabilities  under the zero-momentum Breit frame.  Time flows from right to left and the four momentum conservation is expressed as $p_2 =q_2+q_1+p_1$.}
\label{fig:diagram-4pt2}
\end{figure}

In  Ref.~\cite{Wilcox:2021rtt}, a formula is derived for electric polarizability of a charged pion,
\fontsize{9}{9}
\beq
\alpha_E= {\alpha \langl  r_E^2\rangl \over 3m_\pi}+\lim_{\bm q\to 0}{2\alpha  \over \bm q^{\,2}} \int_{0}^\infty d t \bigg[Q_{44}(\bm q,t) -Q^{elas}_{44}(\bm q,t) \bigg].
\label{eq:alpha}
\eeq
Here $\alpha=1/137$ is the fine structure constant.
The first term in the formula involves the charge radius and pion mass (we will refer to this term as the elastic contribution). The second term has the elastic contribution $Q^{elas}_{44}$ subtracted from the total (we will refer to this term as the  inelastic contribution). 
The formula will be used in discrete Euclidean spacetime but we keep the Euclidean time axis continuous for notational convenience.
Special kinematics (called zero-momentum Breit frame)
are employed in the formula to mimic low-energy Compton scattering.
The process is illustrated in Fig~\ref{fig:diagram-4pt2}, 
where the initial ($p_1$) and final  ($p_2$)  pions are at rest and the photons have purely spacelike momentum,
\beq
p_1 =(\bm 0,m_\pi),
q_1 =(\bm q,0),
q_2 =(-\bm q,0),
p_2 =(\bm 0,m_\pi).
\label{eq:Breit2}
\eeq

The $Q_{44}$ is defined as the $\mu=4=\nu$ component of the Fourier transforms,
\beq
Q_{\mu\nu}(\bm q,t_2,t_1)  \equiv \sum_{\bm x_2,\bm x_1} e^{-i\bm q\cdot \bm x_2} e^{i\bm q\cdot \bm x_1} 
P_{\mu\nu}(\bm x_2,\bm x_1,t_3,t_2,t_1,t_0),
\label{eq:Qmn}
\eeq
where $P_{\mu\nu}$ is a four-point function defined in position space ($\Omega$ denotes the vacuum),
\beqs
&P_{\mu\nu}(\bm x_2,\bm x_1,t_3,t_2,t_1,t_0)  \equiv  \\
& {\sum_{\bm x_3,\bm x_0}\langl \Omega | \psi (x_3) :j^L_\mu(x_2) j^L_\nu(x_1):  \psi^\dagger (x_0) |\Omega \rangl
\over 
 \sum_{\bm x_3,\bm x_0} \langl \Omega  | \psi (x_3) \psi^\dagger (x_0) |\Omega \rangl 
}.
 \label{eq:P1}
\eeqs
Here $\psi$ is the interpolating field of the pion and $j^{L}_\mu$  the lattice version of 
the electromagnetic current density. The two-point function in the denominator is for normalization. 
Normal ordering is used to include the required subtraction of vacuum expectation values (VEV) on the lattice. 
The sums over $\bm x_0$ and $\bm x_3$ enforce zero-momentum pions at the source ($t_0$) and sink ($t_3$). 
The two currents are inserted at $t_1$ and $t_2$ with two possibilities of time ordering implied in the normal ordering.
The field operators for $\psi$  and $j^L_\mu$ used in this work, along with conservation properties of $Q_{44}$ at $\bm q=0$, are given in Appendix~\ref{sec:op}.
\begin{figure}
\includegraphics[scale=0.45]{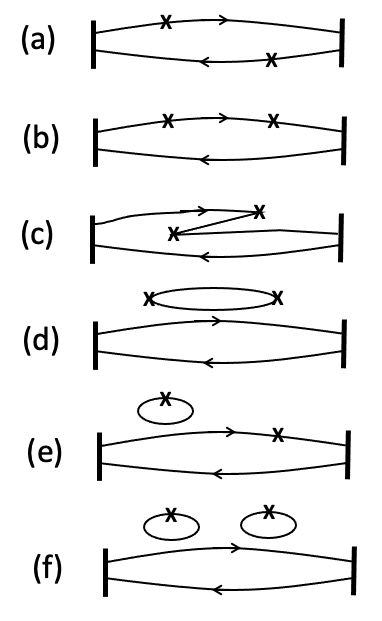}
\caption{Skeleton diagrams of a four-point function contributing to polarizabilities of a meson: (a) connected insertion: different flavor, (b) connected insertion: same flavor, (c) connected insertion: same flavor Z-graph, (d) disconnected insertion: single loop, double current, (e) disconnected insertion: single loop, (f) disconnected insertion: double loop.  In each diagram, flavor permutations are assumed as well as gluon lines that connect the quark lines. The zero-momentum pion interpolating fields are represented by vertical bars (wall sources). Time flows from right to left.}
\label{fig:4pt}
\end{figure}
To see the structure of the four-point function in Eq.\eqref{eq:P1}, we insert a complete set of states in the numerator (twice) and in the denominator (once).
When the times are well separated (defined by the time limits $t_3\gg t_{1,2} \gg t_0$) the correlator is dominated by
the ground state, 
\beqs
& P_{\mu\nu}(\bm x_2,\bm x_1,t_3,t_2,t_1,t_0)  \to \\ & 
{N_s^2|\langl \pi(\bm 0) | \psi(0)|\Omega \rangl|^2 e^{-m_\pi t_3} \langl \pi(\bm 0)| :j^L_\mu(x_2) j^L_\nu(x_1): |\pi(\bm 0)\rangl
\over 
N_s^2|\langl \pi(\bm 0) | \psi(0)|\Omega \rangl|^2 e^{-m_\pi t_3}
} \\
&\to \langl \pi(\bm 0) | : j^{L}_\mu(x_2) j^{L}_\nu(x_1): |  \pi(\bm 0)\rangl \\ &
=\langl \pi(\bm 0) | T j^{L}_\mu(x_2) j^{L}_\nu(x_1) |  \pi(\bm 0)\rangl  
 -  \langl \Omega | T j^{L}_\mu(x_2) j^{L}_\nu(x_1) | \Omega\rangl.
 \eeqs
Here $N_s=N_xN_yN_z$ is the number of spatial sites on the lattice.
The role of the two-point function as normalization and the inclusion of VEV subtraction is evident in the limit.

Assuming time separation $t=t_2-t_1>0 $ and inserting a complete set of intermediate states,  
the diagonal component of $Q_{\mu\nu}$ develops the time dependence  in the same limits,
\beqs
& Q_{\mu\mu}(\bm q,t)  = \\ & N^2_s \sum_{n} 
| \langl \pi(\bm 0) | j^L_\mu(0) | n(\bm q) \rangl|^2  e^{-a(E_n-m_\pi) t}  \\  &
 - N^2_s \sum_{n}  |\langl \Omega | j^L_\mu(0) | n(\bm q) \rangl |^2  e^{-aE_n t}.
\label{eq:Q1} 
\eeqs
At large time separations, it is dominated by the elastic contribution ($n=\pi$ term in the first sum),
\beq
Q^{elas}_{\mu\mu}(\bm q,t)   \equiv N^2_s 
|\langl \pi(\bm 0) | j^L_\mu(0) | \pi(\bm q) \rangl |^2   e^{-a(E_\pi-m_\pi) t}.
\label{eq:Q2} 
\eeq
We see that the elastic piece in the four-point function has information on the form factor of the pion through the amplitude squared.
The form factor  $F_\pi$   can be determined from $Q_{44}$ at large time separations,
\beq
Q^{elas}_{44}(\bm q,t)= {(E_\pi+m_\pi)^2\over 4 E_\pi m_\pi} F^2_\pi(\bm q^2) \, e^{-a(E_\pi-m_\pi) t}.
\label{eq:Q44elas}
\eeq
The charge radius $ \langl  r_E^2\rangl $ in the formula can then be extracted from  $F_\pi$.
A salient feature here is that the elastic contribution in four-point functions is positive definite.

Aside from the charge radius term in Eq.~\eqref{eq:alpha}, $\alpha_E$ is proportional to the difference in the areas under the $Q_{44}$ and  $Q^{elas}_{44}$  curves. 
It is this difference that is responsible for the sign of $\alpha^\pi_E$.
On a finite lattice the time integral does not really extend to $\infty$, but are limited to the available 
time slices between the two current insertions. In practice, one should check if the largest time separation is enough to establish the elastic limit.
Equivalent directions for $\bm q$ can be used to improve the signal-to-noise ratio.
Note that  $\alpha_E$ has the expected physical unit of $a^3$ (fm$^3$) since $1/\bm q^{\,2}$ scales like $a^2$, the integral scales like $a$, and  $Q_{44}$ is dimensionless in our notation.

\vspace*{5mm}
\section{Correlation functions}
\label{sec:cfun}
In this section, we detail how to simulate Eq.\eqref{eq:P1} and its Fourier transform Eq.\eqref{eq:Qmn} at the quark level.
Wick contractions of quark-antiquark pairs in the unsubtracted part lead to topologically 
distinct quark-line diagrams shown in Fig.~\ref{fig:4pt}.
The raw correlation functions can be found  in Appendix~\ref{sec:wick}.

Diagrams a, b, and c are connected. Diagram d has a loop that is disconnected from the hadron, but connected between the two currents. Diagram e has one disconnected loop and diagram f has two such loops. 
Furthermore, diagrams d, e and f must have associated VEV subtracted.
However, if conserved lattice current density is used, there is no need for subtraction in diagram e
since the VEV vanishes in the configuration average~\cite{Draper:1988bp}.
In this work, we focus on the connected contributions (diagrams a,b,c). The disconnected contributions (diagrams d,e,f) are more challenging and are left for future work.
In particular, we will explain how to use the sequential source technique (SST) to simplify the evaluations.

\begin{widetext}
\subsection{Two-point functions}
\label{sec:2pt}
First, we show how to evaluate the two-point function in Eq.\eqref{eq:P1} which serves as normalization for the four-point functions. 
It has the following Wick contraction using the interpolating operator in Eq.\eqref{eq:op}, 
%
\beqs
& \sum_{\bm x_3,\bm x_0}  \langl \Omega | \psi (x_3) \psi^\dagger (x_0) |\Omega \rangl \\&
 =\sum_{\bm x_3,\bm x_0}  \Tr_{s,c} \bigg[ \gamma_5 S_d(x_0,x_3) \gamma_5  S_u(x_3,x_0) \bigg],
 \label{eq:2pt}
 \eeqs
 where $S_q$ denotes a quark propagator that carries the full space-time and spin and color information between two points~\footnote{In this work, all correlation functions in such expressions are understood as path integral expectation values in lattice QCD. They are evaluated as averages over gauge configurations in Monte Carlo simulations.}. The double sum projects to zero momentum both at the source $x_0$ and the sink $x_3$ as required by the special kinematics. The full evaluation involves essentially all-to-all propagation which is computationally prohibitive.
Instead, we employ wall sources without gauge fixing as an approximation, with the expectation that gauge-dependent contributions to  the final observables will vanish in the configuration average~\cite{WallSource1993,Fu_2012}.
Only terms in the double sum where the quarks are at the same location form the signal, the rest contribute to noise.
Details of our implementation of the wall source can be found in Appendix~\ref{sec:wall}.

If we insert the wall at time slice $t_0$ and project to zero momentum at $t_3$ in Eq.\eqref{eq:2pt}, we have
\beqs
  \sum_{\bm x_3,\bm x_0} \langl \Omega | \psi (x_3) \psi^\dagger (x_0) |\Omega\rangl & 
  =  \Tr_{s,c} \bigg[ {\cal W}^T \gamma_5 S_d(x_0,x_3) \gamma_5  S_u(x_3,x_0)  {\cal W}  \bigg] \\&
  =\Tr_{s,c} \bigg[ {\cal W}^T  \gamma_5 P(t_0)  M_d^{-1} P(t_3)^T \;\gamma_5  P(t_3) M_u^{-1} P(t_0)^T {\cal W}  \bigg].
\eeqs
The symbols $\cal{W}$ and $P(t)$ are defined in Appendix~\ref{sec:wall}.
We introduce two zero-momentum quark propagators called $a1$ and $a2$ emanating from the walls at $t_0$ and $t_3$, respectively,
\beq
V_{a1}^{(q)} \equiv  M_q^{-1} P(t_0)^T {\cal W}, \quad
V_{a2}^{(q)} \equiv  M_q^{-1} P(t_3)^T {\cal W}.
\label{eq:a1a2}
\eeq
We use ``V" to emphasize that the wall-to-point quark propagators so defined are column vectors in the $(\bm x,s,c)$ space. 
 Using ${a1}$, the two-point function can be written as,
\beqs
 \sum_{\bm x_3,\bm x_0} \langl \Omega | \psi(x_3) \psi^\dagger (x_0) |\Omega\rangl 
  =
\Tr_{s,c} \Bigg[ \bigg(P(t_3)\gamma_5 V_{a1}^{(d)}\bigg)^\dagger \bigg(P(t_3) \gamma_5 V_{a1}^{(u)}\bigg) \Bigg] 
  =
\Tr_{s,c} \Bigg[ \bigg(P(t_3)V_{a1}^{(d)}\bigg)^\dagger \bigg(P(t_3)V_{a1}^{(u)}\bigg) \Bigg] \text{\bf (Type 1)} 
\label{eq:2pt1}
\eeqs
In the last step the $\gamma_5$-hermiticity of $M_q^{-1}$ is used to eliminate $\gamma_5$.
Similarly, if we insert the wall at time slice $t_3$ and project to zero momentum at $t_0$, we get in terms of ${a2}$,
\beqs
\sum_{\bm x_3,\bm x_0} \langl \Omega | \psi (x_3) \psi^\dagger (x_0) |\Omega\rangl  
   =
\Tr_{s,c} \Bigg[ \bigg(P(t_0)\gamma_5 V_{a2}^{(u)}\bigg)^\dagger \bigg(P(t_0) \gamma_5 V_{a2}^{(d)}\bigg) \Bigg] 
= \Tr_{s,c} \Bigg[  \bigg(P(t_0)V_{a2}^{(u)}\bigg)^\dagger \bigg(P(t_0)V_{a2}^{(d)}\bigg) \Bigg] \text{\bf (Type 2)} 
\label{eq:2pt2}
\eeqs
If we insert two walls, one at $t_0$, one at $t_3$, we obtain additional expressions,
\beqs
&  \sum_{\bm x_3,\bm x_0} \langl \Omega | \psi (x_3) \psi^\dagger (x_0) |\Omega\rangl 
 = \Tr_{s,c} \bigg[ {\cal W}^T \gamma_5  S_d(x_0,x_3) {\cal W}{\cal W}^T \gamma_5  S_u(x_3,x_0)  {\cal W}  \bigg] \\ &
  =
\Tr_{s,c} \Bigg[ 
\bigg( {\cal W}^T P(t_3) V_{a1}^{(d)} \bigg)^\dagger 
\bigg({\cal W}^T P(t_3)  V_{a1}^{(u)}  \bigg) \Bigg]  
 =
\Tr_{s,c} \Bigg[ 
\bigg( {\cal W}^T P(t_0) V_{a2}^{(u)} \bigg)^\dagger 
\bigg({\cal W}^T P(t_0)  V_{a2}^{(d)}  \bigg) \Bigg] \text{\bf (Type 3)} 
\label{eq:2pt3}
\eeqs
The expressions in the above three equations (which we denote as Type 1, 2, 3 as indicated) are different estimators of the wall-to-wall two-point function with zero momentum for both initial and final pions. They are expected to approach the same value in the limit of infinite number of configurations.   In the following, we use our notation to evaluate the connected four-point functions in Fig.~\ref{fig:4pt}.

\subsection{Four-point functions}
We start with local (or point) current insertions of four-point functions  which have relatively  simple Wick contractions.
The results in this work will be based on conserved (or point-split) currents which
avoids the issue of computing the renormalization constant $Z_V$ for vector currents.  
Below we detail how to evaluate the  connected contributions using both local and conserved currents. 

\subsubsection{Diagram a (different flavor)}

There are two terms, $d_{4}$ and $d_{2}$ in Eq.\eqref{eq:dPC12}, that are contributing to the connected part of diagram a. They are characterized by the charge factor $q_uq_{\bar{d}}=2/9$. 
The two terms are related by a flavor permutation ($1 \leftrightarrow 2$ switch).
Under isospin symmetry in u and d quarks, the two terms have equal contributions.
Including the Fourier transforms and setting $\mu=4=\nu$ for electric polarizability, the correlation function can be written as\footnote{We use $Q_{\mu\nu}$ for normalized correlation functions as defined in Eq.\eqref{eq:Qmn} and Eq.\eqref{eq:P1}, and tilded $\tilde{Q}_{\mu\nu}$ for unnormalized, {\it i.e.}, without the denominator Eq.\eqref{eq:P1}.},
\fontsize{10}{10}
\beqs
\tilde{Q}^{(a, PC)}_{44}
&=-{4\over 9}Z_V^2\kappa^2
   \Tr_{s,c} \bigg[ \gamma_5S(t_0, t_2)\gamma_{4}e^{-i\bm q} 
  S(t_2, t_3)\gamma_5 S(t_3, t_1)\gamma_{4}e^{i\bm q}S(t_1, t_0)
\bigg].
\label{eq:cQ44a-raw}
\eeqs
We evaluate the correlation function by inserting two walls, one at $t_0$ and one at $t_3$, 
\beqs
\tilde{Q}^{(a, PC)}_{44}(\bm q,t_1,t_2)
&=-{4\over 9}Z_V^2\kappa^2 \Tr_{s,c} \bigg[ 
 {\cal W}^T\gamma_{5} S(t_0,t_2)e^{-i\bm q} \gamma_{4} S(t_2,t_3) {\cal W}
 {\cal W}^T \gamma_5 S(t_3,t_1)e^{i\bm q} \gamma_{4}S(t_1,t_0) {\cal W}
\bigg]\\
&=-{4\over 9}Z_V^2\kappa^2 \Tr_{s,c} \bigg[ 
 {\cal W}^T\gamma_{5} P(t_0) M_q^{-1} P(t_2)^T e^{-i\bm q} \gamma_{4} P(t_2) M_q^{-1} P(t_3)^T {\cal W} \\
& \qquad\qquad\qquad {\cal W}^T \gamma_5 P(t_3) M_q^{-1} P(t_1)^T e^{i\bm q} \gamma_{4} P(t_1) M_q^{-1} P(t_0)^T {\cal W}
\bigg].
\label{eq:cQ44aW}
\eeqs
The notation makes it clear that all spatial sums are automatically incorporated into the matrix multiplications.
Using the $V_{a1}$ and $V_{a2}$ propagators defined in Eq.\eqref{eq:a1a2} and the $\gamma_5$-hermiticity of $M^{-1}$, the final expression for diagram a can be written as,
\beq
\tilde{Q}^{(a, PC)}_{44}(\bm q,t_1,t_2)
={4\over 9}Z_V^2\kappa^2  \Tr_{s,c} \Bigg[ \bigg(
 \left[P(t_2)V_{a2}\right]^\dagger \gamma_5 \gamma_{4} e^{i\bm q}  P(t_2) V_{a1}  
\bigg)^\dagger \bigg(
 \left[P(t_1)V_{a2}\right]^\dagger \gamma_5 \gamma_{4} e^{i\bm q}  P(t_1) V_{a1}  
\bigg) \Bigg].
\label{eq:Q44aPC}
\eeq
There is an overall sign change from taking the dagger. The first parenthesis  corresponds to the current insertion at $t_2$ on one of the quark lines in the pion; the second parenthesis  the current insertion at $t_1$ on the other quark line. Both $t_1$ and $t_2$ are free to vary between $t_0$ and $t_3$.

In the case of conserved current,
there are 8 terms contributing to diagram a in Eq.\eqref{eq:dPS1}.
Their sum under isospin symmetry, along with the Fourier transforms and wall-source insertions,  can be written in similar form,
\fontsize{9}{9}
\beqs 
\begin{aligned}
& \tilde{Q}^{(a,PS)}_{44}(\bm q,t_1,t_2) = {1\over 9} \kappa^2
   \big( d_{16}+d_{18}+d_{20}+d_{22}+d_{8}+d_{10}+d_{12}+d_{14} \big) \\&
={4\over 9} \kappa^2 \Tr_{s,c} \Bigg[  \bigg(
 \left[P(t_2)V_{a2}\right]^\dagger \gamma_5(1-\gamma_{4})e^{i\bm q}  U_{4}(t_2,t_2+1) P(t_2+1) V_{a1}
-\left[P(t_2+1)V_{a2}\right]^\dagger \gamma_5(1+\gamma_{4}) U^\dagger_{4}(t_2+1,t_2) e^{i\bm q} P(t_2) V_{a1}
\bigg)^\dagger \\ 
&\qquad\qquad  \bigg(
 \left[P(t_1)V_{a2}\right]^\dagger \gamma_5(1-\gamma_{4})e^{i\bm q}  U_{4}(t_1,t_1+1) P(t_1+1) V_{a1}
-\left[P(t_1+1)V_{a2}\right]^\dagger \gamma_5(1+\gamma_{4}) U^\dagger_{4}(t_1+1,t_1) e^{i\bm q} P(t_1) V_{a1}
\bigg) \Bigg],
\label{eq:Q44aPS}
\end{aligned} 
\eeqs
with local current replaced by its point-split form in the parentheses.
\comment{
The corresponding expressions for magnetic polarizability can be obtained by taking the $\mu=1=\nu$ components,
\beq
\tilde{Q}^{(a, PC)}_{11}(\bm q,t_1,t_2)
={4\over 9}Z_V^2\kappa^2 \Tr_{s,c} \Bigg[ \bigg(
 \left[P(t_2)V_{a2}\right]^\dagger \gamma_5 \gamma_{1} e^{i\bm q}  P(t_2) V_{a1}  
\bigg)^\dagger \bigg(
 \left[P(t_1)V_{a2}\right]^\dagger \gamma_5 \gamma_{1} e^{i\bm q}  P(t_1) V_{a1}  
\bigg) \Bigg],
\eeq
for point current and
\fontsize{9}{9}
\beqs\boxed{ \begin{aligned}
& \tilde{Q}^{(a,PS)}_{11}(\bm q,t_1,t_2)
={4\over 9} \kappa^2 \Tr_{s,c} \Bigg[ \\ & \bigg(
 \left[P(t_2)V_{a2}\right]^\dagger \gamma_5(1-\gamma_{1})e^{i\bm q}  U_{1}(t_2,t_2)  P(t_2) V_{a1}
-\left[P(t_2)V_{a2}\right]^\dagger \gamma_5(1+\gamma_{1}) U^\dagger_{1}(t_2,t_2) e^{i\bm q}  P(t_2) V_{a1}
\bigg)^\dagger \\ 
& \bigg(
 \left[P(t_1)V_{a2}\right]^\dagger \gamma_5(1-\gamma_{1})e^{i\bm q}  U_{1}(t_1,t_1)  P(t_1) V_{a1}
-\left[P(t_1)V_{a2}\right]^\dagger \gamma_5(1+\gamma_{1}) U^\dagger_{1}(t_1,t_1) e^{i\bm q} P(t_1) V_{a1}
\bigg) \Bigg],
\end{aligned} } \eeqs
for conserved current. Note that $U_1$ shifts the field  in the x direction as defined in Eq.\eqref{eq:U}. Consequently, $U_1$ and $U_1^\dagger$ do not commute with $e^{i\bm q}$, unlike $U_4$ and $U_4^\dagger$ in Eq.\eqref{eq:Q44aPS}. 
}

\subsubsection{Diagram b (same flavor) and SST}
For local current, there are 2 terms, $d_{1}$ and $d_{7}$ in Eq.\eqref{eq:dPC12}, that are contributing to the connected part of same-flavor correlations. They are characterized by the charge factors $q_uq_u=4/9$ or $q_{\bar{d}}q_{\bar{d}}=1/9$.
The $d_{1}$ diagram is clock-wise propagation $t_0\to t_3 \to t_2 \to t_1 \to t_0$ where the two currents couple to the same u quark, while the $d_{7}$ diagram is counter clock-wise propagation $t_0\to t_1 \to t_2 \to t_3 \to t_0$ where the two currents couple to the same d quark.  Under isospin symmetry, the total contribution from uu and dd correlations has a total charge factor of $4/9+1/9=5/9$.

Including the Fourier transforms, setting $\mu=4=\nu$ for electric polarizability, and inserting the wall sources, the correlation function can be written as,
\beqs
\tilde{Q}^{(b,PC)}_{44} &=
{5\over 9}Z_V^2\kappa^2  \Tr_{s,c} \bigg[ 
{\cal W}^T\gamma_{5} S(t_0,t_1)\gamma_{4} e^{i\bm q} S(t_1,t_2)  \gamma_4 e^{-i\bm q} S(t_2,t_3){\cal W}{\cal W}^T\gamma_{5} S(t_3,t_0) {\cal W}
\bigg].
\label{eq:Q44bW}
\eeqs
This expression involves numerous quark propagators:  $t_0$ and $t_3$ are fixed, but $t_1$ and $t_2$ are free to vary. To cut down the computational cost, we fix the current at $t_1$. Then only one new inversion between $t_1$ and $t_2$ is required. Since the current insertions take place between the hadron source ($t_0$) and sink ($t_3$),
a method called SST (Sequential Source Technique) can be employed for the propagator.
To see how SST arises in this context,  we first define the product that involves $t_0\rightarrow t_3\rightarrow t_2$ propagation as,
\beqs
 \gamma_4 e^{-i\bm q} S(t_2,t_3) {\cal W}{\cal W}^T \gamma_5 S(t_3,t_0) {\cal W} 
 &= \gamma_4 e^{-i\bm q} P(t_2) M_q^{-1} P(t_3)^T {\cal W}{\cal W}^T \gamma_5 P(t_3) M_q^{-1} P(t_0)^T {\cal W} \\
& = \gamma_4 e^{-i\bm q} P(t_2) V_{a2} {\cal W}^T P(t_3) \gamma_5  V_{a1},
\label{eq:a3b}
\eeqs
which is built directly from the two previously-computed propagators $V_{a1}$ and $V_{a2}$ along with other factors. This does not require a new inversion.  
Next, we define the rest in Eq.\eqref{eq:Q44bW} as,
\beqs
   {\cal W}^T\gamma_{5} S(t_0,t_1)\gamma_{4} e^{i\bm q} S(t_1,t_2) 
 &=  {\cal W}^T \gamma_{5}  P(t_0) M_q^{-1} P(t_1)^T \gamma_{4} e^{i\bm q} P(t_1) M_q^{-1} P(t_2)^T \\
 &= \bigg[ P(t_2) \gamma_{5}M_q^{-1}\gamma_{5}P(t_1)^T \gamma_{4} e^{-i\bm q}
 P(t_1) \gamma_{5}M_q^{-1}\gamma_{5}P(t_0)^T \gamma_{5}  {\cal W} \bigg]^\dagger\\
 &= - \bigg[ P(t_2) \gamma_{5}  M_q^{-1} P(t_1)^T \gamma_{4} e^{-i\bm q}
 P(t_1) V_{a1} \bigg]^\dagger = - \bigg[ P(t_2) \gamma_{5} V_{a3}^{(4,PC)} \bigg]^\dagger,
\eeqs
where we have introduced a  SST propagator called $a3$ (specialized to $\mu=4$ here), 
\beq 
V_{a3}^{(\mu,PC)}(\bm q) \equiv 
 M_q^{-1}P(t_1)^T \big[ \gamma_{\mu} e^{-i\bm q}  P(t_1) V_{a1} \big].
\label{eq:Va3PC}
\eeq
This expression indicates that $V_{a3}^{(4,PC)}$ can be obtained by a standard inversion $M x=b$ with a ``spatially extended source" $b=\big[ \gamma_{4} e^{-i\bm q}  P(t_1) V_{a1} \big]$ at $t_1$. This source is constructed from a previously defined quark propagator $V_{a1}$ and the current insertion, hence the name ``sequential source". 
Using $(a1, a2)$ and the newly-defined propagator $a3$, the final expression for diagram b takes the form,
\beq 
\tilde{Q}^{(b, PC)}_{44}(\bm q,t_2)
=-{5\over 9}Z_V^2\kappa^2 \Tr_{s,c} \Bigg[ 
\left[P(t_2)\gamma_{5} V_{a3}^{(4,PC)}(\bm q)\right]^\dagger  \gamma_4 e^{-i\bm q} P(t_2) V_{a2} {\cal W}^T P(t_3) \gamma_5  V_{a1} \Bigg].
\label{eq:Q44bPC}
\eeq
Fig.~\ref{fig:diagram-b} is a schematic depiction of how the propagators form the full correlation function  in Eq.\eqref{eq:Q44bPC}.
\begin{figure}[ht]
\includegraphics[scale=0.6]{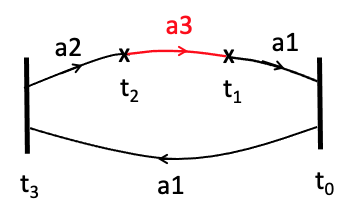}
\caption{Diagram (b) in terms of quark propagators: one part is $V_{a1}$ to the pion wall at $t_3$, then $V_{a2}$ to the current insertion at $t_2$; the other is a SST propagator $V_{a3}$ (red) built from $V_{a1}$ and the current insertion at $t_1$.}
\label{fig:diagram-b}
\end{figure}

For conserved current, there are 8 terms contributing to diagram b in Eq.\eqref{eq:dPS1}.
\comment{
Their sum under isospin symmetry, along with the Fourier transforms and wall-source insertions,  can be written in the computationally convenient form, 
\fontsize{9}{9}
\beqs
& \tilde{Q}^{(b,PS)}_{44}(\bm q,t_2) = {1\over 9} \kappa^2
   \big( d_{1}+d_{3}+d_{5}+d_{7}+d_{25}+d_{31}+d_{37}+d_{43} \big) \\
&= {5\over 9} \kappa^2
    \Tr_{s,c}  \Bigg[ {\cal W}^T \gamma_5  \bigg(
 S(t_0,t_1)(1-\gamma_{4}) e^{i\bm q}  U_{4}(t_1,t_1+1) S(t_1+1,t_2)   
- S(t_0,t_1+1)(1+\gamma_{4})  U^\dagger_{4}(t_1+1,t_1)  e^{i\bm q}S(t_1,t_2) 
\bigg)   \\  &
(1-\gamma_{4}) e^{-i\bm q}  U_{4}(t_2,t_2+1) S(t_2+1,t_3) {\cal W}{\cal W}^T\gamma_5 S(t_3,t_0) {\cal W} \\&
-  {\cal W}^T \gamma_5 \bigg(
 S(t_0,t_1)(1-\gamma_{4})  e^{i\bm q} U_{4}(t_1,t_1+1)  S(t_1+1,t_2+1)   
- S(t_0,t_1+1)(1+\gamma_{4}) U^\dagger_{4}(t_1+1,t_1)  e^{i\bm q}S(t_1,t_2+1) \bigg) \\&
(1+\gamma_{4})  U^\dagger_{4}(t_2+1,t_2)e^{-i\bm q} S(t_2,t_3) {\cal W}{\cal W}^T \gamma_5 S(t_3,t_0)
\bigg) {\cal W} \Bigg].
\label{eq:Q44bW}
\eeqs
The $(\cdots)$ has the current split at $t_1$ and momentum factor $e^{i\bm q}$, while the difference of the two ${\cal W}^T\cdots {\cal W}$ terms has the current split at $t_2$ and momentum factor $e^{-i\bm q}$.
}
Following the same procedure as for point current, the final expression for diagram b from point-split current can be written as,
\normalsize
\beqs  
\begin{aligned}
 \tilde{Q}^{(b,PS)}_{44}(\bm q,t_2)& = {1\over 9} \kappa^2
   \big( d_{1}+d_{3}+d_{5}+d_{7}+d_{25}+d_{31}+d_{37}+d_{43} \big) \\
  & =-{5\over 9} \kappa^2 \Tr_{s,c} \Bigg[
  [P(t_2)\gamma_5 V_{a3}^{(4,PS)}({\bf q})]^\dagger (1-\gamma_{4})  e^{-i\bm q}  U_{4}(t_2,t_2+1) P(t_2+1)  V_{a2} {\cal W}^T P(t_3)\gamma_5 V_{a1}  \\& \qquad \qquad
  -[P(t_2+1)\gamma_5 V_{a3}^{(4,PS)}({\bf q})]^\dagger (1+\gamma_{4})  U^\dagger_{4}(t_2+1,t_2) e^{-i\bm q} P(t_2) V_{a2} {\cal W}^T P(t_3)\gamma_5 V_{a1}  \Bigg],
\label{eq:Q44b}
\end{aligned} 
\eeqs
where a new inversion is needed for the SST propagator,
\fontsize{9}{9}
\beq 
\begin{aligned}
V_{a3}^{(4, PS)}(\bm q) \equiv M_q^{-1} \bigg[
P^T(t_1)(1-\gamma_{4})   e^{-i\bm q} U_{4}(t_1,t_1+1) P(t_1+1)  V_{a1} 
-   P^T(t_1+1) (1+\gamma_{4})  U^\dagger_{4}(t_1+1,t_1)e^{-i\bm q}  P(t_1) V_{a1} 
   \bigg].
\label{eq:Va3PS}
\end{aligned} 
\eeq
This is the point-split version of  Eq.\eqref{eq:Va3PC} with $\mu=4$.
Since the current is  split in the $t$ direction, $U_4$ and $U_4^\dagger$ commute with $e^{-i\bm q}$  in these two equations.

\comment{
The corresponding expressions for magnetic polarizability are obtained from the $\mu=1$ components. For point current,
\normalsize
\beqs
\begin{aligned}
\tilde{Q}^{(b, PC)}_{11}(\bm q,t_2)
&=-{5\over 9}Z_V^2\kappa^2  \Tr_{s,c} \Bigg[ 
\left[P(t_2)\gamma_5 V_{a3}^{(1,PC)}(\bm q)\right]^\dagger  \gamma_1 e^{-i\bm q} P(t_2) V_{a2} {\cal W}^T P(t_3) \gamma_5  V_{a1} \Bigg], \\
\end{aligned} 
\eeqs
where $V_{a3}^{(1,PC)}$ is given in Eq.\eqref{eq:Va3PC} by setting $\mu=1$.

For conserved current, 
\fontsize{9}{9}
\beqs  \boxed{
 \tilde{Q}^{(b, PS)}_{11}(\bm q,t_2)
   =-{5\over 9} \kappa^2 \Tr_{s,c} \Bigg[
  \big[P(t_2)\gamma_5 V_{a3}^{(1,PS)}({\bf q})\big]^\dagger \big[ (1-\gamma_{1}) e^{-i\bm q}  U_{1}(t_2,t_2) 
  -(1+\gamma_{1})  U^\dagger_{1}(t_2,t_2)e^{-i\bm q} \big] P(t_2) V_{a2} {\cal W}^T P(t_3)V_{a1}  \Bigg],
\label{eq:Q11b}
} \eeqs
where a new inversion is needed for the SST propagator, 
\fontsize{9}{9}
\beq \boxed{
V_{a3}^{(1, PS)}(\bm q) \equiv M_q^{-1} \bigg[
   P^T(t_1) (1+\gamma_{1}) e^{-i\bm q} U_{1}(t_1,t_1) P(t_1) V_{a1} 
   -P^T(t_1)(1-\gamma_{1}) U^\dagger_{1}(t_1,t_1)e^{-i\bm q}   P(t_1) V_{a1} \bigg]  .
} \eeq
Since the current is split in the $x$ direction, $U_1$ and $U_1^\dagger$ do not commute with $e^{- i\bm q}$. 
}

\subsubsection{Diagram c (same flavor Z-graph) and SST}


For local current, there are 2 terms, $d_{0}$ and $d_{9}$ in Eq.\eqref{eq:dPC12}, that are contributing to the connected part of same-flavor correlations. They are characterized by the same charge factors $q_uq_u=4/9$ or $q_{\bar{d}}q_{\bar{d}}=1/9$.
The $d_{0}$ diagram is a clock-wise propagation $t_0\to t_3 \to t_1 \to t_2 \to t_0$ where the two currents couple to the u quark, while the $d_{9}$ diagram is a counter clock-wise propagation $t_0\to t_2 \to t_1 \to t_3 \to t_0$ where the two currents couple to the d quark. They are essentially the Z-graph of diagram b with the current insertions 1 and 2 switched, whose correlation function can be written as,
\beqs
\tilde{Q}^{(c,PC)}_{44} &
={5\over 9}Z_V^2\kappa^2  \Tr_{s,c} \bigg[ \gamma_{5}
{\cal W}^TS(t_0,t_2)\gamma_{4} e^{-i\bm q} S(t_2,t_1) 
\gamma_4 e^{i\bm q}  S(t_1,t_3){\cal W} {\cal W}^T\gamma_{5} S(t_3,t_0)  {\cal W} 
\bigg].
\label{eq:Q44cPSraw}
\eeqs
First we isolate the $t_3 \to t_1 \to t_2$ propagation,
\beqs
 S(t_2,t_1) \gamma_4 e^{i\bm q} S(t_1,t_3) {\cal W}
 &=P(t_2) M_q^{-1} P(t_1)^T \gamma_4 e^{i\bm q} P(t_1) M_q^{-1} P(t_3)^T {\cal W} \\
& = P(t_2) M_q^{-1} P(t_1)^T \gamma_4 e^{i\bm q} P(t_1) V_{a2} 
\equiv P(t_2) V^{(4,PC)}_{a4}(\bm q), 
\eeqs
where a new  SST propagator is introduced (specialized to $\mu=4$ here),
\beq 
V_{a4}^{(\mu,PC)}(\bm q) \equiv 
 M_q^{-1}P(t_1)^T \big[ \gamma_{\mu} e^{i\bm q}  P(t_1) V_{a2}  \big]. 
\label{eq:Va4PC}
\eeq
Using $a1$ and $a4$, the final expression for diagram c using point current takes the  form,
\beq 
\tilde{Q}^{(c, PC)}_{44}(\bm q,t_2)
={5\over 9}Z_V^2\kappa^2 \Tr_{s,c} \Bigg[ 
\left[\gamma_4 e^{i\bm q}  P(t_2)\gamma_5 V_{a1}\right]^\dagger  P(t_2) V^{(4,PC)}_{a4}(\bm q) {\cal W}^T P(t_3) \gamma_5  V_{a1} \Bigg].
\label{eq:cQ44b}
\eeq
Fig.~\ref{fig:diagram-c} is a schematic depiction of how the propagators form this correlation function.
\begin{figure}[ht]
\includegraphics[scale=0.6]{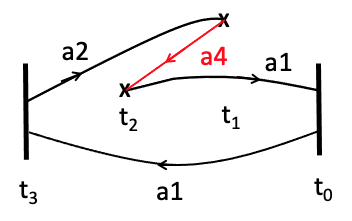}
\caption{
Diagram (c) in terms of quark propagators: ${a1}$ from $t_0$ to $t_3$, SST quark propagator ${a4}$ (red) with sequential source built from $a2$ and current insertion at $t_1$, and $a1$ from $t_2$ to $t_0$.  This is the Z-graph of diagram b.}
\label{fig:diagram-c}
\end{figure}

For conserved current, there are 8 terms contributing to diagram c in Eq.\eqref{eq:dPS1}.
\comment{
Their sum under isospin symmetry, along with the Fourier transforms and wall-source insertions,  can be written in the computationally convenient form,
\fontsize{9}{9}
\beqs
& \tilde{Q}^{(c,PS)}_{44}(\bm q,t_2) = {1\over 9} \kappa^2
   \big( d_{0}+d_{2}+d_{4}+d_{6}+d_{27}+d_{33}+d_{39}+d_{45} \big)  \\
&  ={5\over 9} \kappa^2 \Tr_{s,c} \Bigg[ 
  {\cal W}^T \gamma_5
 S(t_0,t_2)(1-\gamma_{4})e^{-i\bm q} U_{4}(t_2,t_2+1)  
\bigg( S(t_2+1,t_1) 
(1-\gamma_{4})   e^{i\bm q}  U_{4}(t_1,t_1+1)S(t_1+1,t_3){\cal W}{\cal W}^T \gamma_5 S(t_3,t_0) \\&
- S(t_2+1,t_1+1) 
(1+\gamma_{4})   U^\dagger_{4}(t_1+1,t_1)e^{i\bm q}S(t_1,t_3) {\cal W}{\cal W}^T\gamma_5 S(t_3,t_0)
\bigg) {\cal W} \\ &
- {\cal W}^T \gamma_5 
S(t_0,t_2+1)(1+\gamma_{4})  U^\dagger_{4}(t_2+1,t_2) e^{-i\bm q}
 \bigg( 
S(t_2,t_1) 
(1-\gamma_{4})  e^{i\bm q} U_{4}(t_1,t_1+1) S(t_1+1,t_3) {\cal W}{\cal W}^T\gamma_5 S(t_3,t_0) \\&
-S(t_2,t_1+1) 
(1+\gamma_{4})  U^\dagger_{4}(t_1+1,t_1)e^{i\bm q}  S(t_1,t_3) {\cal W}{\cal W}^T\gamma_5 S(t_3,t_0)
\bigg) {\cal W} \Bigg].
\label{eq:Q44cW}
\eeqs
The $(\cdots)$ has the current split at $t_1$ and momentum factor $e^{i\bm q}$. The ${\cal W}^T\cdots {\cal W}$ difference has the current split at $t_2$ and momentum factor $e^{-i\bm q}$.
}
Following a similar procedure as for local current, the final expression for diagram (c) from point-split current can be written as,
\normalsize
\beqs 
\begin{aligned}
  \tilde{Q}^{(c,PS)}_{44}(\bm q,t_2) &= {1\over 9} \kappa^2
   \big( d_{0}+d_{2}+d_{4}+d_{6}+d_{27}+d_{33}+d_{39}+d_{45} \big)  \\
   ={5\over 9} \kappa^2\Tr_{s,c} \Bigg[ 
 &   [P(t_2)\gamma_5 V_{a1}]^\dagger (1-\gamma_{4})  e^{-i{\bf q}} U_{4}(t_2,t_2+1)  P(t_2+1)V_{a4}^{(4,PS)}({\bf q}) {\cal W}^T P(t_3) \gamma_5  V_{a1}\\ 
- &  [P(t_2+1)\gamma_5 V_{a1}]^\dagger (1+\gamma_{4}) U^\dagger_{4}(t_2+1,t_2)e^{-i{\bf q}} P(t_2)V_{a4}^{(4,PS)}({\bf q}) {\cal W}^T P(t_3) \gamma_5  V_{a1}\Bigg],
\label{eq:Q44c}
\end{aligned} 
\eeqs
where
\fontsize{9}{9}
\beqs 
\begin{aligned}
V_{a4}^{(4,PS)}(\bm q) \equiv 
M_q^{-1} \bigg[P^T(t_1)(1-\gamma_{4})  e^{i\bm q} U_{4}(t_1,t_1+1)  P(t_1+1) V_{a2}    -  P(t_1+1)^T (1+\gamma_{4})  U^\dagger_{4}(t_1+1,t_1) e^{i\bm q} P(t_1) V_{a2}  \bigg].
\label{eq:Va4}
\end{aligned} 
\eeqs
Compare to Eq.\eqref{eq:Va3PS} for diagram b, this expression has $a2$ instead of $a1$,  $\bm q$ instead of $-\bm q$.

\comment{
The corresponding expressions for magnetic polarizability are, 
for point current,
\beq 
\tilde{Q}^{(c, PC)}_{11}(\bm q,t_2)
={5\over 9}Z_V^2\kappa^2 \Tr_{s,c} \Bigg[ 
\left[\gamma_1 e^{i\bm q}  P(t_2)\gamma_5 V_{a1}\right]^\dagger  P(t_2) V^{(1,PC)}_{a4}(\bm q) {\cal W}^T P(t_3) \gamma_5  V_{a1}  \Bigg].
\label{eq:cQ44b}
\eeq
 where $V^{(1,PC)}_{a4}$
is Eq.\eqref{eq:Va4PC} specialized to $\mu=1$. 
For conserved current,
\normalsize
\beqs\boxed{ \begin{aligned}
 \tilde{Q}^{(c, PS)}_{11}(\bm q,t_2)
   ={5\over 9} \kappa^2 \Tr_{s,c} \Bigg[ 
 &   [P(t_2)\gamma_5 V_{a1}]^\dagger (1-\gamma_{1})  e^{-i{\bf q}} U_{1}(t_2,t_2)  P(t_2)V_{a4}^{(1,PS)}({\bf q}){\cal W}^T P(t_3) \gamma_5  V_{a1} \\ 
- &  [P(t_2+1)\gamma_5 V_{a1}]^\dagger (1+\gamma_{1}) U^\dagger_{1}(t_2,t_2)e^{-i{\bf q}} P(t_2)V_{a4}^{(1,PS)}({\bf q}){\cal W}^T P(t_3) \gamma_5  V_{a1} \Bigg],
\label{eq:Q11c}
\end{aligned} } \eeqs
where
\normalsize
\beqs\boxed{ \begin{aligned}
V_{a4}^{(1,PS)}(\bm q) \equiv 
M_q^{-1} \bigg[P^T(t_1)(1-\gamma_{1}) e^{i\bm q} U_{1}(t_1,t_1)  P(t_1) V_{a2}   
   -  P(t_1)^T (1+\gamma_{1})  U^\dagger_{1}(t_1,t_1) e^{i\bm q} P(t_1) V_{a2} \bigg].
\label{eq:Va4PS1}
\end{aligned} } \eeqs
The current split is in the $x$ direction implicit in the gauge links. Consequently, $U_1$ and $U_1^\dagger$ do not commute with $e^{\pm i\bm q}$.

Table~\ref{tab:props} gives a summary of all the quark propagators introduced and how they are used for each polarizability and connected diagram. 
}

\end{widetext}
The total connected contribution to the polarizabilities in Eq.\eqref{eq:alpha} 
is simply the sum of the individual normalized terms in Fig.~\ref{fig:4pt},
\beqs
Q_{44}(\bm q,t_2,t_1)&=Q^{(a)}_{44}+ Q^{(b)}_{44}+Q^{(c)}_{44},
\label{eq:Q44abc}
\eeqs
for either point current or conserved current. The charge factors and flavor-equivalent contributions have been included in each diagram.
 
\comment{

\begin{widetext}
\begin{eqnarray}
H &=& -\vec{p}\cdot \vec{E}-\vec{\mu}\cdot \vec{B}-{1\over 2}\alpha E^2 - {1\over 2}\beta B^2
 \nonumber \\ &&
  -\frac12\,\Big( \gamma_{E1} \vec\sigma \cdot \vec E \times \vec{\dot E}
   + \gamma_{M1} \vec\sigma \cdot \vec B \times \vec{\dot B}
   -2 \gamma_{E2} E_{ij}\sigma_i B_j
   +2 \gamma_{M2} B_{ij}\sigma_i E_j \Big)
 \nonumber \\ &&
  -\frac12\, (\alpha_{E\nu} \vec{\dot E}^2 + \beta_{M\nu} \vec{\dot B}^2)
    -\frac1{12}\, 4\pi (\alpha_{E2} E_{ij}^2 + \beta_{M2} B_{ij}^2) + \cdots.
\label{Hpolar}
\end{eqnarray}
\end{widetext}

\begin{table} [ht]
\caption{Breakdown of quark propagators required for each polarizability and connected diagram. The $(a1,a2)$ are wall-source propagators at pion source and sink, respectively. The $(a3,a4)$ are SST propagators that are built from $(a1,a2)$. Point current requires renormalization on the lattice, while point-split current is conserved.}
\label{tab:props}
\begin{tabular}{c}
$      
\renewcommand{\arraystretch}{1.5}
\begin{array}{cc | l | l}
\toprule
\text{Polarizability} & \text{diagram} & \text{Local current} & \text{Conserved current} \\
 \midrule
\alpha^\pi_E & (a)  & V_{a1}, V_{a2}  &  V_{a1}, V_{a2} \\
             & (b)  &  V_{a3}^{(4,PC)}(\bm q) &   V_{a3}^{(4,PS)}(\bm q) \\
             & (c)  & V_{a4}^{(4,PC)}(\bm q) &  V_{a4}^{(4,PS)}(\bm q) \\
\hline
\beta^\pi_M & (a)  & V_{a1}, V_{a2}  &  V_{a1}, V_{a2} \\
             & (b)  & V_{a3}^{(1,PC)}(\bm q) & V_{a3}^{(1,PS)}(\bm q) \\
             & (c)  &  V_{a4}^{(1,PC)}(\bm q) &   V_{a4}^{(1,PS)}(\bm q) \\
\bottomrule
\end{array}
$      
\end{tabular}
\end{table}
}

\comment{
Taken together the connected part of the electric polarizability $\alpha_E$ requires four propagators per configuration ($a1,a2,a3,a4$) for fixed $\bm q$.
If we consider several momenta in the correlation functions,  more inversions are needed. 
The physics payout is fairly decent, without the issue of acceleration or Landau levels for a charged pion in the background field method, 
\begin{enumerate}
\item
charge radius
\item
electric polarizability
\item
magnetic polarizability
\item
renormalization factor $Z_V$ if we also consider point current propagators.
Traditionally, this problem is solved by considering three-point functions. 
Here we can obtain $Z_V$ from four-point functions by considering the large time behavior of only one of the diagrams. 
This could become a separate publication, perhaps with the addition of $Z_A$.
These constants have not been determined for the nHYP action.
\end{enumerate}
Due to the large number of propagators and configurations involved, it is impractical to save quark propagators. 
We will compute them on the fly and only  save correlation functions  to disk.
}
\section{Simulation details and results}
\label{sec:results}
Having laid out the methodology and detailed the correlation functions, we now discuss how to numerically evaluate them in a Monte Carlo simulation in order to extract the polarizability. 
As a proof-of-principle test, we use quenched Wilson action with $\beta=6.0$ and $\kappa=0.1520,\;0.1543,\; 0.1555,\; 0.1565$ on the lattice $24^3\times 48$. The pion mass corresponding to the kappas will be determined in our simulation.
We analyzed 1000 configurations for each of the kappas.
The scale of this action has been determined in Ref.~\cite{CABASINO1991195}, with inverse lattice spacing $1/a=2.312$ GeV and kappa critical $\kappa_c=0.15708$.
It also gives the pion mass as a function of kappa,
\beq
(m_\pi a)^2 = 2.09 \times \frac{1}{2}\bigg(\frac{1}{\kappa} - \frac{1}{\kappa_c}\bigg),
\label{eq:scale}
\eeq
which will be compared with the measured $m_\pi$.
Dirichlet (or open) boundary condition is imposed in the time direction, while periodic boundary conditions are used in spatial dimensions.
The pion source is placed at $t_0=7$ and sink at $t_3=42$ (time is labeled from 1 to 48). One current is inserted at a fixed time $t_1$, while the other current $t_2$ is free to vary.
We use integers $\{n_x,n_y,n_z\}$ to label the discrete momentum on the lattice,
\beqs
 \bm q &=\big\{{2\pi n_x\over L_x}, {2\pi n_y\over L_y}, {2\pi n_z\over L_z} \big\}, \\& 
 \quad n_x,n_y,n_z =0, \pm 1, \pm 2, \cdots,
 \label{eq:q}
\eeqs
and consider five different combinations $\{0,0,0\},\,\{0,0,1\},\,\{0,1,1\},\, \{1,1,1\},\,\{0,0,2\}$.
In lattice units they correspond to  the values ${\bm q}^2a^2=0,\, 0.068,\, 0.137,\, 0.206,\, 0.274$, or in physical units to ${\bm q}^2=0,\, 0.366,\, 0.733,\, 1.100,\, 1.465$ (GeV$^2$).
In order to evaluate the connected diagrams, we need four inversions of the quark matrix with varying sources: two wall-sourced propagators $V_{a1}$ and $V_{a2}$, and two SST propagators $V_{a3}^{(4,PS)}(\bm q)$ and $V_{a4}^{(4,PS)}(\bm q)$ at a fixed $\bm q$. So the count for five momenta is $2+2\times 5= 12$ per kappa per configuration.  It takes longer to do the inversions for larger kappas due to critical slowing down. 

\subsection{Raw correlation functions}

\begin{figure}[th!]
\includegraphics[scale=0.4]{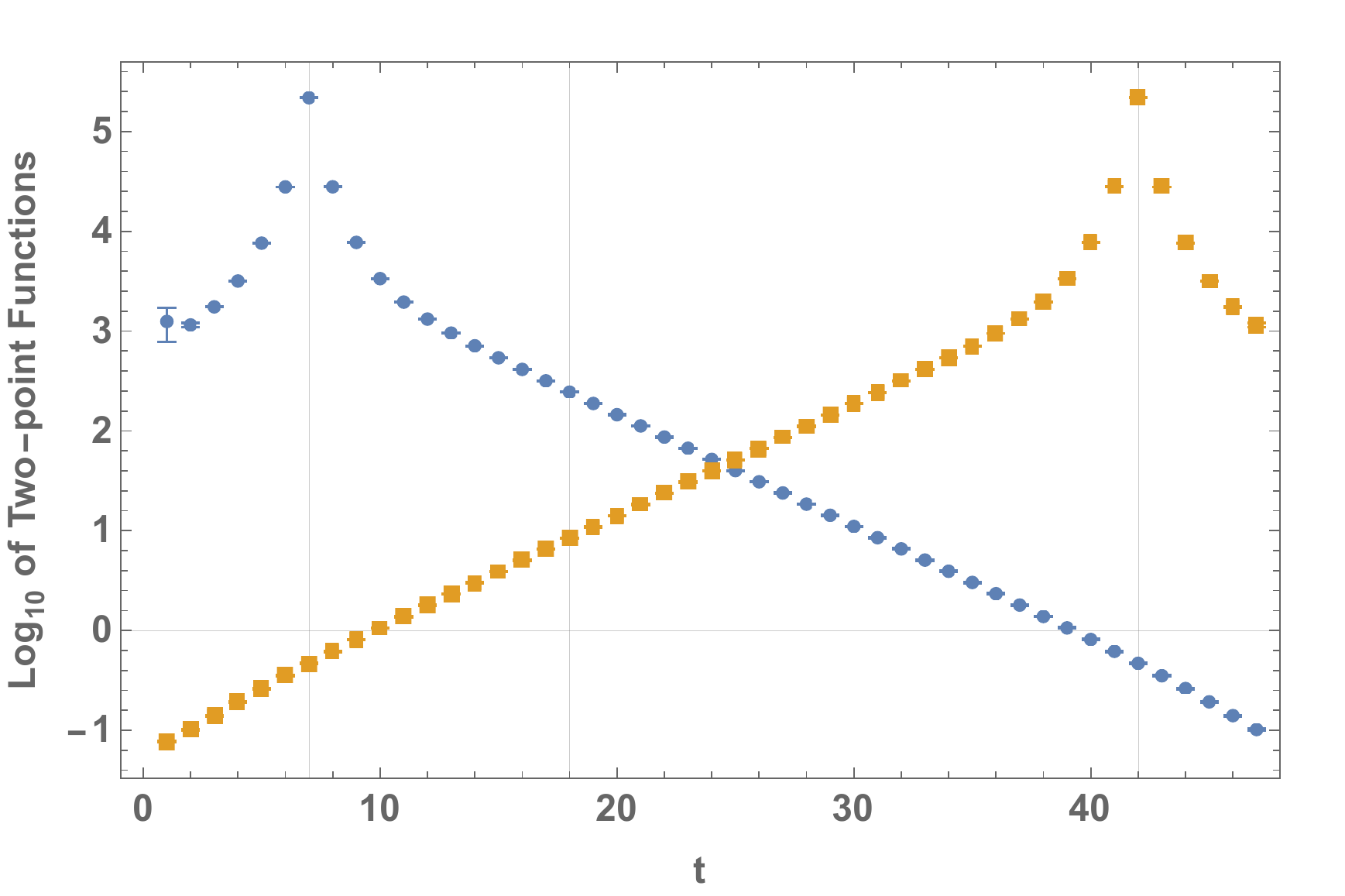}
\includegraphics[scale=0.4]{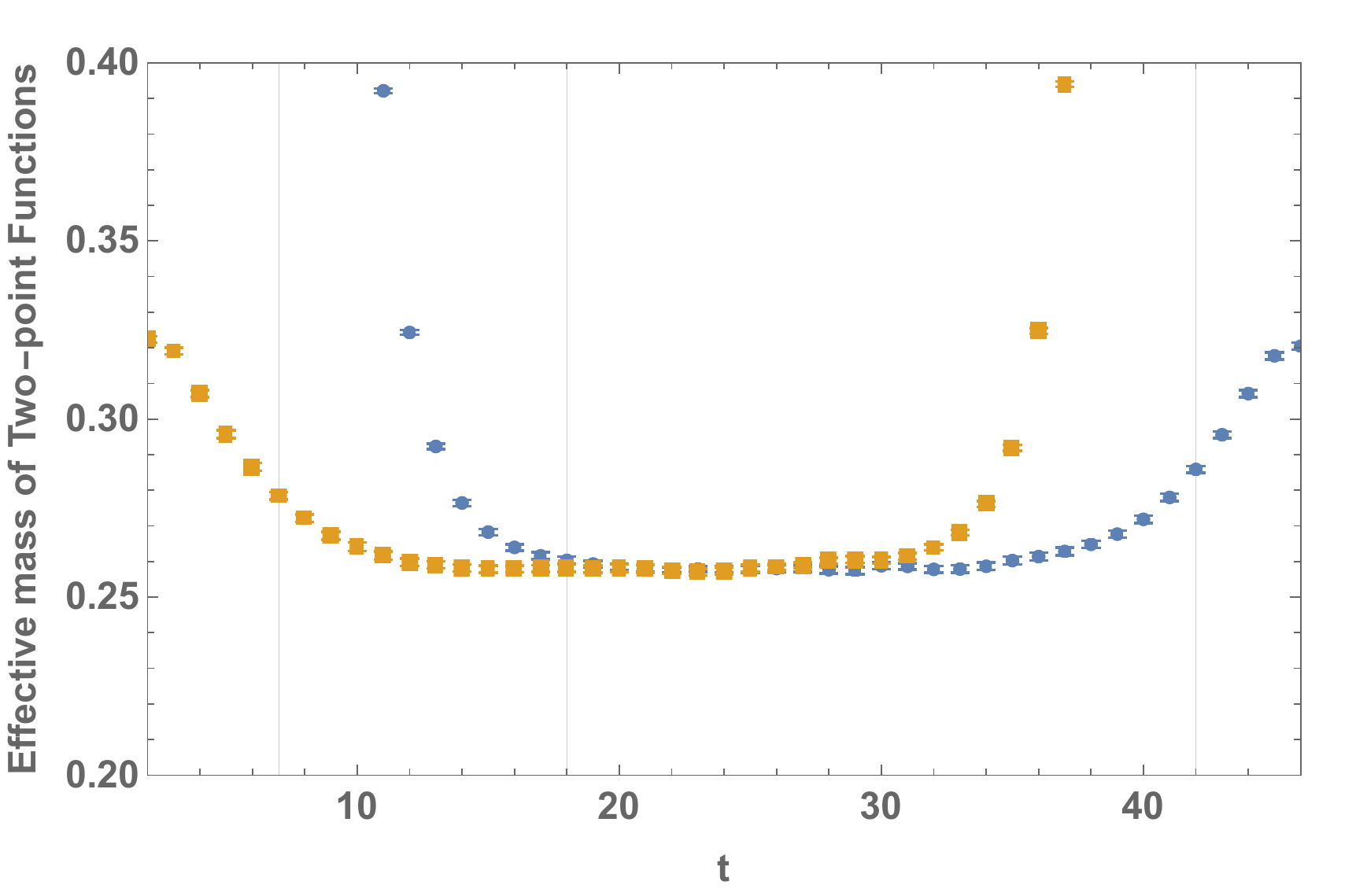}
\caption{
Moving sink zero-momentum pion correlator Type 1 (blue)  and Type 2 (orange)  and their effective mass functions at $m_\pi=600$ MeV.
They are constructed from either $a1$ or $a2$ quark propagators as explained in the text. 
The vertical gridlines indicate the three fixed time points in the study. 
These functions can be used to extract the pion mass in single-exponential fashion.
The value at $t_3=42$ in Type 1 or at $t_0=7$ in Type 2 can be used for normalization of four-point functions.
}
\label{fig:2pt}
\end{figure}
\begin{figure*}[ht!]
\includegraphics[scale=0.55]{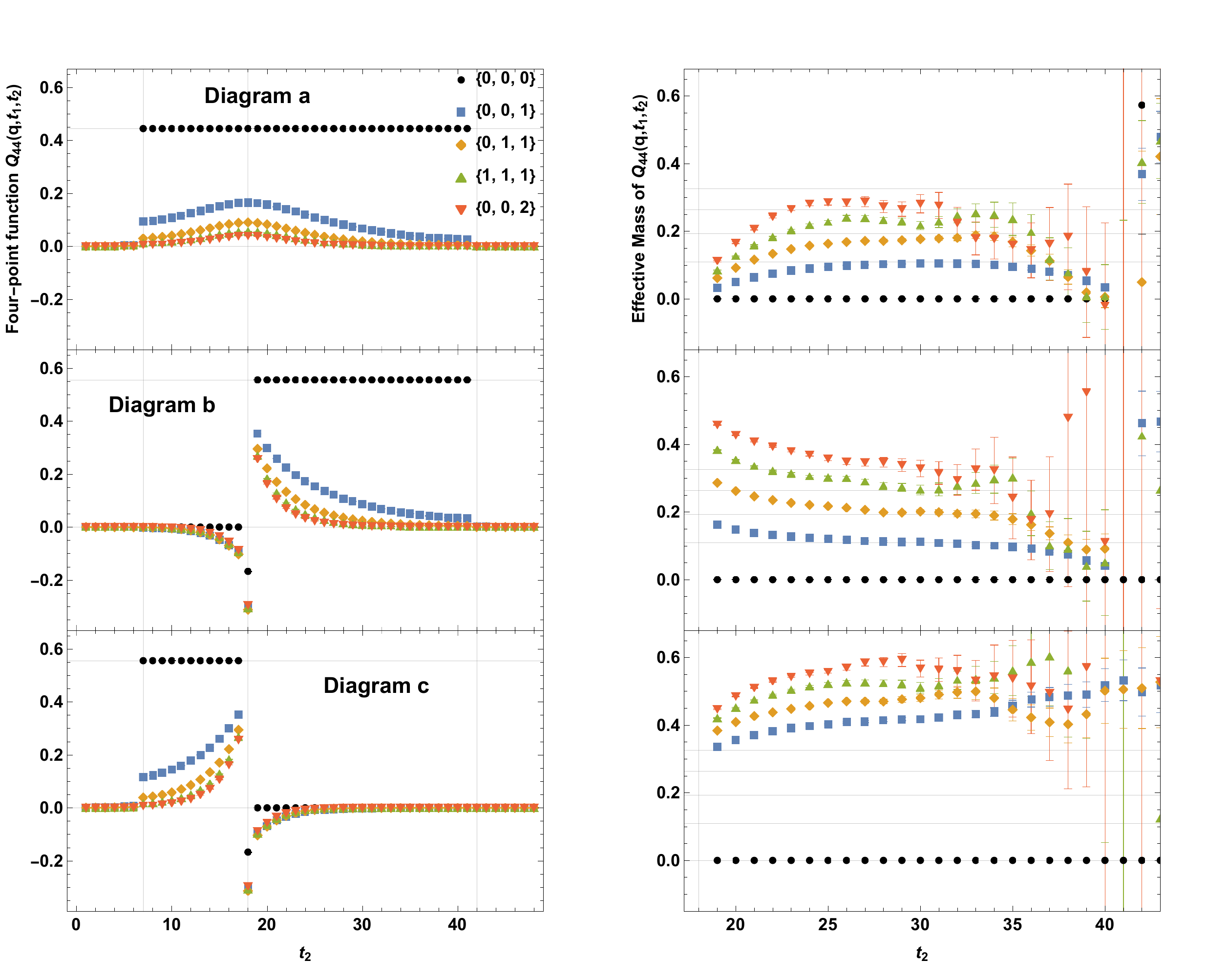}
\caption{
Normalized four-point functions (left panel) and their effective mass functions (right panel) from the connected diagrams as a function of current separation at $m_\pi=600$ MeV. 
The $\bm q=0$ results serve as a check of current conservation. 
The results for non-zero $\bm q$ between $t_2=18$ and $t_2=41$ will become the basis for our analysis.
The vertical gridlines indicate the pion walls ($t_0=7$ and $t_3=42$) and the  fixed current insertion ($t_1=18$).
The horizontal gridlines in the effective mass functions indicate the value of $E_\pi-m_\pi$ where the continuum dispersion relation  $E_\pi=\sqrt{\bm q^2+m_\pi^2}$ is used.
}
\label{fig:Q44PS}
\end{figure*}

First, we discuss how to determine pion mass from the  various two-point functions in Sec.~\ref{sec:2pt}.
In Fig.~\ref{fig:2pt} we show the wall-to-wall pion correlations based on  Eq.\eqref{eq:2pt1} (Type 1) and Eq.\eqref{eq:2pt2} (Type 2) at $\kappa=0.1555$. 
Type 1 only depends on the $a1$ quark propagator originating from the wall source at $t_0=7$. Instead of ending at fixed $t_3=42$, we allow it to vary in the entire range of $t$ on the lattice. One can visualize it as a moving wall sink. In this way, we get to observe a plateau in the effective mass function which we use to extract the mass.  Similarly, Type 2 only depends on the $a2$ quark propagator originating from the wall source at $t_3=42$. Instead of ending at fixed $t_0=7$, we allow it to vary in the entire range of $t$ on the lattice. We flip the sign of its effective mass function so a direct comparison of the plateaus for the two types can be made. We use Type 1 with a varying sink to extract pion and rho masses at the four kappa values. 
We obtain approximately 1100, 800, 600, and 370 MeV for pion mass at $\kappa=0.1520,\; 0.1543,\; 0.1555,\; 0.1565$, respectively.
These values agree well with those predicted from the relation in Eq.\eqref{eq:scale}. From this point on, we will refer to pion mass rather than kappa values.
The rho meson is considered in this work to judge the efficacy of vector meson dominance in charge radius extraction. More precise numbers  for $m_\pi$ and $m_\rho$ with uncertainties will be given in the summary table at the end (Table~\ref{tab:final}).
Another benefit of plotting the Type 1 and Type 2 correlators with a varying sink is we get to see the limited ``window of opportunity" in the effective mass where ground state dominates. This is the window in which we study the current-current correlations. We utilize this information to fix one of the two currents in the four-point function calculation so it mainly couple to the zero-momentum ground state.  Having examined the plots, we settle on $t_1=18,\,18,\, 18,\, 14$ for $m_\pi=1100,\,800,\, 600,\, 370$ MeV, respectively.

Next, we discuss normalization constant for four-point functions. This is the zero-momentum wall-to-wall two-point function in the denominator of  Eq.\eqref{eq:P1}.  We have three options, corresponding to the three types in Eq.\eqref{eq:2pt1}, Eq.\eqref{eq:2pt2}, and Eq.\eqref{eq:2pt3}.
Type 1 normalization constant is simply the special value at $t=t_3=42$ in the blue curve of Fig.~\ref{fig:2pt}, and
Type 2  the special value at $t=t_0=7$ in the orange curve of Fig.~\ref{fig:2pt}. Type 3 normalization constant is computed separately.
The three types are not expected to agree configuration by configuration since they originate from different wall sources, but they should approach the same value in the configuration average within statistics. We found the numerical values  $0.4683(6)$, $0.4672(6)$, $0.468(7)$, from Type1, Type2, and Type 3, respectively, at this pion mass.
We see that Type 3 has larger statistical uncertainties than in Type 1 and Type 2. This is expected since Type 3 is constructed from two wall sources, while the other two from one.
We will use Type 3 as normalization for the reason to be discussed below.

\begin{figure}[ht!]
\includegraphics[scale=0.4]{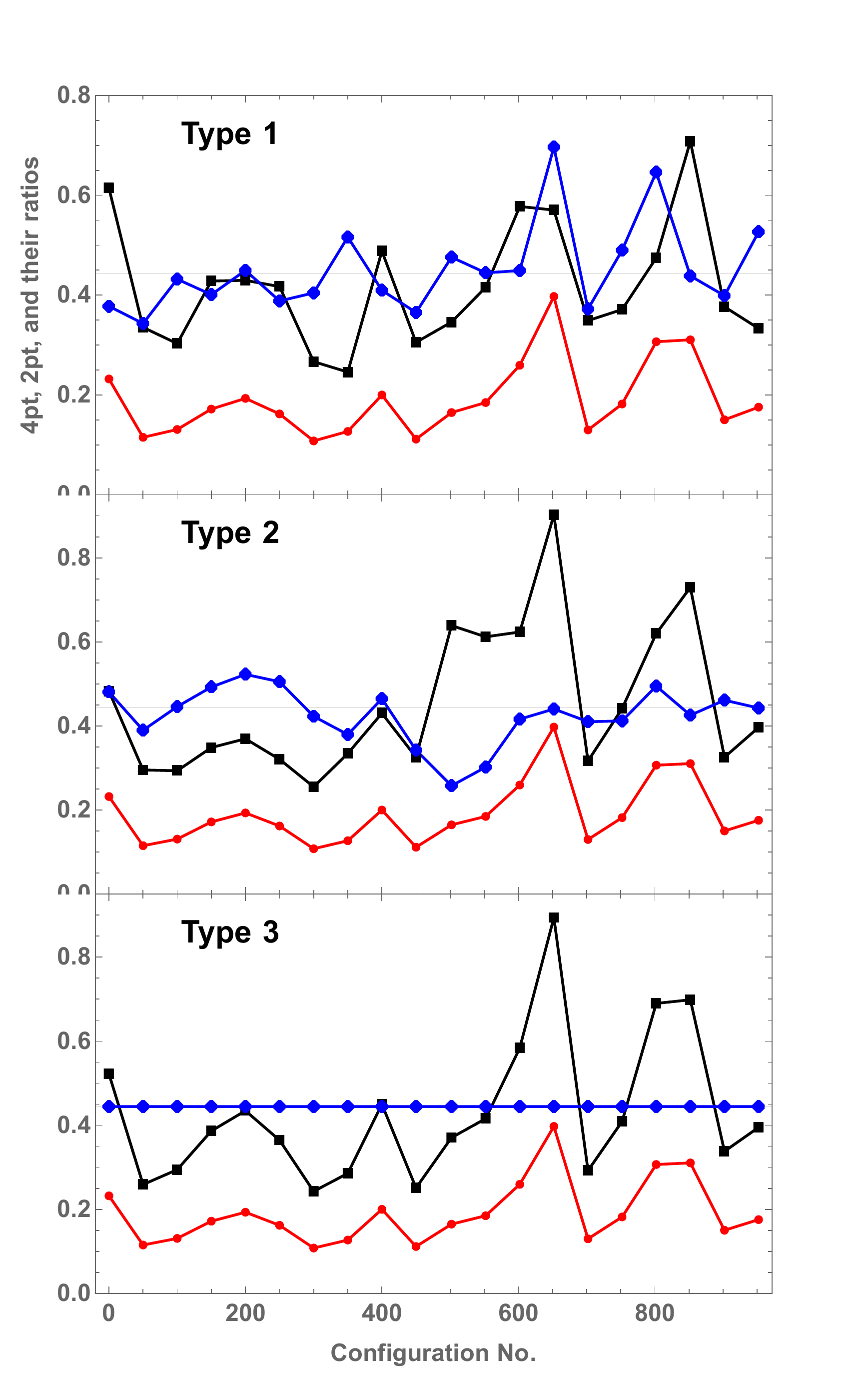}
\caption{
Statistical fluctuations are shown in the unnormalized four-point function (red), three types of two-point functions (black), and their ratios (blue) at 20 randomly-selected configurations.
For this figure, diagram a at $\bm q=0$ and $m_\pi=600$ MeV is used as an example. Neighboring points are connected by straight lines to facilitate visualization.
The faint horizontal gridline indicates the expected ratio $4/9$ for this diagram and conserved currents.
}
\label{fig:ratio}
\end{figure}

Having determined the two-point functions, we present in Fig.~\ref{fig:Q44PS} the raw normalized four-point functions $Q_{44}$ at five different values of momentum $\bm q$ and at $m_\pi=600$ MeV.  For comparison purposes, all points in $Q_{44}$ are  displayed on the same linear scale. For the effective mass function $\ln Q_{44}(t)/Q_{44}(t+1)$,  only points between the pion walls are displayed for clarity.
The results are based on conserved currents and only the connected diagrams a, b and c. 
There are a number of interesting features in these plots. 

First, 
the results for $\bm q=0$ confirms the current conservation property discussed in Eq.\eqref{eq:Q4pt}.
Basically, for conserved current, we expect the ratio of four-point function to two-point function to approach the charge factor $2q_uq_{\bar{d}}=4/9$ for diagram a in the isospin limit, 
independent of current insertion points $t_1$ and $t_2$.
For diagrams b and c, the factor  is $q_uq_u+q_{\bar{d}} q_{\bar{d}}=5/9$. Indeed, this is confirmed in all three diagrams (black dots). 
In diagram a, current conservation is limited between $t_2=7$ (on the pion wall source) and $t_2=41$ (one step inside the pion wall sink) because the two currents independently couple to two different quarks in this range. In diagram b, where they couple to the same quark, current conservation  emerges only starting from $t_2=19$.  In diagram c, it is limited between $t_2=7$ and $t_2=17$ because it is the Z-graph of b (different time-ordering).  If diagrams b and c are added,  then current conservation extends to the whole range, just like diagram a, except for the special point of $t_1=t_2$ to be discussed below. 
Outside the regions of current conservation,  the $\bm q=0$ signal is exactly zero, while the $\bm q\neq 0$ signal gradually goes to zero towards the Dirichlet wall.

 Second, we found that although we have three options for two-point functions to be used as normalization, they have different statistical fluctuations. This is demonstrated in Fig.~\ref{fig:ratio} where we plot the three types for a select few configurations out of the 1000, using diagram (a) at zero momentum and a fixed time slice in the conserved region ($7<t_2<41)$ as an example. 
 For each type, we plot separately the unnormalized four-point function, two-point function, and their ratios.  We see that the ratio from Type 3 gives the expected value (4/9) exactly whereas Type 1 and Type 2 fluctuate around it. The reason is that Type 3, despite being more noisy than Type 1 and Type 2, is exactly correlated with the four-point function configuration by configuration, both being constructed from the same two wall sources.  We rely on this perfect correlation in Type 3 to serve as a strong numerical validation that the wall sources and the conserved currents are correctly implemented in our study.  At nonzero momentum ($\bm q\neq 0$), however,  we found that all three normalization types produce comparable statistical uncertainties for the normalized four-point functions. Fig.~\ref{fig:Q44PS} is plotted using Type 3 normalization. 
 
 Third, the special point of $t_1=t_2$ is regular in diagram a, but gives irregular results in diagram b and c for all values of $\bm q$. This is the contact term in the discussion surrounding Eq.\eqref{eq:Q4pt}. We avoid this point in our analysis.  
 
 Fourth, we observe that the results about $t_1=18$ in diagram b and c are mirror images of each other,  simply due to the fact that they are from the two different  time orderings of the same diagram. In principle, this property could be exploited to reduce the cost of simulations. In this study, however, we computed all three diagrams separately, and add them between $t_1=19$ and $t_3=41$ as the signal. We also note in passing that the $Q_{44}$ signal in diagram c is negative definite whereas it is positive definite in diagrams a and b.
 
 Finally, the effective mass function of $Q_{44}$ for diagram b approaches the value of $E_\pi-m_\pi$ at large separation times between $t_1$ and $t_2$. This is an indication that  the four-point function for diagram b is dominated by the elastic contribution with a fall-off rate of  $E_\pi-m_\pi$  according to Eq.\eqref{eq:Q2}. The same is true for diagram a, although deviations are slightly larger at higher momentum.
The situation for diagram c, however,  is completely different. The fall-off rates approach high above their respective $E_\pi-m_\pi$ values, suggesting they are dominated by inelastic contributions. In other words, the intermediate state is not a pion, but some four-quark state at higher mass and energy. 

We also used local current as a guide to develop our formalism and algorithms.  
If we take four-point function ratio at zero momentum, we expect $\tilde{Q}_{44}^{(PS)}(\bm 0)/\tilde{Q}_{44}^{(PC) }(\bm 0) \to Z_V^2$ where $\tilde{Q}_{44}^{(PC) }$ is computed without the $Z_V$ factor in the formulas.
For example, we obtain an estimate of $Z_V\approx 1.35$ at $m_\pi= 600$ MeV, which is consistent with literature~\cite{MARTINELLI1988865}. 
Since our results are based exclusively on conserved current, we will not discuss local current further.

\subsection{Elastic form factor}

\begin{figure}[htb!]
\includegraphics[scale=0.4]{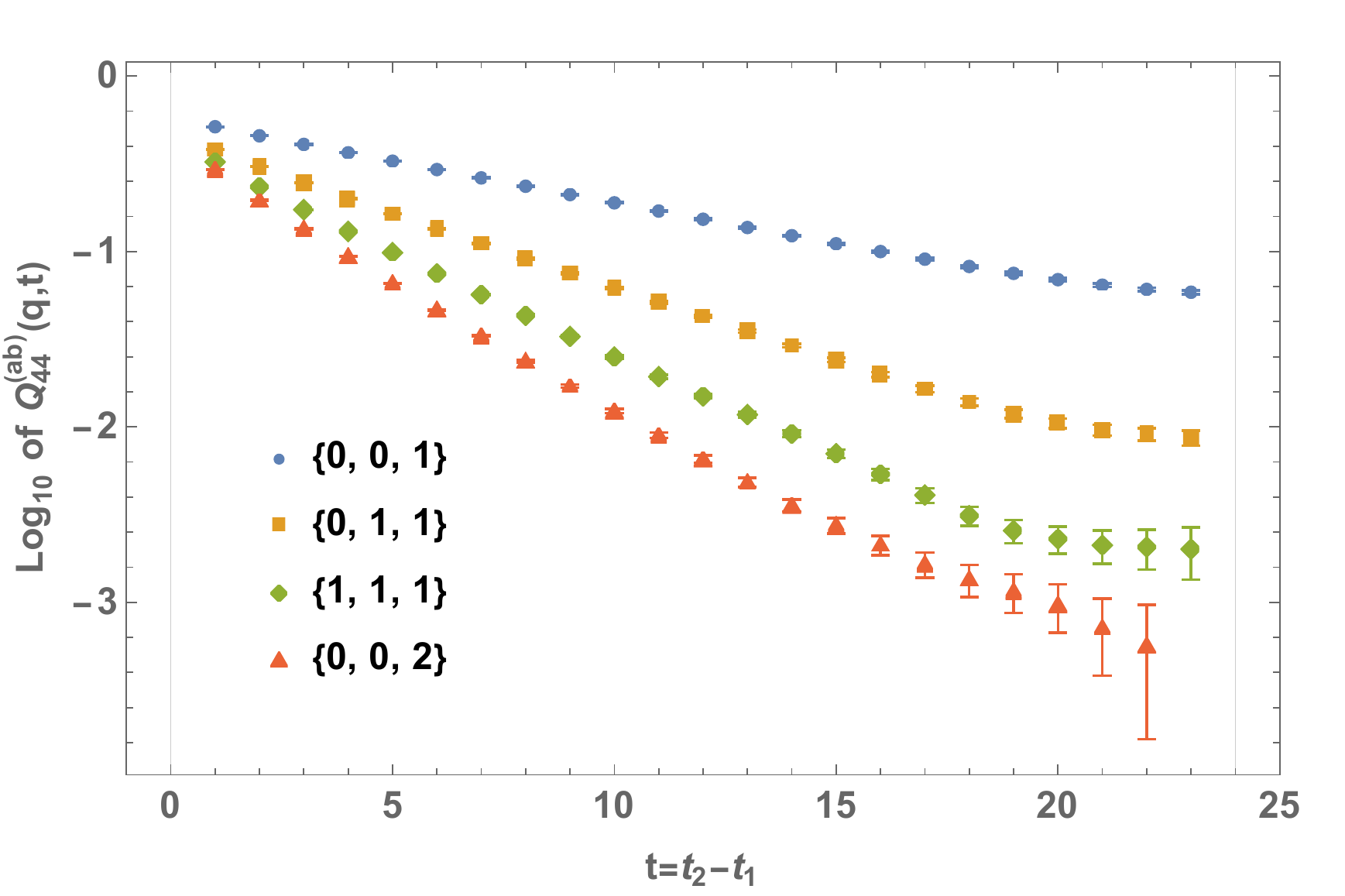}
\includegraphics[scale=0.4]{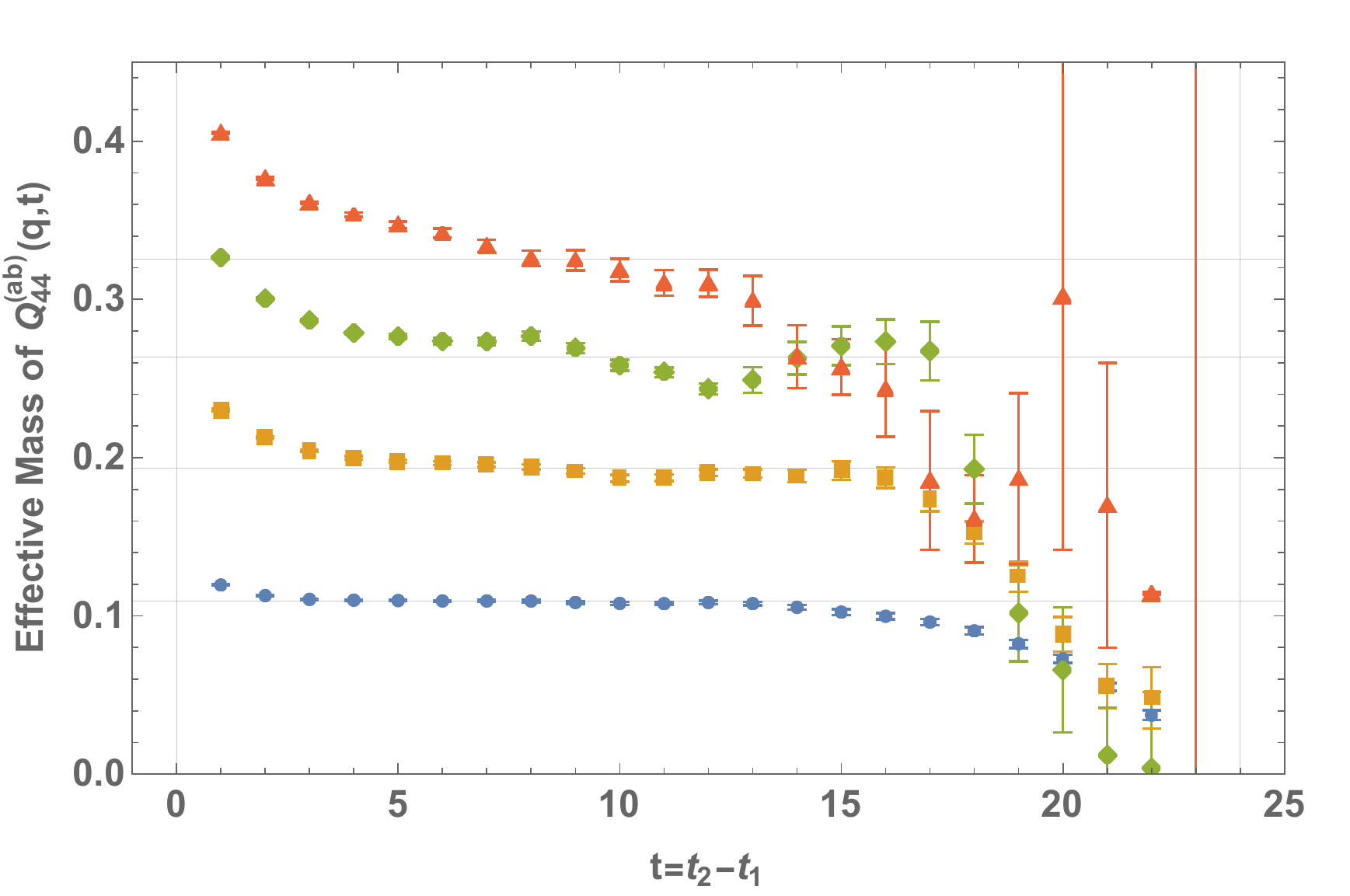}
\caption{
Normalized four-point functions from diagrams a and b in log plot and their effective mass functions at different values of $\bm q$ and $m_\pi=600$ MeV.
They are plotted as functions of  time separation $t=t_2-t_1$ between the two currents relative to fixed $t_1=18$.
The horizontal gridlines in the effective mass are $E_\pi-m_\pi$ using continuum dispersion relation for $E_\pi$ with measured $m_\pi$.
 These functions are used to extract the elastic contributions $Q^{elas}_{44}$.
}
\label{fig:Q44ab}
\end{figure}

The formula for electric polarizability in Eq.\eqref{eq:alpha}  involves the charge radius  $r_E$ and  the elastic contribution $Q^{elas}_{44}$, both of which can be extracted from the large-time behavior of  four-point functions $Q_{44}$.  According to Eq.\eqref{eq:Q44elas},  $Q^{elas}_{44}$ is expected to exhibit single-exponential behavior  with a fall-off rate of $E_\pi-m_\pi$.  The form factor $F_\pi$ is contained in the amplitude of this fall-off.  Based on the discussion about Fig.~\ref{fig:Q44PS}, diagrams a and b have the expected  fall-off whereas diagram c does not. As far as elastic contribution is concerned, we can drop diagram c and focus only on diagrams a and b. This improves the form factor analysis by eliminating the inelastic `contamination' from diagram c.  
It can be regarded as a form of optimization in the analysis. 
Fig.\ref{fig:Q44ab} shows an example of the four-point functions $Q^{ab}_{44}$ including only  diagrams a and b, along with their effective mass functions. We focus in the region of signal between $t_1$ and $t_3$ and plot them as a function of time separation $t=t_2-t_1$ between the two currents. Note that we exclude the $t=0$ point from the analysis due to contact terms, as discussed earlier.
We see that there is a region where the effective mass functions coincide with the $E_\pi-m_\pi$  gridlines, indicating that $Q^{ab}_{44}$ is dominated by elastic contributions. The agreement is better at  smaller momentum values. The signal at large times is noisy and increasingly so at higher momentum. We also see the effect of the Dirichlet wall which forces the effective mass to curve down.  In this context, the inclusion of diagram c would push the elastic limit into larger times where the signal is lost.
To account for possible violation of the continuum dispersion relation, we perform a fit to  the functional form of $Q^{elas}_{44}$ in Eq.\eqref{eq:Q44elas}, treating both 
$\{F_\pi, E_\pi\}$ as free parameters with $m_\pi$ fixed at the measured values from two-point functions.
Details of the fits at all four pion masses are given in Table~\ref{tab:ff} in Appendix~\ref{sec:ff4}.  From this table, we observe that the $E_\pi$ from the fit  largely agrees with that from the continuum dispersion relation. Deviations become more apparent  at higher momentum.

\begin{figure}[h!]
\includegraphics[scale=0.45]{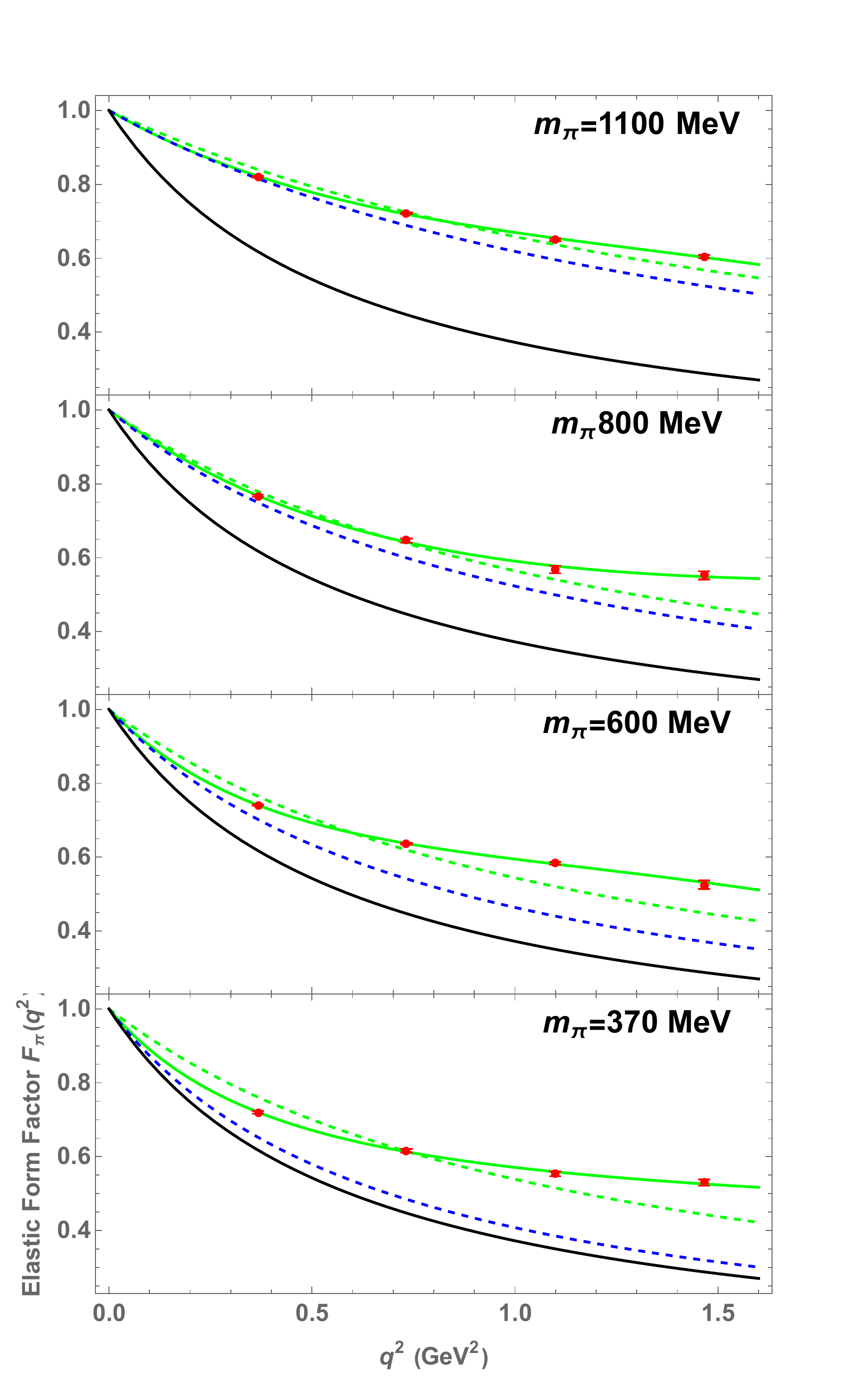}
\caption{
Pion elastic form factors extracted from four-point functions. The red data points are the measured values in Table~\ref{tab:ff}. The green solid line is a fit to the z-expansion in Eq.~\eqref{eq:z}. The green dashed line is a fit to the monopole form in Eq.~\eqref{eq:vmd}.  The blue dashed line is the same monopole form plotted with the measured rho mass, and  the black solid line with the physical rho mass.
}
\label{fig:ff}
\end{figure}
\begin{figure}[h!]
\includegraphics[scale=0.35]{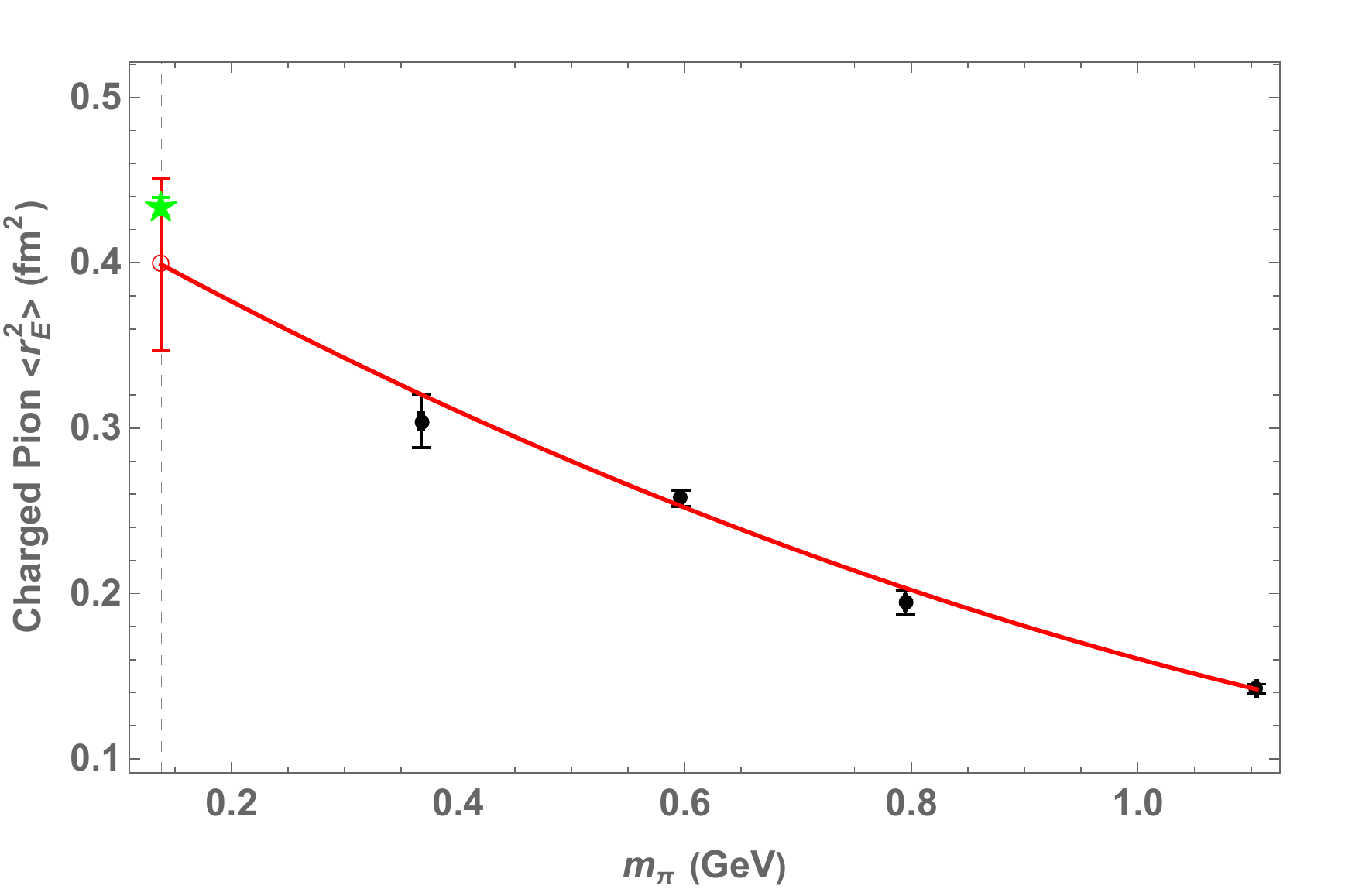}
\includegraphics[scale=0.35]{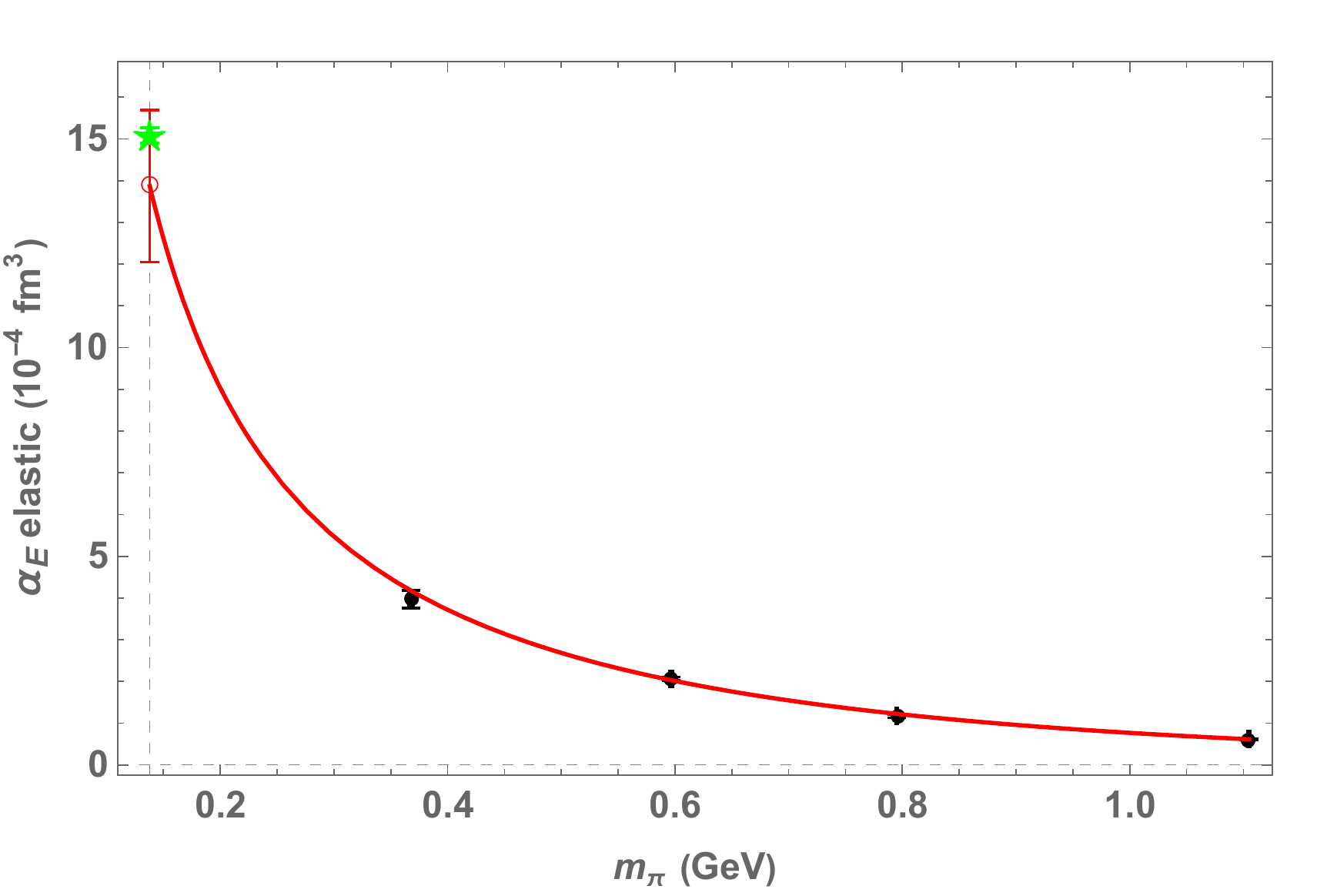}
\caption{
Chiral extrapolation of charge radius (top) and the elastic part of electric polarizability (bottom). The green stars are derived from PDG values.
}
\label{fig:re2}
\end{figure}
\begin{figure}[b!]
\includegraphics[scale=0.4]{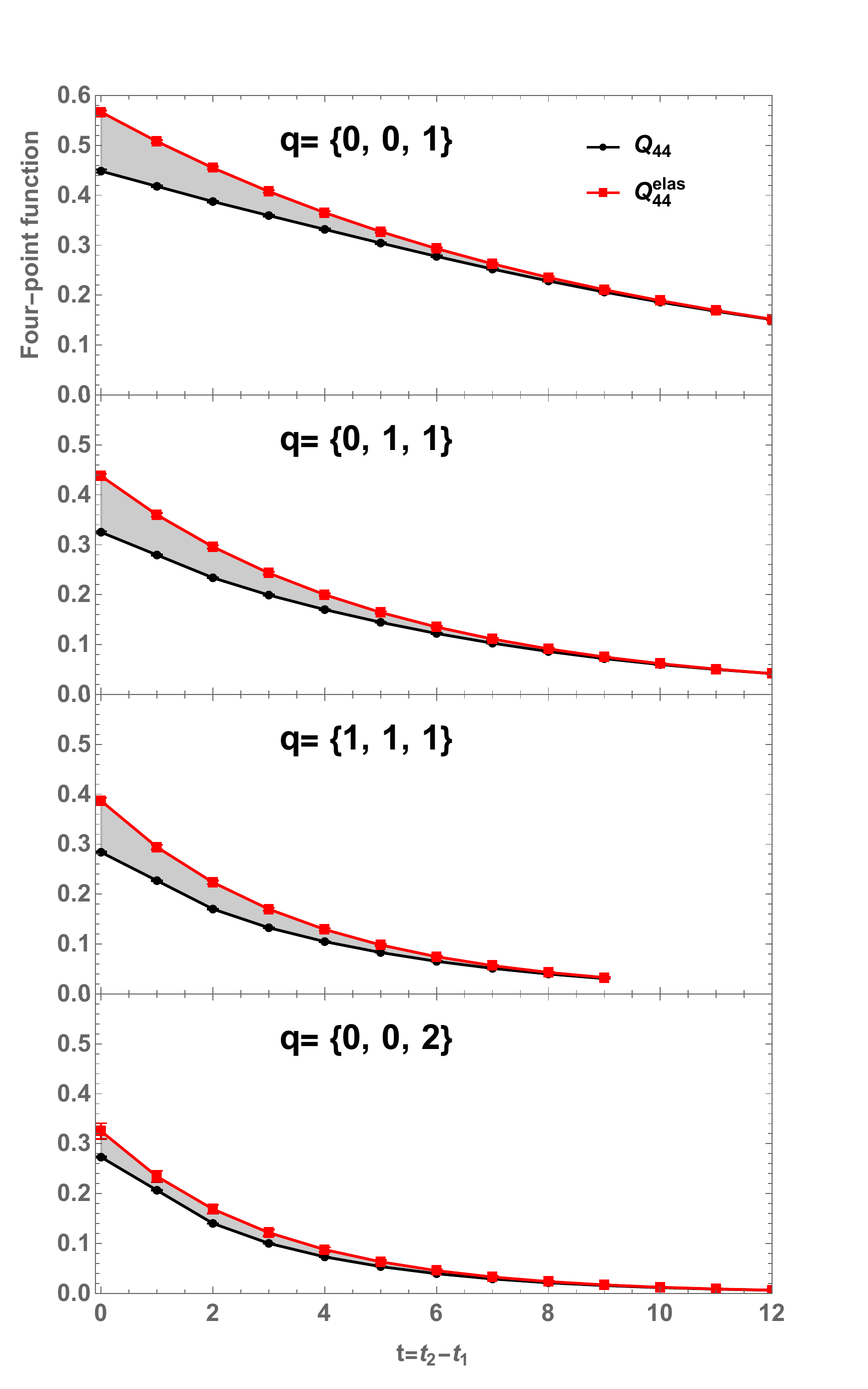}
\caption{
Total $Q_{44}$ and elastic $Q_{44}^{elas}$  at different values of $\bm q$ at $m_\pi=600$ MeV.
 The shaded area, $(1/a)\int dt \big[ Q_{44}(\bm q,t)-Q_{44}^{elas}(\bm q,t)\big]$,  is the dimensionless signal contributing to polarizability.
}
\label{fig:QQ}
\end{figure}
\begin{figure}[b!]
\includegraphics[scale=0.4]{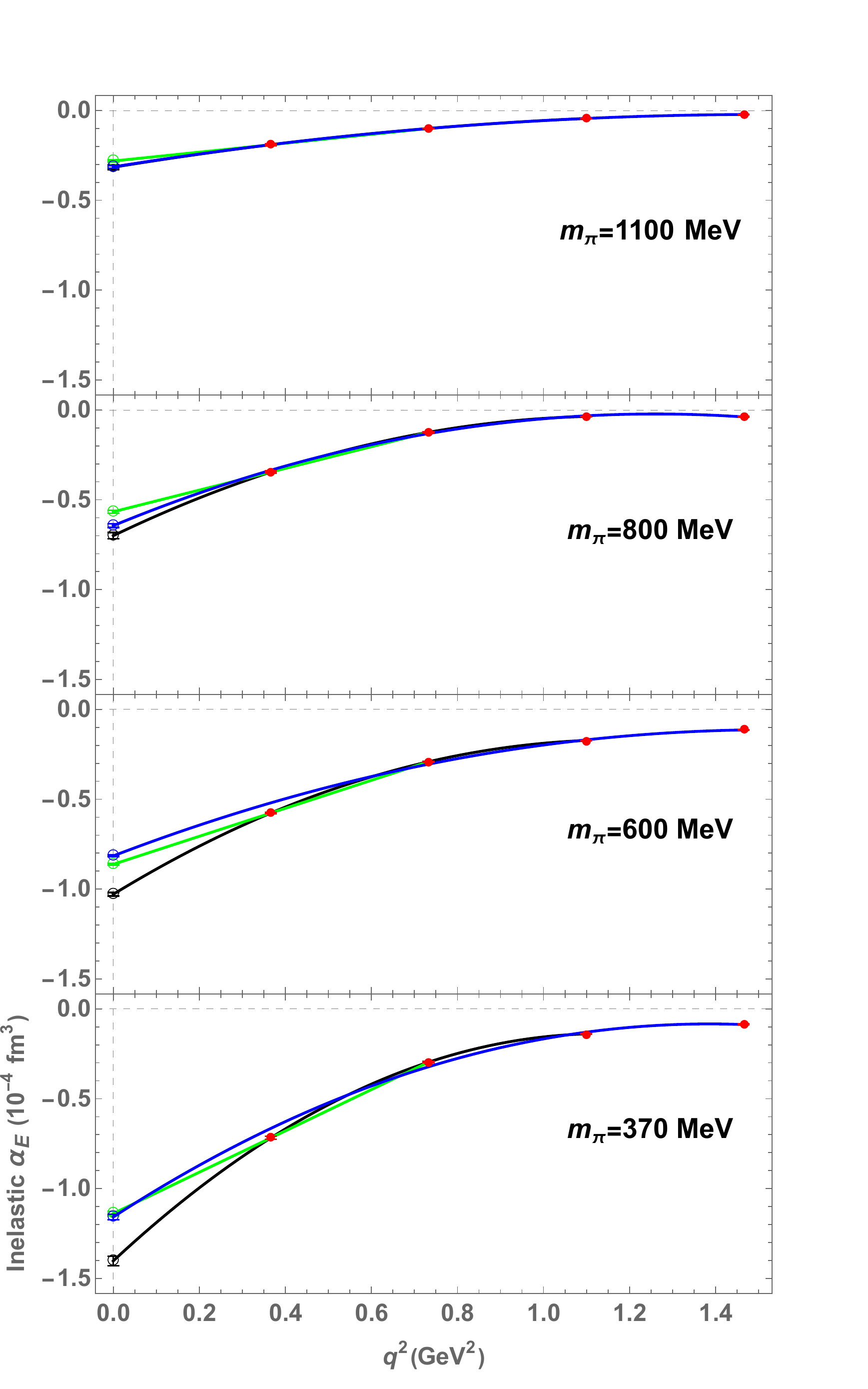}
\caption{
Momentum dependence of the inelastic term in Eq.~\eqref{eq:alpha} and its extrapolation to $\bm q^2=0$ at all pion masses.
Red points are based on the shaded areas in Fig.~\ref{fig:QQ}. Blue curve is a quadratic extrapolation using all points. Black curve is the same quadratic extrapolation using the three lowest points.  Green curve is a linear extrapolation based on the two lowest points. Empty points indicate the corresponding extrapolated values contributing to $\alpha_E$.
}
\label{fig:Qzero}
\end{figure}
After the form factor data are obtained, we fit them to the monopole form,
\beq
F_\pi(\bm q^2) = \frac{1}{1+\bm q^2/m_V^2},
\label{eq:vmd}
\eeq
which is  the well-known vector meson dominance (VMD)  commonly considered  in pion form factor studies. 
The results are illustrated in  Fig.~\ref{fig:ff}. 
We see that the monopole form does not fit the data well, especially at higher momentum and lower pion mass. We will not consider the monopole fit further. Instead,  we opt for  the $z$-expansion parametrization~\cite{Lee_2015}
\beqs
& F_\pi(\bm q^2) =1+\sum_{k=1}^{k_{max}}  a_k \, z^k,  \\
& \text{ where }z \equiv \frac{ \sqrt{ t_{cut}-t} - \sqrt{ t_{cut}-t_0}  }{ \sqrt{ t_{cut}-t} + \sqrt{ t_{cut}-t_0} } \\
& \text{ and } t=-\bm q^2, \; t_{cut}=4 m_\pi^2,
\label{eq:z}
\eeqs
where $a_k$ are free parameters and $t_{cut}$ is the two-pion production threshold.
We take $t_0=0$ so the form goes through $F_\pi(0)=1$ by construction. 
Using this form, we can find a good fit with $k_{max}=3$ in all cases.  For comparison, we also plot the monopole function with the measured rho mass $m_\rho$ and the physical rho mass of $m_\rho^{phys}=0.77$ GeV.
We observe significant differences between the fitted monopole form ($m_V$) and the VMD form ($m_\rho$). The difference grows with increasing momentum and decreasing pion mass.
Similar behavior has been observed in previous studies~~\cite{Draper:1988bp,van_der_Heide_2004}.  
This issue of form factors in the four-point function formalism deserves further study with more advanced setups, such as dynamical ensembles, smaller pion masses, and wider momentum coverage.
Once the functional form of form factor  is determined, the charge radius is  obtained by 
\beq
\langl  r_E^2\rangl= -6 \frac{dF_\pi (\bm q^2)}{d \bm q^2 } \Big|_{\bm q^2\to 0}.
\label{eq:re2}
\eeq

From the extracted charge radius, we attempt a chiral extrapolation using a quadratic form $a+b\, m_\pi + c\, m_\pi^2$.
We also perform a chiral extrapolation of the elastic part of $\alpha_E$ using the form ${a\over m_\pi}+b + c\, m_\pi$.
The result is shown in Fig.~\ref{fig:re2}. The extrapolated charge radius at the physics point is consistent with PDG value albeit our results suffer from relatively large statistical errors. 
The same is true for the elastic part of $\alpha_E$ in Eq.~\eqref{eq:alpha}.
Their values in physical units can be found in Table \ref{tab:final}.

\begin{table*}[t!]
\caption{Summary of results in physical units from two-point and four-point functions. 
Charge radius is chirally extrapolated to the physical point, as well as 
 $\alpha_E$ elastic and  $\alpha_E$ total. The $\alpha_E$  inelastic at the physical point is taken as the difference of the two.
Known values from ChPT and PDG are listed for comparison purposes.
All polarizabilities are in units of $10^{-4}\;\text{fm}^3$.
}
\label{tab:final}
\begin{tabular}{c}
$      
\renewcommand{\arraystretch}{1.2}
\begin{array}{l|ccccc|c}
\hline
  & \text{$\kappa $=0.1520} & \text{$\kappa $=0.1543} & \text{$\kappa $=0.1555} & \text{$\kappa $=0.1565} & \text{physical point} & \text{known value}
   \\
\hline
 m_{\pi }\text{ (MeV)} & 1104.7\pm 1.2 & 795.0\pm 1.1 & 596.8\pm 1.4 & 367.7\pm 2.2 & 138 & 138\\
 m_{\rho }\text{ (MeV)} & 1273.1\pm 2.5 & 1047.3\pm 3.4 & 930.\pm 7. & 830.\pm 17. & 770 & 770\\
 \hline
 \langl r_E^2\text{$\rangl$ (}\text{fm}^2\text{) } & 0.1424\pm 0.0029 & 0.195\pm 0.007 & 0.257\pm 0.005 & 0.304\pm 0.016  & 0.40\pm 0.05 & 0.434\pm 0.005 \text{ (PDG)}\\
     \hline
 \alpha _E\text{ elastic}  & 0.618\pm 0.012 & 1.17\pm 0.04 & 2.07\pm 0.04 & 3.97\pm 0.21 & 13.9\pm 1.8 & 15.08\pm 0.13 \text{ (PDG)}\\
  \alpha _E\text{ inelastic }  & -0.299\pm 0.019 & -0.672\pm 0.030 & -0.92\pm 0.11 & -1.27\pm 0.13 & -9.7\pm 1.9 \text{ to} -5.1\pm 2.0 &\\
 \alpha _E\text{ total } & 0.319\pm 0.023 & 0.50\pm 0.05 & 1.15\pm 0.11 & 2.70\pm 0.25 & 4.2\pm  0.5 \text{ to } 8.8\pm 0.9& 2.93\pm 0.05 \text{ (ChPT)}\\
      &   &   &   &  & & 2.0\pm 0.6 \pm 0.7\text{ (PDG)}\\
  \hline
\end{array}
$   
\end{tabular}
\end{table*}
%

\subsection{Electric polarizability}

Having obtained the elastic contribution $Q_{44}^{elas}$, we now turn to the inelastic part of $\alpha_E$ from Eq.\eqref{eq:alpha}. In Fig.~\ref{fig:QQ} we show separately the total contribution $Q_{44}$  (from all three diagrams) and  $Q_{44}^{elas}$ as a function of current separation $t=t_2-t_1$. We use $m_\pi=600$ MeV as an example; the graphs at the other pion masses look similar. Note that although $Q_{44}^{elas}$ is obtained in the large time region, the subtraction is done in the whole region according to the functional form in Eq.\eqref{eq:Q44elas}. Most of the contribution is in the small time region where inelastic contributions are significant. We observe that $Q_{44}^{elas}$ is consistently larger than $Q_{44}$, suggesting that the inelastic  term in the formula  is negative. The time integral is simply the negative of the shaded area between the two curves. One detail to notice is that the curves include the $t=0$ point which has unphysical contributions in $Q_{44}$ as mentioned earlier. We would normally avoid this point and only start the integral from $t=1$. However, as one can see, the chunk of area between $t=0$ and $t=1$ is the largest piece in the integral. To include this contribution, we linearly extrapolated the  $Q_{44}$ term back to $t=0$ using the two points at $t=1$ and $t=2$. This will incur a systematic effect on the order of $O(a^2)$ since the error itself is order of $O(a)$. As the continuum limit is approached, the systematic effect will vanish (the chunk will shrink to zero).
There is no issue to include this point in  $Q_{44}^{elas}$ using its functional form. 

The inelastic term can now be constructed by multiplying ${2\alpha / \bm q^{\,2}}$ and the time integral,  and the whole term is a function of momentum.  
Since $\alpha_E$ is a static property, we  extrapolate it to $\bm q^2=0$ smoothly. We consider three fits, a quadratic fit $a+b\, x+c\, x^2$  ($x=\bm q^2$) using all four data points, the same quadratic fit using the lowest three points, and a linear fit using the two lowest points. The results are shown in Fig.~\ref{fig:Qzero} for all pion masses. One observes a  spread in the extrapolated values. The fits with four or two points do not capture the curvature in the data; only the one with three points does. We treat the spread as a systematic effect as follows. We take the average of the largest spread out of the three values at each pion mass, and it comes with a statistical uncertainty.  We then take half value of the spread as a systematic uncertainty. The statistical and  systematic uncertainties are then propagated in quadrature  to the analysis of  $\alpha_E$.
For our data, the statistical uncertainties are relatively small, so the systematic uncertainties from the extrapolation are dominant in the inelastic contribution.

Finally, we assemble the two terms in the formula in Eq.\eqref{eq:alpha} to obtain $\alpha_E$ in physical units.
To see how the trend continues to smaller pion masses, we take the total values for $\alpha_E$ at the four pion masses and perform a smooth extrapolation to the physical point. Since our pion masses are relatively large, we consider two forms to cover the range of uncertainties in the extrapolation: 
 a polynomial form $a+b\, m_\pi+c\, m_\pi^3$ and a form ${a\over m_\pi}+b\, m_\pi+c\, m_\pi^3$  inspired by ChPT~\cite{Gasser_2006}. 
The spread can be considered as a systematic effect. Since ChPT for pions has no $m_\pi^2$ term, we choose to leave it out in the forms.
The leading $1/m_\pi$ term is divergent at the chiral limit. 
The extrapolated value of $4.2\pm  0.5 \text{ to } 8.8\pm 0.9$ is higher than the  known value from ChPT at two-loop~\cite{Moinester:2019sew} which gives $\alpha_E=2.93(5)$, and from PDG~\cite{Workman:2022ynf} which quotes a value $\alpha_E=2.0(6)(7)$ from experiment with large uncertainties. 
Combining the chirally extrapolated total and the previously chirally extrapolated elastic term from Fig.~\ref{fig:re2}, we obtain the inelastic term by taking the difference of the two.
This yields a prediction  of  $-9.7\pm 1.9 \text{ to} -5.1\pm 2.0$  for the inelastic value at the physics point.  
We should mention that the range is slightly smaller in magnitude than the inelastic contribution obtained in another lattice study~\cite{Feng:2022rkr}  near physical pion mass. It employs a formula derived from a different method but has a similar structure. 

\begin{figure}[h!]
\includegraphics[scale=0.5]{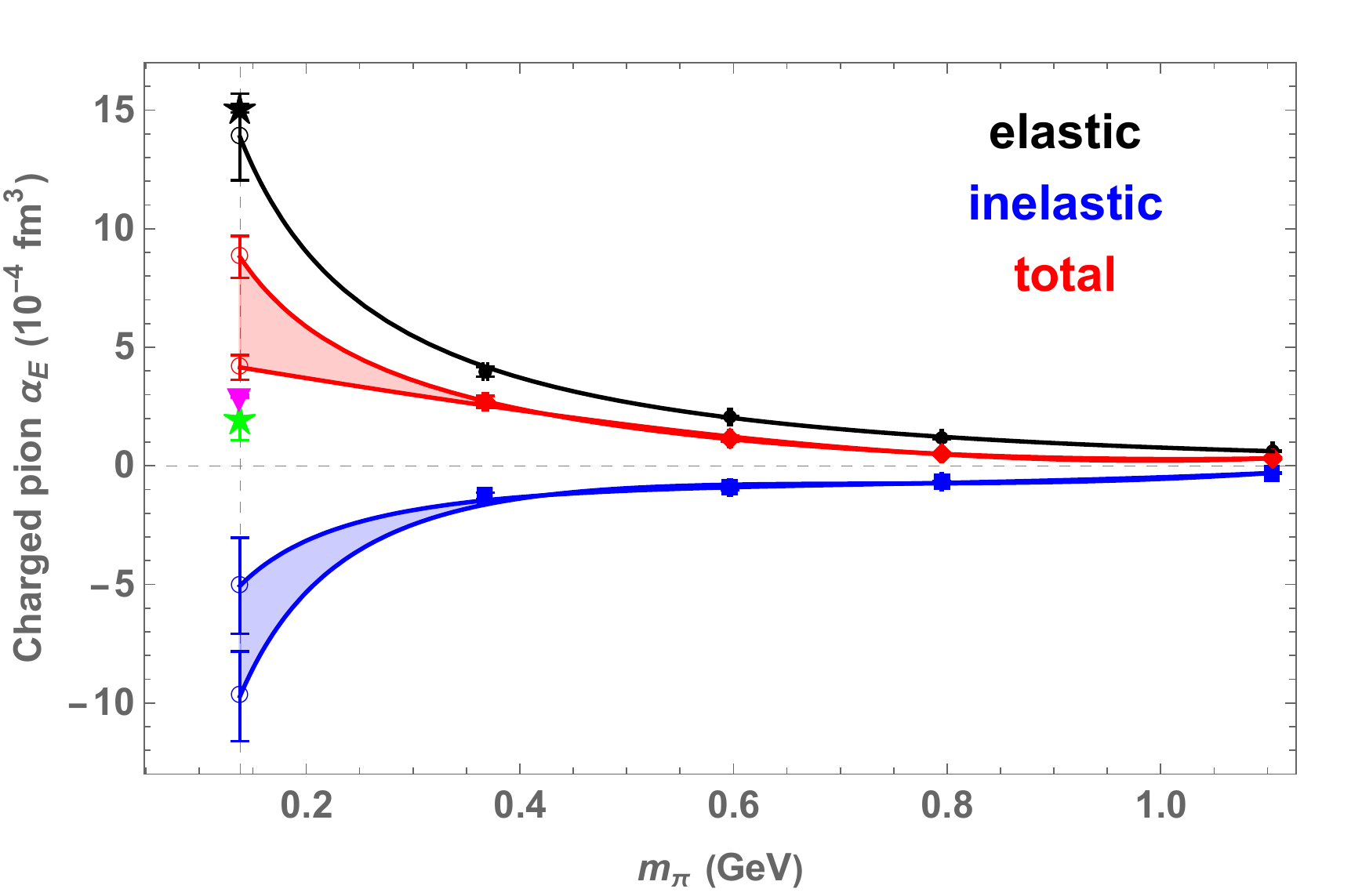}
\caption{
Pion mass dependence of electric polarizability  of a charged pion from  four-point functions in lattice QCD.
Elastic and inelastic contributions correspond to the two terms in the formula in Eq.\eqref{eq:alpha}. 
Elastic and total are chirally extrapolated to the physical point. Inelastic is the difference of the two.
Empty circles are extrapolated values at the physical point.
Magenta triangle is known value from ChPT.
Black and green stars are PDG values for elastic and total, respectively.
}
\label{fig:chiral}
\end{figure}

We summarize  the results  in Fig.~\ref{fig:chiral} and in Table~\ref{tab:final}.  
At the pion masses explored, our lattice results show a clear pattern for electric polarizability:  the elastic term makes a positive contribution, whereas the inelastic term makes a negative and smaller in magnitude contribution. The cancellation leads to a positive value in the total. 
The cancellation appears to continue in the approach to the physical point, but it is less conclusive quantitatively, as indicated by the uncertainty bands from extrapolations. This points to the importance of exploring smaller pion masses in future simulations.

\comment{
For production, we use the two-flavor nHYP ensembles~\cite{Niyazi:2020erg} summarized in Table~\ref{tab:ensembles}.

\begin{table}[b]
\caption{Details of the nHYP ensembles at GW~\cite{Niyazi:2020erg}. Here $N_{cfg}$ and $N_{src}$ label the number of configurations and number of sources on each ensemble, respectively. The top four ensembles correspond to $m_\pi=315(1)\MeV$ and the bottom four $m_\pi=224(2)\MeV$.}
\label{tab:ensembles}
\begin{tabular}{c}
$      
\renewcommand{\arraystretch}{1.5}
\begin{array}{cccccc}
\toprule
Label &N_z\times N_y \times N_x \times N_t &  a (fm) &  \kappa& N_{\text{cfg}}& N_{src}\\
 \midrule
EN1 &16\times16^2 \times 32  & 0.1245& 0.12820  &  230&11\\
EN2 &24\times 24^2 \times 48 & 0.1245& 0.12820  &  300& 25\\
EN3 &30 \times 24^2 \times 48& 0.1245& 0.12820 &  300& 29\\
EN4 &48 \times 24^2 \times 48& 0.1245& 0.12820 &  270& 37\\
\cmidrule[0pt]{1-6}
EN5 &16\times16^2 \times 32& 0.1215& 0.12838  & 230 & 16\\
EN6 &24\times 24^2 \times 64& 0.1215& 0.12838 & 450& 23 \\
EN7 &28\times 24^2 \times 64& 0.1215& 0.12838 & 670& 33\\
EN8 &32\times 24^2 \times 64& 0.1215& 0.12838 & 500& 37\\
\bottomrule
\end{array}
$      
\end{tabular}
\end{table}
}

\section{Summary and outlook}
\label{sec:con}
We investigated the feasibility of using four-point functions in lattice QCD to extract charged pion electric polarizability. 
The approach is based on low-energy Compton scattering tensor constructed with quark and gluon fields in Euclidean spacetime~\cite{Wilcox:2021rtt}. 
The central object is the formula given in Eq.\eqref{eq:alpha} which consists of two terms. One is an elastic contribution involving charge radius $\langl r_E^2\rangl$ and pion mass. The other an inelastic contribution in the form of a subtracted time integral. In addition to four-point functions, it requires two-point functions for pion mass and normalization, but not three-point functions. The elastic contribution can be obtained from the same four-point function in the elastic limit. 

We laid out a detailed formalism and notation using standard Wilson fermion as a baseline.
Although we use both local current and conserved current on the lattice to develop and test the formalism, our results are based on conserved current on the lattice. 
It sidesteps the renormalization issue ($Z_V=1$), but comes with increased complexity in implementation. 
To apply the special kinematics (zero-momentum Breit frame) in the formula, we employ wall sources without gauge-fixing for the creation and annihilation of pions.
We show how to construct the four-point functions using SST quark propagation, develop efficient algorithms for numerical evaluation, and use a high-performance solver~\cite{Alexandru:2011ee}.

We carried out a proof-of-concept simulation using quenched Wilson action with pion mass ranging from 1100 to  370 MeV.
We only considered the connected contributions in this work. 
 We discussed three types of wall-to-wall two-point functions for normalization.  We found a perfect correlation between the four-point function $Q_{44}(\bm q^2=0)$ and Type 3 two-point function imposed by current conservation, configuration by configuration. This property provides a strong check of our implementation.

The analysis procedure used to determine $\alpha_E$ in Eq.\eqref{eq:alpha} involves multiple steps which we summarize here.
1) Fit Type 1 two-point function to obtain  $m_\pi$ (and $m_\rho$).
2) Fit  four-point function $Q^{(ab)}_{44}$ from diagrams a and b to $Q^{elas}_{44}$ at large times for elastic form factor $F_\pi$.
3) Fit $F_\pi$  data to a functional form, then extract charge radius $\langl r_E^2\rangl$ which is then chirally extrapolated.
4) Perform subtraction  $Q^{(abc)}_{44}(\bm q)-Q^{elas}_{44}(\bm q)$ at small times  using all three diagrams a,b,c. Do the time integration. Extrapolate back to $t=0$ to include the  missing chunk due to contact terms. 
5) Extrapolate the inelastic term to $\bm q^2=0$ to obtain the static limit, then assemble everything in physical units for $\alpha_E$.
6) Extrapolate the elastic and total $\alpha_E$ in pion mass to the physical point, obtain the inelastic by taking the difference.

Our results at the pion masses explored so far reveal a clear physical picture for charged pion $\alpha_E$: it is the result of a cancellation between a positive elastic contribution and a negative inelastic contribution. It would be interesting to see how the cancellation plays out in the approach to the physical point. 
Nevertheless, the simulation demonstrates that the four-point function methodology can be a viable alternative to the background method for polarizabilities of charged hadrons.
We caution that the picture is subject to a number of systematic effects not incorporated at this stage, such as the quenched approximation, finite-volume effects, and disconnected loops.
Other sources of uncertainty in the present analysis include fitting the elastic form factor,  the contact term at  $t=0$ in the inelastic term,  and the extrapolation of the inelastic term to $\bm q^2=0$.  
All these open issues deserve further study in future simulations.

Going forward, the investigation can proceed in multiple directions.
First, the quenched approximation should be removed by employing dynamical fermions. Work is underway to use our collection of two-flavor nHYP-clover ensembles~\cite{Niyazi:2020erg} which have been successfully used in a number of physics projects. They have smaller pion masses (about 315 MeV and 227 MeV) that can be used to check the expected chiral behavior and facilitate a chiral extrapolation study. The elongated geometries in these ensembles offer a cost-effective way of studying finite-volume effects and reaching smaller $\bm q$ values.  It would be interesting to see how the charge radius is affected by the change of action.
Second, a simulation of charged pion magnetic polarizability  ($\beta_M$) is straightforward. The formula has been derived in Ref.~\cite{Wilcox:2021rtt}. One just needs to replace $Q_{44}$ with $Q_{11}$ in the formalism. It would be interesting to check the well-known prediction $\alpha_E+\beta_M\approx 0$ from ChPT.
Third,  the disconnected contributions should be included. This is a challenging task. Although disconnected loops generally give smaller contributions than connected ones, they must be dealt with for a complete picture from lattice QCD.
Fourth, the methodology can be equally applied to neutral particles (for example $\pi^0$ and the neutron). The advantage it offers over the background field method is the natural treatment of disconnected loops (or sea quarks)~\cite{Freeman:2013eta,Freeman:2014kka}. 
Our ultimate target is the proton for which a formula is also available~\cite{Wilcox:2021rtt}.   A first-principles-based calculation of its  polarizabilities will be a valuable addition to the  Compton scattering effort in nuclear physics.

\vspace*{5mm}
\begin{acknowledgments}
This work was supported in part by U.S. Department of Energy under  Grant~No.~DE-FG02-95ER40907 (FL, AA) and UK Research and Innovation grant {MR/S015418/1} (CC).
AA would like to acknowledge support from University of Maryland. 
WW would like to acknowledge support from the Baylor College of Arts and Sciences SRA program.
\end{acknowledgments}

\bibliography{x4ptfun}

\begin{thebibliography}{66}%
\makeatletter
\providecommand \@ifxundefined [1]{%
 \@ifx{#1\undefined}
}%
\providecommand \@ifnum [1]{%
 \ifnum #1\expandafter \@firstoftwo
 \else \expandafter \@secondoftwo
 \fi
}%
\providecommand \@ifx [1]{%
 \ifx #1\expandafter \@firstoftwo
 \else \expandafter \@secondoftwo
 \fi
}%
\providecommand \natexlab [1]{#1}%
\providecommand \enquote  [1]{``#1''}%
\providecommand \bibnamefont  [1]{#1}%
\providecommand \bibfnamefont [1]{#1}%
\providecommand \citenamefont [1]{#1}%
\providecommand \href@noop [0]{\@secondoftwo}%
\providecommand \href [0]{\begingroup \@sanitize@url \@href}%
\providecommand \@href[1]{\@@startlink{#1}\@@href}%
\providecommand \@@href[1]{\endgroup#1\@@endlink}%
\providecommand \@sanitize@url [0]{\catcode `\\12\catcode `\$12\catcode
  `\&12\catcode `\#12\catcode `\^12\catcode `\_12\catcode `\%12\relax}%
\providecommand \@@startlink[1]{}%
\providecommand \@@endlink[0]{}%
\providecommand \url  [0]{\begingroup\@sanitize@url \@url }%
\providecommand \@url [1]{\endgroup\@href {#1}{\urlprefix }}%
\providecommand \urlprefix  [0]{URL }%
\providecommand \Eprint [0]{\href }%
\providecommand \doibase [0]{http://dx.doi.org/}%
\providecommand \selectlanguage [0]{\@gobble}%
\providecommand \bibinfo  [0]{\@secondoftwo}%
\providecommand \bibfield  [0]{\@secondoftwo}%
\providecommand \translation [1]{[#1]}%
\providecommand \BibitemOpen [0]{}%
\providecommand \bibitemStop [0]{}%
\providecommand \bibitemNoStop [0]{.\EOS\space}%
\providecommand \EOS [0]{\spacefactor3000\relax}%
\providecommand \BibitemShut  [1]{\csname bibitem#1\endcsname}%
\let\auto@bib@innerbib\@empty
\bibitem [{\citenamefont {Fiebig}\ \emph {et~al.}(1989)\citenamefont {Fiebig},
  \citenamefont {Wilcox},\ and\ \citenamefont {Woloshyn}}]{Fiebig:1988en}%
  \BibitemOpen
  \bibfield  {author} {\bibinfo {author} {\bibfnamefont {H.~R.}\ \bibnamefont
  {Fiebig}}, \bibinfo {author} {\bibfnamefont {W.}~\bibnamefont {Wilcox}}, \
  and\ \bibinfo {author} {\bibfnamefont {R.~M.}\ \bibnamefont {Woloshyn}},\
  }\bibfield  {title} {\enquote {\bibinfo {title} {{A Study of Hadron Electric
  Polarizability in Quenched Lattice QCD}},}\ }\href {\doibase
  10.1016/0550-3213(89)90180-6} {\bibfield  {journal} {\bibinfo  {journal}
  {Nucl. Phys. B}\ }\textbf {\bibinfo {volume} {324}},\ \bibinfo {pages}
  {47--66} (\bibinfo {year} {1989})}\BibitemShut {NoStop}%
\bibitem [{\citenamefont {Lujan}\ \emph {et~al.}(2016)\citenamefont {Lujan},
  \citenamefont {Alexandru}, \citenamefont {Freeman},\ and\ \citenamefont
  {Lee}}]{Lujan:2016ffj}%
  \BibitemOpen
  \bibfield  {author} {\bibinfo {author} {\bibfnamefont {M.}~\bibnamefont
  {Lujan}}, \bibinfo {author} {\bibfnamefont {A.}~\bibnamefont {Alexandru}},
  \bibinfo {author} {\bibfnamefont {W.}~\bibnamefont {Freeman}}, \ and\
  \bibinfo {author} {\bibfnamefont {F.~X.}\ \bibnamefont {Lee}},\ }\bibfield
  {title} {\enquote {\bibinfo {title} {{Finite volume effects on the electric
  polarizability of neutral hadrons in lattice QCD}},}\ }\href {\doibase
  10.1103/PhysRevD.94.074506} {\bibfield  {journal} {\bibinfo  {journal} {Phys.
  Rev.}\ }\textbf {\bibinfo {volume} {D94}},\ \bibinfo {pages} {074506}
  (\bibinfo {year} {2016})},\ \Eprint {http://arxiv.org/abs/1606.07928}
  {arXiv:1606.07928 [hep-lat]} \BibitemShut {NoStop}%
\bibitem [{\citenamefont {Lujan}\ \emph {et~al.}(2014)\citenamefont {Lujan},
  \citenamefont {Alexandru}, \citenamefont {Freeman},\ and\ \citenamefont
  {Lee}}]{Lujan:2014kia}%
  \BibitemOpen
  \bibfield  {author} {\bibinfo {author} {\bibfnamefont {Michael}\ \bibnamefont
  {Lujan}}, \bibinfo {author} {\bibfnamefont {Andrei}\ \bibnamefont
  {Alexandru}}, \bibinfo {author} {\bibfnamefont {Walter}\ \bibnamefont
  {Freeman}}, \ and\ \bibinfo {author} {\bibfnamefont {Frank}\ \bibnamefont
  {Lee}},\ }\bibfield  {title} {\enquote {\bibinfo {title} {{Electric
  polarizability of neutral hadrons from dynamical lattice QCD ensembles}},}\
  }\href {\doibase 10.1103/PhysRevD.89.074506} {\bibfield  {journal} {\bibinfo
  {journal} {Phys. Rev.}\ }\textbf {\bibinfo {volume} {D89}},\ \bibinfo {pages}
  {074506} (\bibinfo {year} {2014})},\ \Eprint {http://arxiv.org/abs/1402.3025}
  {arXiv:1402.3025 [hep-lat]} \BibitemShut {NoStop}%
\bibitem [{\citenamefont {Freeman}\ \emph {et~al.}(2014)\citenamefont
  {Freeman}, \citenamefont {Alexandru}, \citenamefont {Lujan},\ and\
  \citenamefont {Lee}}]{Freeman:2014kka}%
  \BibitemOpen
  \bibfield  {author} {\bibinfo {author} {\bibfnamefont {Walter}\ \bibnamefont
  {Freeman}}, \bibinfo {author} {\bibfnamefont {Andrei}\ \bibnamefont
  {Alexandru}}, \bibinfo {author} {\bibfnamefont {Michael}\ \bibnamefont
  {Lujan}}, \ and\ \bibinfo {author} {\bibfnamefont {Frank~X.}\ \bibnamefont
  {Lee}},\ }\bibfield  {title} {\enquote {\bibinfo {title} {{Sea quark
  contributions to the electric polarizability of hadrons}},}\ }\href {\doibase
  10.1103/PhysRevD.90.054507} {\bibfield  {journal} {\bibinfo  {journal} {Phys.
  Rev. D}\ }\textbf {\bibinfo {volume} {90}},\ \bibinfo {pages} {054507}
  (\bibinfo {year} {2014})},\ \Eprint {http://arxiv.org/abs/1407.2687}
  {arXiv:1407.2687 [hep-lat]} \BibitemShut {NoStop}%
\bibitem [{\citenamefont {Freeman}\ \emph {et~al.}(2013)\citenamefont
  {Freeman}, \citenamefont {Alexandru}, \citenamefont {Lee},\ and\
  \citenamefont {Lujan}}]{Freeman:2013eta}%
  \BibitemOpen
  \bibfield  {author} {\bibinfo {author} {\bibfnamefont {Walter}\ \bibnamefont
  {Freeman}}, \bibinfo {author} {\bibfnamefont {Andrei}\ \bibnamefont
  {Alexandru}}, \bibinfo {author} {\bibfnamefont {Frank~X.}\ \bibnamefont
  {Lee}}, \ and\ \bibinfo {author} {\bibfnamefont {Mike}\ \bibnamefont
  {Lujan}},\ }\bibfield  {title} {\enquote {\bibinfo {title} {{Update on the
  Sea Contributions to Hadron Electric Polarizabilities through
  Reweighting}},}\ }in\ \href@noop {} {\emph {\bibinfo {booktitle} {{31st
  International Symposium on Lattice Field Theory}}}}\ (\bibinfo {year}
  {2013})\ \Eprint {http://arxiv.org/abs/1310.4426} {arXiv:1310.4426 [hep-lat]}
  \BibitemShut {NoStop}%
\bibitem [{\citenamefont {Tiburzi}(2008)}]{Tiburzi:2008ma}%
  \BibitemOpen
  \bibfield  {author} {\bibinfo {author} {\bibfnamefont {Brian~C.}\
  \bibnamefont {Tiburzi}},\ }\bibfield  {title} {\enquote {\bibinfo {title}
  {{Hadrons in Strong Electric and Magnetic Fields}},}\ }\href {\doibase
  10.1016/j.nuclphysa.2008.10.010} {\bibfield  {journal} {\bibinfo  {journal}
  {Nucl. Phys.}\ }\textbf {\bibinfo {volume} {A814}},\ \bibinfo {pages}
  {74--108} (\bibinfo {year} {2008})},\ \Eprint
  {http://arxiv.org/abs/0808.3965} {arXiv:0808.3965 [hep-ph]} \BibitemShut
  {NoStop}%
\bibitem [{\citenamefont {Detmold}\ \emph
  {et~al.}(2009{\natexlab{a}})\citenamefont {Detmold}, \citenamefont
  {Tiburzi},\ and\ \citenamefont {Walker-Loud}}]{Detmold:2009fr}%
  \BibitemOpen
  \bibfield  {author} {\bibinfo {author} {\bibfnamefont {William}\ \bibnamefont
  {Detmold}}, \bibinfo {author} {\bibfnamefont {Brian~C.}\ \bibnamefont
  {Tiburzi}}, \ and\ \bibinfo {author} {\bibfnamefont {Andre}\ \bibnamefont
  {Walker-Loud}},\ }\bibfield  {title} {\enquote {\bibinfo {title} {{Lattice
  QCD in Background Fields}},}\ }\bibfield  {booktitle} {\emph {\bibinfo
  {booktitle} {{Proceedings, 10th Workshop on Non-Perturbative Quantum
  Chromodynamics : Paris, France, June 8-12, 2009}}},\ }\href@noop {} {\
  (\bibinfo {year} {2009}{\natexlab{a}})},\ \Eprint
  {http://arxiv.org/abs/0908.3626} {arXiv:0908.3626 [hep-lat]} \BibitemShut
  {NoStop}%
\bibitem [{\citenamefont {Alexandru}\ and\ \citenamefont
  {Lee}(2008)}]{Alexandru:2008sj}%
  \BibitemOpen
  \bibfield  {author} {\bibinfo {author} {\bibfnamefont {Andrei}\ \bibnamefont
  {Alexandru}}\ and\ \bibinfo {author} {\bibfnamefont {Frank~X.}\ \bibnamefont
  {Lee}},\ }\bibfield  {title} {\enquote {\bibinfo {title} {{The Background
  field method on the lattice}},}\ }\href {\doibase 10.22323/1.066.0145}
  {\bibfield  {journal} {\bibinfo  {journal} {PoS}\ }\textbf {\bibinfo {volume}
  {LATTICE2008}},\ \bibinfo {pages} {145} (\bibinfo {year} {2008})},\ \Eprint
  {http://arxiv.org/abs/0810.2833} {arXiv:0810.2833 [hep-lat]} \BibitemShut
  {NoStop}%
\bibitem [{\citenamefont {Lee}\ \emph {et~al.}(2006)\citenamefont {Lee},
  \citenamefont {Zhou}, \citenamefont {Wilcox},\ and\ \citenamefont
  {Christensen}}]{Lee:2005dq}%
  \BibitemOpen
  \bibfield  {author} {\bibinfo {author} {\bibfnamefont {Frank~X.}\
  \bibnamefont {Lee}}, \bibinfo {author} {\bibfnamefont {Leming}\ \bibnamefont
  {Zhou}}, \bibinfo {author} {\bibfnamefont {Walter}\ \bibnamefont {Wilcox}}, \
  and\ \bibinfo {author} {\bibfnamefont {Joseph~C.}\ \bibnamefont
  {Christensen}},\ }\bibfield  {title} {\enquote {\bibinfo {title} {{Magnetic
  polarizability of hadrons from lattice QCD in the background field
  method}},}\ }\href {\doibase 10.1103/PhysRevD.73.034503} {\bibfield
  {journal} {\bibinfo  {journal} {Phys. Rev. D}\ }\textbf {\bibinfo {volume}
  {73}},\ \bibinfo {pages} {034503} (\bibinfo {year} {2006})},\ \Eprint
  {http://arxiv.org/abs/hep-lat/0509065} {arXiv:hep-lat/0509065} \BibitemShut
  {NoStop}%
\bibitem [{\citenamefont {Lee}\ \emph {et~al.}(2005)\citenamefont {Lee},
  \citenamefont {Kelly}, \citenamefont {Zhou},\ and\ \citenamefont
  {Wilcox}}]{Lee:2005ds}%
  \BibitemOpen
  \bibfield  {author} {\bibinfo {author} {\bibfnamefont {F.~X.}\ \bibnamefont
  {Lee}}, \bibinfo {author} {\bibfnamefont {R.}~\bibnamefont {Kelly}}, \bibinfo
  {author} {\bibfnamefont {L.}~\bibnamefont {Zhou}}, \ and\ \bibinfo {author}
  {\bibfnamefont {W.}~\bibnamefont {Wilcox}},\ }\bibfield  {title} {\enquote
  {\bibinfo {title} {{Baryon magnetic moments in the background field
  method}},}\ }\href {\doibase 10.1016/j.physletb.2005.08.106} {\bibfield
  {journal} {\bibinfo  {journal} {Phys. Lett. B}\ }\textbf {\bibinfo {volume}
  {627}},\ \bibinfo {pages} {71--76} (\bibinfo {year} {2005})},\ \Eprint
  {http://arxiv.org/abs/hep-lat/0509067} {arXiv:hep-lat/0509067} \BibitemShut
  {NoStop}%
\bibitem [{\citenamefont {Engelhardt}(2007)}]{Engelhardt:2007ub}%
  \BibitemOpen
  \bibfield  {author} {\bibinfo {author} {\bibfnamefont {Michael}\ \bibnamefont
  {Engelhardt}},\ }\bibfield  {title} {\enquote {\bibinfo {title} {{Neutron
  electric polarizability from unquenched lattice QCD using the background
  field approach}},}\ }\href {\doibase 10.1103/PhysRevD.76.114502} {\bibfield
  {journal} {\bibinfo  {journal} {Phys. Rev. D}\ }\textbf {\bibinfo {volume}
  {76}},\ \bibinfo {pages} {114502} (\bibinfo {year} {2007})},\ \Eprint
  {http://arxiv.org/abs/0706.3919} {arXiv:0706.3919 [hep-lat]} \BibitemShut
  {NoStop}%
\bibitem [{\citenamefont {Bignell}\ \emph
  {et~al.}(2020{\natexlab{a}})\citenamefont {Bignell}, \citenamefont {Kamleh},\
  and\ \citenamefont {Leinweber}}]{Bignell:2020xkf}%
  \BibitemOpen
  \bibfield  {author} {\bibinfo {author} {\bibfnamefont {Ryan}\ \bibnamefont
  {Bignell}}, \bibinfo {author} {\bibfnamefont {Waseem}\ \bibnamefont
  {Kamleh}}, \ and\ \bibinfo {author} {\bibfnamefont {Derek}\ \bibnamefont
  {Leinweber}},\ }\bibfield  {title} {\enquote {\bibinfo {title} {{Magnetic
  polarizability of the nucleon using a Laplacian mode projection}},}\ }\href
  {\doibase 10.1103/PhysRevD.101.094502} {\bibfield  {journal} {\bibinfo
  {journal} {Phys. Rev. D}\ }\textbf {\bibinfo {volume} {101}},\ \bibinfo
  {pages} {094502} (\bibinfo {year} {2020}{\natexlab{a}})},\ \Eprint
  {http://arxiv.org/abs/2002.07915} {arXiv:2002.07915 [hep-lat]} \BibitemShut
  {NoStop}%
\bibitem [{\citenamefont {Deshmukh}\ and\ \citenamefont
  {Tiburzi}(2018)}]{Deshmukh:2017ciw}%
  \BibitemOpen
  \bibfield  {author} {\bibinfo {author} {\bibfnamefont {Amol}\ \bibnamefont
  {Deshmukh}}\ and\ \bibinfo {author} {\bibfnamefont {Brian~C.}\ \bibnamefont
  {Tiburzi}},\ }\bibfield  {title} {\enquote {\bibinfo {title} {{Octet Baryons
  in Large Magnetic Fields}},}\ }\href {\doibase 10.1103/PhysRevD.97.014006}
  {\bibfield  {journal} {\bibinfo  {journal} {Phys. Rev. D}\ }\textbf {\bibinfo
  {volume} {97}},\ \bibinfo {pages} {014006} (\bibinfo {year} {2018})},\
  \Eprint {http://arxiv.org/abs/1709.04997} {arXiv:1709.04997 [hep-ph]}
  \BibitemShut {NoStop}%
\bibitem [{\citenamefont {Bali}\ \emph
  {et~al.}(2018{\natexlab{a}})\citenamefont {Bali}, \citenamefont {Brandt},
  \citenamefont {Endr\H{o}di},\ and\ \citenamefont
  {Gl\"a\ss{}le}}]{Bali:2017ian}%
  \BibitemOpen
  \bibfield  {author} {\bibinfo {author} {\bibfnamefont {Gunnar~S.}\
  \bibnamefont {Bali}}, \bibinfo {author} {\bibfnamefont {Bastian~B.}\
  \bibnamefont {Brandt}}, \bibinfo {author} {\bibfnamefont {Gergely}\
  \bibnamefont {Endr\H{o}di}}, \ and\ \bibinfo {author} {\bibfnamefont
  {Benjamin}\ \bibnamefont {Gl\"a\ss{}le}},\ }\bibfield  {title} {\enquote
  {\bibinfo {title} {{Meson masses in electromagnetic fields with Wilson
  fermions}},}\ }\href {\doibase 10.1103/PhysRevD.97.034505} {\bibfield
  {journal} {\bibinfo  {journal} {Phys. Rev. D}\ }\textbf {\bibinfo {volume}
  {97}},\ \bibinfo {pages} {034505} (\bibinfo {year} {2018}{\natexlab{a}})},\
  \Eprint {http://arxiv.org/abs/1707.05600} {arXiv:1707.05600 [hep-lat]}
  \BibitemShut {NoStop}%
\bibitem [{\citenamefont {Bruckmann}\ \emph {et~al.}(2017)\citenamefont
  {Bruckmann}, \citenamefont {Endrodi}, \citenamefont {Giordano}, \citenamefont
  {Katz}, \citenamefont {Kovacs}, \citenamefont {Pittler},\ and\ \citenamefont
  {Wellnhofer}}]{Bruckmann:2017pft}%
  \BibitemOpen
  \bibfield  {author} {\bibinfo {author} {\bibfnamefont {F.}~\bibnamefont
  {Bruckmann}}, \bibinfo {author} {\bibfnamefont {G.}~\bibnamefont {Endrodi}},
  \bibinfo {author} {\bibfnamefont {M.}~\bibnamefont {Giordano}}, \bibinfo
  {author} {\bibfnamefont {S.~D.}\ \bibnamefont {Katz}}, \bibinfo {author}
  {\bibfnamefont {T.~G.}\ \bibnamefont {Kovacs}}, \bibinfo {author}
  {\bibfnamefont {F.}~\bibnamefont {Pittler}}, \ and\ \bibinfo {author}
  {\bibfnamefont {J.}~\bibnamefont {Wellnhofer}},\ }\bibfield  {title}
  {\enquote {\bibinfo {title} {{Landau levels in QCD}},}\ }\href {\doibase
  10.1103/PhysRevD.96.074506} {\bibfield  {journal} {\bibinfo  {journal} {Phys.
  Rev. D}\ }\textbf {\bibinfo {volume} {96}},\ \bibinfo {pages} {074506}
  (\bibinfo {year} {2017})},\ \Eprint {http://arxiv.org/abs/1705.10210}
  {arXiv:1705.10210 [hep-lat]} \BibitemShut {NoStop}%
\bibitem [{\citenamefont {Parreno}\ \emph {et~al.}(2017)\citenamefont
  {Parreno}, \citenamefont {Savage}, \citenamefont {Tiburzi}, \citenamefont
  {Wilhelm}, \citenamefont {Chang}, \citenamefont {Detmold},\ and\
  \citenamefont {Orginos}}]{Parreno:2016fwu}%
  \BibitemOpen
  \bibfield  {author} {\bibinfo {author} {\bibfnamefont {Assumpta}\
  \bibnamefont {Parreno}}, \bibinfo {author} {\bibfnamefont {Martin~J.}\
  \bibnamefont {Savage}}, \bibinfo {author} {\bibfnamefont {Brian~C.}\
  \bibnamefont {Tiburzi}}, \bibinfo {author} {\bibfnamefont {Jonas}\
  \bibnamefont {Wilhelm}}, \bibinfo {author} {\bibfnamefont {Emmanuel}\
  \bibnamefont {Chang}}, \bibinfo {author} {\bibfnamefont {William}\
  \bibnamefont {Detmold}}, \ and\ \bibinfo {author} {\bibfnamefont {Kostas}\
  \bibnamefont {Orginos}},\ }\bibfield  {title} {\enquote {\bibinfo {title}
  {{Octet baryon magnetic moments from lattice QCD: Approaching experiment from
  a three-flavor symmetric point}},}\ }\href {\doibase
  10.1103/PhysRevD.95.114513} {\bibfield  {journal} {\bibinfo  {journal} {Phys.
  Rev. D}\ }\textbf {\bibinfo {volume} {95}},\ \bibinfo {pages} {114513}
  (\bibinfo {year} {2017})},\ \Eprint {http://arxiv.org/abs/1609.03985}
  {arXiv:1609.03985 [hep-lat]} \BibitemShut {NoStop}%
\bibitem [{\citenamefont {Luschevskaya}\ \emph {et~al.}(2016)\citenamefont
  {Luschevskaya}, \citenamefont {Solovjeva},\ and\ \citenamefont
  {Teryaev}}]{Luschevskaya:2015cko}%
  \BibitemOpen
  \bibfield  {author} {\bibinfo {author} {\bibfnamefont {E.~V.}\ \bibnamefont
  {Luschevskaya}}, \bibinfo {author} {\bibfnamefont {O.~E.}\ \bibnamefont
  {Solovjeva}}, \ and\ \bibinfo {author} {\bibfnamefont {O.~V.}\ \bibnamefont
  {Teryaev}},\ }\bibfield  {title} {\enquote {\bibinfo {title} {{Magnetic
  polarizability of pion}},}\ }\href {\doibase 10.1016/j.physletb.2016.08.054}
  {\bibfield  {journal} {\bibinfo  {journal} {Phys. Lett. B}\ }\textbf
  {\bibinfo {volume} {761}},\ \bibinfo {pages} {393--398} (\bibinfo {year}
  {2016})},\ \Eprint {http://arxiv.org/abs/1511.09316} {arXiv:1511.09316
  [hep-lat]} \BibitemShut {NoStop}%
\bibitem [{\citenamefont {Chang}\ \emph {et~al.}(2015)\citenamefont {Chang},
  \citenamefont {Detmold}, \citenamefont {Orginos}, \citenamefont {Parreno},
  \citenamefont {Savage}, \citenamefont {Tiburzi},\ and\ \citenamefont
  {Beane}}]{Chang:2015qxa}%
  \BibitemOpen
  \bibfield  {author} {\bibinfo {author} {\bibfnamefont {Emmanuel}\
  \bibnamefont {Chang}}, \bibinfo {author} {\bibfnamefont {William}\
  \bibnamefont {Detmold}}, \bibinfo {author} {\bibfnamefont {Kostas}\
  \bibnamefont {Orginos}}, \bibinfo {author} {\bibfnamefont {Assumpta}\
  \bibnamefont {Parreno}}, \bibinfo {author} {\bibfnamefont {Martin~J.}\
  \bibnamefont {Savage}}, \bibinfo {author} {\bibfnamefont {Brian~C.}\
  \bibnamefont {Tiburzi}}, \ and\ \bibinfo {author} {\bibfnamefont {Silas~R.}\
  \bibnamefont {Beane}} (\bibinfo {collaboration} {NPLQCD}),\ }\bibfield
  {title} {\enquote {\bibinfo {title} {{Magnetic structure of light nuclei from
  lattice QCD}},}\ }\href {\doibase 10.1103/PhysRevD.92.114502} {\bibfield
  {journal} {\bibinfo  {journal} {Phys. Rev. D}\ }\textbf {\bibinfo {volume}
  {92}},\ \bibinfo {pages} {114502} (\bibinfo {year} {2015})},\ \Eprint
  {http://arxiv.org/abs/1506.05518} {arXiv:1506.05518 [hep-lat]} \BibitemShut
  {NoStop}%
\bibitem [{\citenamefont {Detmold}\ \emph {et~al.}(2010)\citenamefont
  {Detmold}, \citenamefont {Tiburzi},\ and\ \citenamefont
  {Walker-Loud}}]{Detmold:2010ts}%
  \BibitemOpen
  \bibfield  {author} {\bibinfo {author} {\bibfnamefont {W.}~\bibnamefont
  {Detmold}}, \bibinfo {author} {\bibfnamefont {B.~C.}\ \bibnamefont
  {Tiburzi}}, \ and\ \bibinfo {author} {\bibfnamefont {A.}~\bibnamefont
  {Walker-Loud}},\ }\bibfield  {title} {\enquote {\bibinfo {title} {{Extracting
  Nucleon Magnetic Moments and Electric Polarizabilities from Lattice QCD in
  Background Electric Fields}},}\ }\href {\doibase 10.1103/PhysRevD.81.054502}
  {\bibfield  {journal} {\bibinfo  {journal} {Phys. Rev.}\ }\textbf {\bibinfo
  {volume} {D81}},\ \bibinfo {pages} {054502} (\bibinfo {year} {2010})},\
  \Eprint {http://arxiv.org/abs/1001.1131} {arXiv:1001.1131 [hep-lat]}
  \BibitemShut {NoStop}%
\bibitem [{\citenamefont {Davoudi}\ and\ \citenamefont
  {Detmold}(2015)}]{Davoudi:2015cba}%
  \BibitemOpen
  \bibfield  {author} {\bibinfo {author} {\bibfnamefont {Zohreh}\ \bibnamefont
  {Davoudi}}\ and\ \bibinfo {author} {\bibfnamefont {William}\ \bibnamefont
  {Detmold}},\ }\bibfield  {title} {\enquote {\bibinfo {title} {{Implementation
  of general background electromagnetic fields on a periodic hypercubic
  lattice}},}\ }\href {\doibase 10.1103/PhysRevD.92.074506} {\bibfield
  {journal} {\bibinfo  {journal} {Phys. Rev. D}\ }\textbf {\bibinfo {volume}
  {92}},\ \bibinfo {pages} {074506} (\bibinfo {year} {2015})},\ \Eprint
  {http://arxiv.org/abs/1507.01908} {arXiv:1507.01908 [hep-lat]} \BibitemShut
  {NoStop}%
\bibitem [{\citenamefont {Engelhardt}(2011)}]{Engelhardt:2011qq}%
  \BibitemOpen
  \bibfield  {author} {\bibinfo {author} {\bibfnamefont {Michael}\ \bibnamefont
  {Engelhardt}},\ }\bibfield  {title} {\enquote {\bibinfo {title} {{Exploration
  of the electric spin polarizability of the neutron in lattice QCD}},}\ }\href
  {\doibase 10.22323/1.139.0153} {\bibfield  {journal} {\bibinfo  {journal}
  {PoS}\ }\textbf {\bibinfo {volume} {LATTICE2011}},\ \bibinfo {pages} {153}
  (\bibinfo {year} {2011})},\ \Eprint {http://arxiv.org/abs/1111.3686}
  {arXiv:1111.3686 [hep-lat]} \BibitemShut {NoStop}%
\bibitem [{\citenamefont {Lee}\ and\ \citenamefont
  {Alexandru}(2011)}]{Lee:2011gz}%
  \BibitemOpen
  \bibfield  {author} {\bibinfo {author} {\bibfnamefont {Frank~X.}\
  \bibnamefont {Lee}}\ and\ \bibinfo {author} {\bibfnamefont {Andrei}\
  \bibnamefont {Alexandru}},\ }\bibfield  {title} {\enquote {\bibinfo {title}
  {{Spin Polarizabilities on the Lattice}},}\ }\bibfield  {booktitle} {\emph
  {\bibinfo {booktitle} {{Proceedings, 29th International Symposium on Lattice
  field theory (Lattice 2011): Squaw Valley, Lake Tahoe, USA, July 10-16,
  2011}}},\ }\href {\doibase 10.22323/1.139.0317} {\bibfield  {journal}
  {\bibinfo  {journal} {PoS}\ }\textbf {\bibinfo {volume} {LATTICE2011}},\
  \bibinfo {pages} {317} (\bibinfo {year} {2011})},\ \Eprint
  {http://arxiv.org/abs/1111.4425} {arXiv:1111.4425 [hep-lat]} \BibitemShut
  {NoStop}%
\bibitem [{\citenamefont {Detmold}\ \emph {et~al.}(2006)\citenamefont
  {Detmold}, \citenamefont {Tiburzi},\ and\ \citenamefont
  {Walker-Loud}}]{Detmold:2006vu}%
  \BibitemOpen
  \bibfield  {author} {\bibinfo {author} {\bibfnamefont {W.}~\bibnamefont
  {Detmold}}, \bibinfo {author} {\bibfnamefont {B.~C.}\ \bibnamefont
  {Tiburzi}}, \ and\ \bibinfo {author} {\bibfnamefont {Andre}\ \bibnamefont
  {Walker-Loud}},\ }\bibfield  {title} {\enquote {\bibinfo {title}
  {{Electromagnetic and spin polarisabilities in lattice QCD}},}\ }\href
  {\doibase 10.1103/PhysRevD.73.114505} {\bibfield  {journal} {\bibinfo
  {journal} {Phys. Rev.}\ }\textbf {\bibinfo {volume} {D73}},\ \bibinfo {pages}
  {114505} (\bibinfo {year} {2006})},\ \Eprint
  {http://arxiv.org/abs/hep-lat/0603026} {arXiv:hep-lat/0603026 [hep-lat]}
  \BibitemShut {NoStop}%
\bibitem [{\citenamefont {Detmold}\ \emph
  {et~al.}(2009{\natexlab{b}})\citenamefont {Detmold}, \citenamefont
  {Tiburzi},\ and\ \citenamefont {Walker-Loud}}]{Detmold_2009}%
  \BibitemOpen
  \bibfield  {author} {\bibinfo {author} {\bibfnamefont {W.}~\bibnamefont
  {Detmold}}, \bibinfo {author} {\bibfnamefont {B.~C.}\ \bibnamefont
  {Tiburzi}}, \ and\ \bibinfo {author} {\bibfnamefont {A.}~\bibnamefont
  {Walker-Loud}},\ }\bibfield  {title} {\enquote {\bibinfo {title} {Extracting
  electric polarizabilities from lattice {QCD}},}\ }\href {\doibase
  10.1103/physrevd.79.094505} {\bibfield  {journal} {\bibinfo  {journal}
  {Physical Review D}\ }\textbf {\bibinfo {volume} {79}} (\bibinfo {year}
  {2009}{\natexlab{b}}),\ 10.1103/physrevd.79.094505}\BibitemShut {NoStop}%
\bibitem [{\citenamefont {Niyazi}\ \emph {et~al.}(2021)\citenamefont {Niyazi},
  \citenamefont {Alexandru}, \citenamefont {Lee},\ and\ \citenamefont
  {Lujan}}]{niyazi2021charged}%
  \BibitemOpen
  \bibfield  {author} {\bibinfo {author} {\bibfnamefont {Hossein}\ \bibnamefont
  {Niyazi}}, \bibinfo {author} {\bibfnamefont {Andrei}\ \bibnamefont
  {Alexandru}}, \bibinfo {author} {\bibfnamefont {Frank~X.}\ \bibnamefont
  {Lee}}, \ and\ \bibinfo {author} {\bibfnamefont {Michael}\ \bibnamefont
  {Lujan}},\ }\href@noop {} {\enquote {\bibinfo {title} {Charged pion electric
  polarizability from lattice qcd},}\ } (\bibinfo {year} {2021}),\ \Eprint
  {http://arxiv.org/abs/2105.06906} {arXiv:2105.06906 [hep-lat]} \BibitemShut
  {NoStop}%
\bibitem [{\citenamefont {Bignell}\ \emph
  {et~al.}(2020{\natexlab{b}})\citenamefont {Bignell}, \citenamefont {Kamleh},\
  and\ \citenamefont {Leinweber}}]{Bignell_2020}%
  \BibitemOpen
  \bibfield  {author} {\bibinfo {author} {\bibfnamefont {Ryan}\ \bibnamefont
  {Bignell}}, \bibinfo {author} {\bibfnamefont {Waseem}\ \bibnamefont
  {Kamleh}}, \ and\ \bibinfo {author} {\bibfnamefont {Derek}\ \bibnamefont
  {Leinweber}},\ }\bibfield  {title} {\enquote {\bibinfo {title} {Pion magnetic
  polarisability using the background field method},}\ }\href {\doibase
  10.1016/j.physletb.2020.135853} {\bibfield  {journal} {\bibinfo  {journal}
  {Physics Letters B}\ }\textbf {\bibinfo {volume} {811}},\ \bibinfo {pages}
  {135853} (\bibinfo {year} {2020}{\natexlab{b}})}\BibitemShut {NoStop}%
\bibitem [{\citenamefont {He}\ \emph {et~al.}(2021)\citenamefont {He},
  \citenamefont {Leinweber}, \citenamefont {Thomas},\ and\ \citenamefont
  {Wang}}]{He:2021eha}%
  \BibitemOpen
  \bibfield  {author} {\bibinfo {author} {\bibfnamefont {Fangcheng}\
  \bibnamefont {He}}, \bibinfo {author} {\bibfnamefont {Derek~B.}\ \bibnamefont
  {Leinweber}}, \bibinfo {author} {\bibfnamefont {Anthony~W.}\ \bibnamefont
  {Thomas}}, \ and\ \bibinfo {author} {\bibfnamefont {Ping}\ \bibnamefont
  {Wang}},\ }\bibfield  {title} {\enquote {\bibinfo {title} {{Chiral
  extrapolation of the charged-pion magnetic polarizability with Pad\'e
  approximant}},}\ }\href@noop {} {\  (\bibinfo {year} {2021})},\ \Eprint
  {http://arxiv.org/abs/2104.09963} {arXiv:2104.09963 [nucl-th]} \BibitemShut
  {NoStop}%
\bibitem [{\citenamefont {Liang}\ \emph {et~al.}(2020)\citenamefont {Liang},
  \citenamefont {Draper}, \citenamefont {Liu}, \citenamefont {Rothkopf},\ and\
  \citenamefont {Yang}}]{Liang:2019frk}%
  \BibitemOpen
  \bibfield  {author} {\bibinfo {author} {\bibfnamefont {Jian}\ \bibnamefont
  {Liang}}, \bibinfo {author} {\bibfnamefont {Terrence}\ \bibnamefont
  {Draper}}, \bibinfo {author} {\bibfnamefont {Keh-Fei}\ \bibnamefont {Liu}},
  \bibinfo {author} {\bibfnamefont {Alexander}\ \bibnamefont {Rothkopf}}, \
  and\ \bibinfo {author} {\bibfnamefont {Yi-Bo}\ \bibnamefont {Yang}} (\bibinfo
  {collaboration} {XQCD}),\ }\bibfield  {title} {\enquote {\bibinfo {title}
  {{Towards the nucleon hadronic tensor from lattice QCD}},}\ }\href {\doibase
  10.1103/PhysRevD.101.114503} {\bibfield  {journal} {\bibinfo  {journal}
  {Phys. Rev. D}\ }\textbf {\bibinfo {volume} {101}},\ \bibinfo {pages}
  {114503} (\bibinfo {year} {2020})},\ \Eprint
  {http://arxiv.org/abs/1906.05312} {arXiv:1906.05312 [hep-ph]} \BibitemShut
  {NoStop}%
\bibitem [{\citenamefont {Liang}\ and\ \citenamefont
  {Liu}(2020)}]{Liang_2020a}%
  \BibitemOpen
  \bibfield  {author} {\bibinfo {author} {\bibfnamefont {Jian}\ \bibnamefont
  {Liang}}\ and\ \bibinfo {author} {\bibfnamefont {Keh-Fei}\ \bibnamefont
  {Liu}},\ }\bibfield  {title} {\enquote {\bibinfo {title} {Pdfs and
  neutrino-nucleon scattering from hadronic tensor},}\ }\href {\doibase
  10.22323/1.363.0046} {\bibfield  {journal} {\bibinfo  {journal} {Proceedings
  of 37th International Symposium on Lattice Field Theory ---
  PoS(LATTICE2019)}\ } (\bibinfo {year} {2020}),\
  10.22323/1.363.0046}\BibitemShut {NoStop}%
\bibitem [{\citenamefont {Fu}(2012)}]{Fu_2012}%
  \BibitemOpen
  \bibfield  {author} {\bibinfo {author} {\bibfnamefont {Ziwen}\ \bibnamefont
  {Fu}},\ }\bibfield  {title} {\enquote {\bibinfo {title} {Lattice study on
  $\ensuremath{\pi}k$ scattering with moving wall source},}\ }\href {\doibase
  10.1103/PhysRevD.85.074501} {\bibfield  {journal} {\bibinfo  {journal} {Phys.
  Rev. D}\ }\textbf {\bibinfo {volume} {85}},\ \bibinfo {pages} {074501}
  (\bibinfo {year} {2012})}\BibitemShut {NoStop}%
\bibitem [{\citenamefont {Alexandrou}(2004)}]{Alexandrou_2004}%
  \BibitemOpen
  \bibfield  {author} {\bibinfo {author} {\bibfnamefont {C.}~\bibnamefont
  {Alexandrou}},\ }\bibfield  {title} {\enquote {\bibinfo {title} {Hadron
  deformation from lattice qcd},}\ }\href {\doibase
  10.1016/s0920-5632(03)02451-4} {\bibfield  {journal} {\bibinfo  {journal}
  {Nuclear Physics B - Proceedings Supplements}\ }\textbf {\bibinfo {volume}
  {128}},\ \bibinfo {pages} {1--8} (\bibinfo {year} {2004})}\BibitemShut
  {NoStop}%
\bibitem [{\citenamefont {Bali}\ \emph
  {et~al.}(2018{\natexlab{b}})\citenamefont {Bali}, \citenamefont {Bruns},
  \citenamefont {Castagnini}, \citenamefont {Diehl}, \citenamefont {Gaunt},
  \citenamefont {Gl{\"a}{\ss}le}, \citenamefont {Sch{\"a}fer}, \citenamefont
  {Sternbeck},\ and\ \citenamefont {Zimmermann}}]{Bali_2018}%
  \BibitemOpen
  \bibfield  {author} {\bibinfo {author} {\bibfnamefont {Gunnar~S.}\
  \bibnamefont {Bali}}, \bibinfo {author} {\bibfnamefont {Peter~C.}\
  \bibnamefont {Bruns}}, \bibinfo {author} {\bibfnamefont {Luca}\ \bibnamefont
  {Castagnini}}, \bibinfo {author} {\bibfnamefont {Markus}\ \bibnamefont
  {Diehl}}, \bibinfo {author} {\bibfnamefont {Jonathan~R.}\ \bibnamefont
  {Gaunt}}, \bibinfo {author} {\bibfnamefont {Benjamin}\ \bibnamefont
  {Gl{\"a}{\ss}le}}, \bibinfo {author} {\bibfnamefont {Andreas}\ \bibnamefont
  {Sch{\"a}fer}}, \bibinfo {author} {\bibfnamefont {Andr{\'e}}\ \bibnamefont
  {Sternbeck}}, \ and\ \bibinfo {author} {\bibfnamefont {Christian}\
  \bibnamefont {Zimmermann}},\ }\bibfield  {title} {\enquote {\bibinfo {title}
  {Two-current correlations in the pion on the lattice},}\ }\href {\doibase
  10.1007/jhep12(2018)061} {\bibfield  {journal} {\bibinfo  {journal} {Journal
  of High Energy Physics}\ }\textbf {\bibinfo {volume} {2018}} (\bibinfo {year}
  {2018}{\natexlab{b}}),\ 10.1007/jhep12(2018)061}\BibitemShut {NoStop}%
\bibitem [{\citenamefont {Bali}\ \emph {et~al.}(2021)\citenamefont {Bali},
  \citenamefont {Diehl}, \citenamefont {Gl{\"a}{\ss}le}, \citenamefont
  {Sch{\"a}fer},\ and\ \citenamefont {Zimmermann}}]{bali2021double}%
  \BibitemOpen
  \bibfield  {author} {\bibinfo {author} {\bibfnamefont {Gunnar~S.}\
  \bibnamefont {Bali}}, \bibinfo {author} {\bibfnamefont {Markus}\ \bibnamefont
  {Diehl}}, \bibinfo {author} {\bibfnamefont {Benjamin}\ \bibnamefont
  {Gl{\"a}{\ss}le}}, \bibinfo {author} {\bibfnamefont {Andreas}\ \bibnamefont
  {Sch{\"a}fer}}, \ and\ \bibinfo {author} {\bibfnamefont {Christian}\
  \bibnamefont {Zimmermann}},\ }\href@noop {} {\enquote {\bibinfo {title}
  {Double parton distributions in the nucleon from lattice qcd},}\ } (\bibinfo
  {year} {2021}),\ \Eprint {http://arxiv.org/abs/2106.03451} {arXiv:2106.03451
  [hep-lat]} \BibitemShut {NoStop}%
\bibitem [{\citenamefont {Burkardt}\ \emph {et~al.}(1995)\citenamefont
  {Burkardt}, \citenamefont {Grandy},\ and\ \citenamefont
  {Negele}}]{BURKARDT1995441}%
  \BibitemOpen
  \bibfield  {author} {\bibinfo {author} {\bibfnamefont {M.}~\bibnamefont
  {Burkardt}}, \bibinfo {author} {\bibfnamefont {J.M.}\ \bibnamefont {Grandy}},
  \ and\ \bibinfo {author} {\bibfnamefont {J.W.}\ \bibnamefont {Negele}},\
  }\bibfield  {title} {\enquote {\bibinfo {title} {Calculation and
  interpretation of hadron correlation functions in lattice qcd},}\ }\href
  {\doibase https://doi.org/10.1006/aphy.1995.1026} {\bibfield  {journal}
  {\bibinfo  {journal} {Annals of Physics}\ }\textbf {\bibinfo {volume}
  {238}},\ \bibinfo {pages} {441--472} (\bibinfo {year} {1995})}\BibitemShut
  {NoStop}%
\bibitem [{\citenamefont {Wilcox}(1997)}]{Wilcox:1996vx}%
  \BibitemOpen
  \bibfield  {author} {\bibinfo {author} {\bibfnamefont {Walter}\ \bibnamefont
  {Wilcox}},\ }\bibfield  {title} {\enquote {\bibinfo {title} {{Lattice charge
  overlap. 2: Aspects of charged pion polarizability}},}\ }\href {\doibase
  10.1006/aphy.1996.5649} {\bibfield  {journal} {\bibinfo  {journal} {Annals
  Phys.}\ }\textbf {\bibinfo {volume} {255}},\ \bibinfo {pages} {60--74}
  (\bibinfo {year} {1997})},\ \Eprint {http://arxiv.org/abs/hep-lat/9606019}
  {arXiv:hep-lat/9606019} \BibitemShut {NoStop}%
\bibitem [{\citenamefont {Feng}\ \emph {et~al.}(2022)\citenamefont {Feng},
  \citenamefont {Izubuchi}, \citenamefont {Jin},\ and\ \citenamefont
  {Golterman}}]{Feng:2022rkr}%
  \BibitemOpen
  \bibfield  {author} {\bibinfo {author} {\bibfnamefont {Xu}~\bibnamefont
  {Feng}}, \bibinfo {author} {\bibfnamefont {Taku}\ \bibnamefont {Izubuchi}},
  \bibinfo {author} {\bibfnamefont {Luchang}\ \bibnamefont {Jin}}, \ and\
  \bibinfo {author} {\bibfnamefont {Maarten}\ \bibnamefont {Golterman}},\
  }\bibfield  {title} {\enquote {\bibinfo {title} {{Pion electric
  polarizabilities from lattice QCD}},}\ }\href {\doibase 10.22323/1.396.0362}
  {\bibfield  {journal} {\bibinfo  {journal} {PoS}\ }\textbf {\bibinfo {volume}
  {LATTICE2021}},\ \bibinfo {pages} {362} (\bibinfo {year} {2022})},\ \Eprint
  {http://arxiv.org/abs/2201.01396} {arXiv:2201.01396 [hep-lat]} \BibitemShut
  {NoStop}%
\bibitem [{\citenamefont {Wang}\ \emph {et~al.}(2022)\citenamefont {Wang},
  \citenamefont {Feng},\ and\ \citenamefont {Jin}}]{Wang:2021}%
  \BibitemOpen
  \bibfield  {author} {\bibinfo {author} {\bibfnamefont {X.H.}\ \bibnamefont
  {Wang}}, \bibinfo {author} {\bibfnamefont {X.}~\bibnamefont {Feng}}, \ and\
  \bibinfo {author} {\bibfnamefont {L.C.}\ \bibnamefont {Jin}},\ }\bibfield
  {title} {\enquote {\bibinfo {title} {{Lattice QCD calculation of the proton
  electromagnetic polarizability}},}\ }\href@noop {} {\bibfield  {journal}
  {\bibinfo  {journal} {LATTICE2021 proceedings}\ } (\bibinfo {year}
  {2022})}\BibitemShut {NoStop}%
\bibitem [{\citenamefont {Wilcox}\ and\ \citenamefont
  {Lee}(2021)}]{Wilcox:2021rtt}%
  \BibitemOpen
  \bibfield  {author} {\bibinfo {author} {\bibfnamefont {Walter}\ \bibnamefont
  {Wilcox}}\ and\ \bibinfo {author} {\bibfnamefont {Frank~X.}\ \bibnamefont
  {Lee}},\ }\bibfield  {title} {\enquote {\bibinfo {title} {{Towards charged
  hadron polarizabilities from four-point functions in lattice QCD}},}\ }\href
  {\doibase 10.1103/PhysRevD.104.034506} {\bibfield  {journal} {\bibinfo
  {journal} {Phys. Rev. D}\ }\textbf {\bibinfo {volume} {104}},\ \bibinfo
  {pages} {034506} (\bibinfo {year} {2021})},\ \Eprint
  {http://arxiv.org/abs/2106.02557} {arXiv:2106.02557 [hep-lat]} \BibitemShut
  {NoStop}%
\bibitem [{\citenamefont {Ivanov}\ and\ \citenamefont
  {Mizutani}(1992)}]{PhysRevD.45.1580}%
  \BibitemOpen
  \bibfield  {author} {\bibinfo {author} {\bibfnamefont {M.~A.}\ \bibnamefont
  {Ivanov}}\ and\ \bibinfo {author} {\bibfnamefont {T.}~\bibnamefont
  {Mizutani}},\ }\bibfield  {title} {\enquote {\bibinfo {title} {Pion and kaon
  polarizabilities in the quark confinement model},}\ }\href {\doibase
  10.1103/PhysRevD.45.1580} {\bibfield  {journal} {\bibinfo  {journal} {Phys.
  Rev. D}\ }\textbf {\bibinfo {volume} {45}},\ \bibinfo {pages} {1580--1601}
  (\bibinfo {year} {1992})}\BibitemShut {NoStop}%
\bibitem [{\citenamefont {Dorokhov}\ \emph {et~al.}(1997)\citenamefont
  {Dorokhov}, \citenamefont {Volkov}, \citenamefont {H{\"u}fner}, \citenamefont
  {Klevansky},\ and\ \citenamefont {Rehberg}}]{Dorokhov:1997aa}%
  \BibitemOpen
  \bibfield  {author} {\bibinfo {author} {\bibfnamefont {A.~E.}\ \bibnamefont
  {Dorokhov}}, \bibinfo {author} {\bibfnamefont {M.~K.}\ \bibnamefont
  {Volkov}}, \bibinfo {author} {\bibfnamefont {J.}~\bibnamefont {H{\"u}fner}},
  \bibinfo {author} {\bibfnamefont {S.~P.}\ \bibnamefont {Klevansky}}, \ and\
  \bibinfo {author} {\bibfnamefont {P.}~\bibnamefont {Rehberg}},\ }\bibfield
  {title} {\enquote {\bibinfo {title} {Pion polarizabilities at finite
  temperature},}\ }\href {\doibase 10.1007/s002880050454} {\bibfield  {journal}
  {\bibinfo  {journal} {Zeitschrift f{\"u}r Physik C Particles and Fields}\
  }\textbf {\bibinfo {volume} {75}},\ \bibinfo {pages} {127--135} (\bibinfo
  {year} {1997})}\BibitemShut {NoStop}%
\bibitem [{\citenamefont {Bernard}\ and\ \citenamefont
  {Vautherin}(1989)}]{PhysRevD.40.1615}%
  \BibitemOpen
  \bibfield  {author} {\bibinfo {author} {\bibfnamefont {V\'eronique}\
  \bibnamefont {Bernard}}\ and\ \bibinfo {author} {\bibfnamefont
  {D.}~\bibnamefont {Vautherin}},\ }\bibfield  {title} {\enquote {\bibinfo
  {title} {Electromagnetic polarizabilities of pseudoscalar goldstone
  bosons},}\ }\href {\doibase 10.1103/PhysRevD.40.1615} {\bibfield  {journal}
  {\bibinfo  {journal} {Phys. Rev. D}\ }\textbf {\bibinfo {volume} {40}},\
  \bibinfo {pages} {1615--1627} (\bibinfo {year} {1989})}\BibitemShut {NoStop}%
\bibitem [{\citenamefont {Bernard}\ \emph {et~al.}(1988)\citenamefont
  {Bernard}, \citenamefont {Hiller},\ and\ \citenamefont
  {Weise}}]{BERNARD198816}%
  \BibitemOpen
  \bibfield  {author} {\bibinfo {author} {\bibfnamefont {V{\'e}ronique}\
  \bibnamefont {Bernard}}, \bibinfo {author} {\bibfnamefont {Brigitte}\
  \bibnamefont {Hiller}}, \ and\ \bibinfo {author} {\bibfnamefont {Wolfram}\
  \bibnamefont {Weise}},\ }\bibfield  {title} {\enquote {\bibinfo {title} {Pion
  electromagnetic polarizability and chiral models},}\ }\href {\doibase
  https://doi.org/10.1016/0370-2693(88)90391-7} {\bibfield  {journal} {\bibinfo
   {journal} {Physics Letters B}\ }\textbf {\bibinfo {volume} {205}},\ \bibinfo
  {pages} {16--21} (\bibinfo {year} {1988})}\BibitemShut {NoStop}%
\bibitem [{\citenamefont {L'vov}(1993)}]{Lvov:1993fp}%
  \BibitemOpen
  \bibfield  {author} {\bibinfo {author} {\bibfnamefont {A.~I.}\ \bibnamefont
  {L'vov}},\ }\bibfield  {title} {\enquote {\bibinfo {title} {{Theoretical
  aspects of the polarizability of the nucleon}},}\ }\href {\doibase
  10.1142/S0217751X93002095} {\bibfield  {journal} {\bibinfo  {journal} {Int.
  J. Mod. Phys. A}\ }\textbf {\bibinfo {volume} {8}},\ \bibinfo {pages}
  {5267--5303} (\bibinfo {year} {1993})}\BibitemShut {NoStop}%
\bibitem [{\citenamefont {L'vov}\ \emph {et~al.}(2001)\citenamefont {L'vov},
  \citenamefont {Scherer}, \citenamefont {Pasquini}, \citenamefont {Unkmeir},\
  and\ \citenamefont {Drechsel}}]{Lvov_2001}%
  \BibitemOpen
  \bibfield  {author} {\bibinfo {author} {\bibfnamefont {A.~I.}\ \bibnamefont
  {L'vov}}, \bibinfo {author} {\bibfnamefont {S.}~\bibnamefont {Scherer}},
  \bibinfo {author} {\bibfnamefont {B.}~\bibnamefont {Pasquini}}, \bibinfo
  {author} {\bibfnamefont {C.}~\bibnamefont {Unkmeir}}, \ and\ \bibinfo
  {author} {\bibfnamefont {D.}~\bibnamefont {Drechsel}},\ }\bibfield  {title}
  {\enquote {\bibinfo {title} {Generalized dipole polarizabilities and the
  spatial structure of hadrons},}\ }\href {\doibase 10.1103/PhysRevC.64.015203}
  {\bibfield  {journal} {\bibinfo  {journal} {Phys. Rev. C}\ }\textbf {\bibinfo
  {volume} {64}},\ \bibinfo {pages} {015203} (\bibinfo {year}
  {2001})}\BibitemShut {NoStop}%
\bibitem [{\citenamefont {Pasquini}\ \emph {et~al.}(2010)\citenamefont
  {Pasquini}, \citenamefont {Drechsel},\ and\ \citenamefont
  {Scherer}}]{PhysRevC.81.029802}%
  \BibitemOpen
  \bibfield  {author} {\bibinfo {author} {\bibfnamefont {B.}~\bibnamefont
  {Pasquini}}, \bibinfo {author} {\bibfnamefont {D.}~\bibnamefont {Drechsel}},
  \ and\ \bibinfo {author} {\bibfnamefont {S.}~\bibnamefont {Scherer}},\
  }\bibfield  {title} {\enquote {\bibinfo {title} {Reply to ``comment on
  `polarizability of the pion: No conflict between dispersion theory and chiral
  perturbation theory'''},}\ }\href {\doibase 10.1103/PhysRevC.81.029802}
  {\bibfield  {journal} {\bibinfo  {journal} {Phys. Rev. C}\ }\textbf {\bibinfo
  {volume} {81}},\ \bibinfo {pages} {029802} (\bibinfo {year}
  {2010})}\BibitemShut {NoStop}%
\bibitem [{\citenamefont {Fil'kov}\ and\ \citenamefont
  {Kashevarov}(2017)}]{Fil_kov_2017}%
  \BibitemOpen
  \bibfield  {author} {\bibinfo {author} {\bibfnamefont {L.~V.}\ \bibnamefont
  {Fil'kov}}\ and\ \bibinfo {author} {\bibfnamefont {V.~L.}\ \bibnamefont
  {Kashevarov}},\ }\bibfield  {title} {\enquote {\bibinfo {title} {Dipole
  polarizabilities of charged pions},}\ }\href {\doibase
  10.1134/s1063779617010063} {\bibfield  {journal} {\bibinfo  {journal}
  {Physics of Particles and Nuclei}\ }\textbf {\bibinfo {volume} {48}},\
  \bibinfo {pages} {117--123} (\bibinfo {year} {2017})}\BibitemShut {NoStop}%
\bibitem [{\citenamefont {Moinester}\ and\ \citenamefont
  {Scherer}(2019)}]{Moinester:2019sew}%
  \BibitemOpen
  \bibfield  {author} {\bibinfo {author} {\bibfnamefont {Murray}\ \bibnamefont
  {Moinester}}\ and\ \bibinfo {author} {\bibfnamefont {Stefan}\ \bibnamefont
  {Scherer}},\ }\bibfield  {title} {\enquote {\bibinfo {title} {{Compton
  Scattering off Pions and Electromagnetic Polarizabilities}},}\ }\href
  {\doibase 10.1142/S0217751X19300084} {\bibfield  {journal} {\bibinfo
  {journal} {Int. J. Mod. Phys. A}\ }\textbf {\bibinfo {volume} {34}},\
  \bibinfo {pages} {1930008} (\bibinfo {year} {2019})},\ \Eprint
  {http://arxiv.org/abs/1905.05640} {arXiv:1905.05640 [hep-ph]} \BibitemShut
  {NoStop}%
\bibitem [{\citenamefont {Lensky}\ and\ \citenamefont
  {Pascalutsa}(2010)}]{Lensky:2009uv}%
  \BibitemOpen
  \bibfield  {author} {\bibinfo {author} {\bibfnamefont {Vadim}\ \bibnamefont
  {Lensky}}\ and\ \bibinfo {author} {\bibfnamefont {Vladimir}\ \bibnamefont
  {Pascalutsa}},\ }\bibfield  {title} {\enquote {\bibinfo {title} {{Predictive
  powers of chiral perturbation theory in Compton scattering off protons}},}\
  }\href {\doibase 10.1140/epjc/s10052-009-1183-z} {\bibfield  {journal}
  {\bibinfo  {journal} {Eur. Phys. J. C}\ }\textbf {\bibinfo {volume} {65}},\
  \bibinfo {pages} {195--209} (\bibinfo {year} {2010})},\ \Eprint
  {http://arxiv.org/abs/0907.0451} {arXiv:0907.0451 [hep-ph]} \BibitemShut
  {NoStop}%
\bibitem [{\citenamefont {Hagelstein}(2020)}]{Hagelstein:2020vog}%
  \BibitemOpen
  \bibfield  {author} {\bibinfo {author} {\bibfnamefont {Franziska}\
  \bibnamefont {Hagelstein}},\ }\bibfield  {title} {\enquote {\bibinfo {title}
  {{Nucleon Polarizabilities and Compton Scattering as a Playground for Chiral
  Perturbation Theory}},}\ }\href {\doibase 10.3390/sym12091407} {\bibfield
  {journal} {\bibinfo  {journal} {Symmetry}\ }\textbf {\bibinfo {volume}
  {12}},\ \bibinfo {pages} {1407} (\bibinfo {year} {2020})},\ \Eprint
  {http://arxiv.org/abs/2006.16124} {arXiv:2006.16124 [nucl-th]} \BibitemShut
  {NoStop}%
\bibitem [{\citenamefont {McGovern}\ \emph {et~al.}(2013)\citenamefont
  {McGovern}, \citenamefont {Phillips},\ and\ \citenamefont
  {Griesshammer}}]{McGovern:2012ew}%
  \BibitemOpen
  \bibfield  {author} {\bibinfo {author} {\bibfnamefont {J.~A.}\ \bibnamefont
  {McGovern}}, \bibinfo {author} {\bibfnamefont {D.~R.}\ \bibnamefont
  {Phillips}}, \ and\ \bibinfo {author} {\bibfnamefont {H.~W.}\ \bibnamefont
  {Griesshammer}},\ }\bibfield  {title} {\enquote {\bibinfo {title} {{Compton
  scattering from the proton in an effective field theory with explicit Delta
  degrees of freedom}},}\ }\href {\doibase 10.1140/epja/i2013-13012-1}
  {\bibfield  {journal} {\bibinfo  {journal} {Eur. Phys. J. A}\ }\textbf
  {\bibinfo {volume} {49}},\ \bibinfo {pages} {12} (\bibinfo {year} {2013})},\
  \Eprint {http://arxiv.org/abs/1210.4104} {arXiv:1210.4104 [nucl-th]}
  \BibitemShut {NoStop}%
\bibitem [{\citenamefont {Griesshammer}\ \emph {et~al.}(2012)\citenamefont
  {Griesshammer}, \citenamefont {McGovern}, \citenamefont {Phillips},\ and\
  \citenamefont {Feldman}}]{Griesshammer:2012we}%
  \BibitemOpen
  \bibfield  {author} {\bibinfo {author} {\bibfnamefont {H.~W.}\ \bibnamefont
  {Griesshammer}}, \bibinfo {author} {\bibfnamefont {J.~A.}\ \bibnamefont
  {McGovern}}, \bibinfo {author} {\bibfnamefont {D.~R.}\ \bibnamefont
  {Phillips}}, \ and\ \bibinfo {author} {\bibfnamefont {G.}~\bibnamefont
  {Feldman}},\ }\bibfield  {title} {\enquote {\bibinfo {title} {{Using
  effective field theory to analyse low-energy Compton scattering data from
  protons and light nuclei}},}\ }\href {\doibase 10.1016/j.ppnp.2012.04.003}
  {\bibfield  {journal} {\bibinfo  {journal} {Prog. Part. Nucl. Phys.}\
  }\textbf {\bibinfo {volume} {67}},\ \bibinfo {pages} {841--897} (\bibinfo
  {year} {2012})},\ \Eprint {http://arxiv.org/abs/1203.6834} {arXiv:1203.6834
  [nucl-th]} \BibitemShut {NoStop}%
\bibitem [{\citenamefont {Moinester}(2022)}]{Moinester:2022tba}%
  \BibitemOpen
  \bibfield  {author} {\bibinfo {author} {\bibfnamefont {Murray}\ \bibnamefont
  {Moinester}},\ }\bibfield  {title} {\enquote {\bibinfo {title} {{Pion
  Polarizability 2022 Status Report}},}\ \ }(\bibinfo {year} {2022})\ \Eprint
  {http://arxiv.org/abs/2205.09954} {arXiv:2205.09954 [hep-ph]} \BibitemShut
  {NoStop}%
\bibitem [{\citenamefont {Draper}\ \emph {et~al.}(1989)\citenamefont {Draper},
  \citenamefont {Woloshyn}, \citenamefont {Wilcox},\ and\ \citenamefont
  {Liu}}]{Draper:1988bp}%
  \BibitemOpen
  \bibfield  {author} {\bibinfo {author} {\bibfnamefont {Terrence}\
  \bibnamefont {Draper}}, \bibinfo {author} {\bibfnamefont {R.~M.}\
  \bibnamefont {Woloshyn}}, \bibinfo {author} {\bibfnamefont {Walter}\
  \bibnamefont {Wilcox}}, \ and\ \bibinfo {author} {\bibfnamefont {Keh-Fei}\
  \bibnamefont {Liu}},\ }\bibfield  {title} {\enquote {\bibinfo {title} {{The
  Pion Form-factor in Lattice {QCD}}},}\ }\href {\doibase
  10.1016/0550-3213(89)90609-3} {\bibfield  {journal} {\bibinfo  {journal}
  {Nucl. Phys. B}\ }\textbf {\bibinfo {volume} {318}},\ \bibinfo {pages}
  {319--336} (\bibinfo {year} {1989})}\BibitemShut {NoStop}%
\bibitem [{\citenamefont {Kuramashi}\ \emph {et~al.}(1993)\citenamefont
  {Kuramashi}, \citenamefont {Fukugita}, \citenamefont {Mino}, \citenamefont
  {Okawa},\ and\ \citenamefont {Ukawa}}]{WallSource1993}%
  \BibitemOpen
  \bibfield  {author} {\bibinfo {author} {\bibfnamefont {Y.}~\bibnamefont
  {Kuramashi}}, \bibinfo {author} {\bibfnamefont {M.}~\bibnamefont {Fukugita}},
  \bibinfo {author} {\bibfnamefont {H.}~\bibnamefont {Mino}}, \bibinfo {author}
  {\bibfnamefont {M.}~\bibnamefont {Okawa}}, \ and\ \bibinfo {author}
  {\bibfnamefont {A.}~\bibnamefont {Ukawa}},\ }\bibfield  {title} {\enquote
  {\bibinfo {title} {Lattice qcd calculation of full pion scattering
  lengths},}\ }\href {\doibase 10.1103/PhysRevLett.71.2387} {\bibfield
  {journal} {\bibinfo  {journal} {Phys. Rev. Lett.}\ }\textbf {\bibinfo
  {volume} {71}},\ \bibinfo {pages} {2387--2390} (\bibinfo {year}
  {1993})}\BibitemShut {NoStop}%
\bibitem [{\citenamefont {et~al}(1991)}]{CABASINO1991195}%
  \BibitemOpen
  \bibfield  {author} {\bibinfo {author} {\bibfnamefont {S.~Cabasino}\
  \bibnamefont {et~al}},\ }\bibfield  {title} {\enquote {\bibinfo {title}
  {beta=6.0 quenched wilson fermions},}\ }\href@noop {} {\bibfield  {journal}
  {\bibinfo  {journal} {Physics Letters B}\ }\textbf {\bibinfo {volume}
  {258}},\ \bibinfo {pages} {195--201} (\bibinfo {year} {1991})}\BibitemShut
  {NoStop}%
\bibitem [{\citenamefont {Martinelli}\ and\ \citenamefont
  {Sachrajda}(1988)}]{MARTINELLI1988865}%
  \BibitemOpen
  \bibfield  {author} {\bibinfo {author} {\bibfnamefont {G.}~\bibnamefont
  {Martinelli}}\ and\ \bibinfo {author} {\bibfnamefont {C.T.}\ \bibnamefont
  {Sachrajda}},\ }\bibfield  {title} {\enquote {\bibinfo {title} {A lattice
  calculation of the pion's form factor and structure function},}\ }\href
  {\doibase https://doi.org/10.1016/0550-3213(88)90445-2} {\bibfield  {journal}
  {\bibinfo  {journal} {Nuclear Physics B}\ }\textbf {\bibinfo {volume}
  {306}},\ \bibinfo {pages} {865--889} (\bibinfo {year} {1988})}\BibitemShut
  {NoStop}%
\bibitem [{\citenamefont {Lee}\ \emph {et~al.}(2015)\citenamefont {Lee},
  \citenamefont {Arrington},\ and\ \citenamefont {Hill}}]{Lee_2015}%
  \BibitemOpen
  \bibfield  {author} {\bibinfo {author} {\bibfnamefont {Gabriel}\ \bibnamefont
  {Lee}}, \bibinfo {author} {\bibfnamefont {John~R.}\ \bibnamefont
  {Arrington}}, \ and\ \bibinfo {author} {\bibfnamefont {Richard~J.}\
  \bibnamefont {Hill}},\ }\bibfield  {title} {\enquote {\bibinfo {title}
  {Extraction of the proton radius from electron-proton scattering data},}\
  }\href {\doibase 10.1103/physrevd.92.013013} {\bibfield  {journal} {\bibinfo
  {journal} {Physical Review D}\ }\textbf {\bibinfo {volume} {92}} (\bibinfo
  {year} {2015}),\ 10.1103/physrevd.92.013013}\BibitemShut {NoStop}%
\bibitem [{\citenamefont {van~der Heide}\ \emph {et~al.}(2004)\citenamefont
  {van~der Heide}, \citenamefont {Koch},\ and\ \citenamefont
  {Laermann}}]{van_der_Heide_2004}%
  \BibitemOpen
  \bibfield  {author} {\bibinfo {author} {\bibfnamefont {J.}~\bibnamefont
  {van~der Heide}}, \bibinfo {author} {\bibfnamefont {J.~H.}\ \bibnamefont
  {Koch}}, \ and\ \bibinfo {author} {\bibfnamefont {E.}~\bibnamefont
  {Laermann}},\ }\bibfield  {title} {\enquote {\bibinfo {title} {Pion structure
  from improved lattice {QCD}: Form factor and charge radius at low masses},}\
  }\href {\doibase 10.1103/physrevd.69.094511} {\bibfield  {journal} {\bibinfo
  {journal} {Physical Review D}\ }\textbf {\bibinfo {volume} {69}} (\bibinfo
  {year} {2004}),\ 10.1103/physrevd.69.094511}\BibitemShut {NoStop}%
\bibitem [{\citenamefont {Gasser}\ \emph {et~al.}(2006)\citenamefont {Gasser},
  \citenamefont {Ivanov},\ and\ \citenamefont {Sainio}}]{Gasser_2006}%
  \BibitemOpen
  \bibfield  {author} {\bibinfo {author} {\bibfnamefont {J.}~\bibnamefont
  {Gasser}}, \bibinfo {author} {\bibfnamefont {M.A.}\ \bibnamefont {Ivanov}}, \
  and\ \bibinfo {author} {\bibfnamefont {M.E.}\ \bibnamefont {Sainio}},\
  }\bibfield  {title} {\enquote {\bibinfo {title} {Revisiting gamma + gamma to
  pi+ and pi- at low energies},}\ }\href {\doibase
  10.1016/j.nuclphysb.2006.03.022} {\bibfield  {journal} {\bibinfo  {journal}
  {Nuclear Physics B}\ }\textbf {\bibinfo {volume} {745}},\ \bibinfo {pages}
  {84--108} (\bibinfo {year} {2006})}\BibitemShut {NoStop}%
\bibitem [{\citenamefont {Workman}\ and\ \citenamefont
  {Others}(2022)}]{Workman:2022ynf}%
  \BibitemOpen
  \bibfield  {author} {\bibinfo {author} {\bibfnamefont {R.~L.}\ \bibnamefont
  {Workman}}\ and\ \bibinfo {author} {\bibnamefont {Others}} (\bibinfo
  {collaboration} {Particle Data Group}),\ }\bibfield  {title} {\enquote
  {\bibinfo {title} {{Review of Particle Physics}},}\ }\href {\doibase
  10.1093/ptep/ptac097} {\bibfield  {journal} {\bibinfo  {journal} {PTEP}\
  }\textbf {\bibinfo {volume} {2022}},\ \bibinfo {pages} {083C01} (\bibinfo
  {year} {2022})}\BibitemShut {NoStop}%
\bibitem [{\citenamefont {Alexandru}\ \emph {et~al.}(2012)\citenamefont
  {Alexandru}, \citenamefont {Pelissier}, \citenamefont {Gamari},\ and\
  \citenamefont {Lee}}]{Alexandru:2011ee}%
  \BibitemOpen
  \bibfield  {author} {\bibinfo {author} {\bibfnamefont {A.}~\bibnamefont
  {Alexandru}}, \bibinfo {author} {\bibfnamefont {C.}~\bibnamefont
  {Pelissier}}, \bibinfo {author} {\bibfnamefont {B.}~\bibnamefont {Gamari}}, \
  and\ \bibinfo {author} {\bibfnamefont {F.}~\bibnamefont {Lee}},\ }\bibfield
  {title} {\enquote {\bibinfo {title} {{Multi-mass solvers for lattice QCD on
  GPUs}},}\ }\href {\doibase 10.1016/j.jcp.2011.11.003} {\bibfield  {journal}
  {\bibinfo  {journal} {J. Comput. Phys.}\ }\textbf {\bibinfo {volume} {231}},\
  \bibinfo {pages} {1866--1878} (\bibinfo {year} {2012})},\ \Eprint
  {http://arxiv.org/abs/1103.5103} {arXiv:1103.5103 [hep-lat]} \BibitemShut
  {NoStop}%
\bibitem [{\citenamefont {Niyazi}\ \emph {et~al.}(2020)\citenamefont {Niyazi},
  \citenamefont {Alexandru}, \citenamefont {Lee},\ and\ \citenamefont
  {Brett}}]{Niyazi:2020erg}%
  \BibitemOpen
  \bibfield  {author} {\bibinfo {author} {\bibfnamefont {Hossein}\ \bibnamefont
  {Niyazi}}, \bibinfo {author} {\bibfnamefont {Andrei}\ \bibnamefont
  {Alexandru}}, \bibinfo {author} {\bibfnamefont {Frank~X.}\ \bibnamefont
  {Lee}}, \ and\ \bibinfo {author} {\bibfnamefont {Ruair\'\i{}}\ \bibnamefont
  {Brett}},\ }\bibfield  {title} {\enquote {\bibinfo {title} {{Setting the
  scale for nHYP fermions with the L\"uscher-Weisz gauge action}},}\ }\href
  {\doibase 10.1103/PhysRevD.102.094506} {\bibfield  {journal} {\bibinfo
  {journal} {Phys. Rev. D}\ }\textbf {\bibinfo {volume} {102}},\ \bibinfo
  {pages} {094506} (\bibinfo {year} {2020})},\ \Eprint
  {http://arxiv.org/abs/2008.13022} {arXiv:2008.13022 [hep-lat]} \BibitemShut
  {NoStop}%
\bibitem [{\citenamefont {Karsten}\ and\ \citenamefont
  {Smith}(1981)}]{karsten1981lattice}%
  \BibitemOpen
  \bibfield  {author} {\bibinfo {author} {\bibfnamefont {Luuk~H}\ \bibnamefont
  {Karsten}}\ and\ \bibinfo {author} {\bibfnamefont {Jan}\ \bibnamefont
  {Smith}},\ }\bibfield  {title} {\enquote {\bibinfo {title} {Lattice fermions:
  species doubling, chiral invariance and the triangle anomaly},}\ }\href@noop
  {} {\bibfield  {journal} {\bibinfo  {journal} {Nuclear Physics B}\ }\textbf
  {\bibinfo {volume} {183}},\ \bibinfo {pages} {103--140} (\bibinfo {year}
  {1981})}\BibitemShut {NoStop}%
\bibitem [{\citenamefont {Gattringer}\ and\ \citenamefont
  {Lang}(2010)}]{Gattringer:2010zz}%
  \BibitemOpen
  \bibfield  {author} {\bibinfo {author} {\bibfnamefont {Christof}\
  \bibnamefont {Gattringer}}\ and\ \bibinfo {author} {\bibfnamefont
  {Christian~B.}\ \bibnamefont {Lang}},\ }\href {\doibase
  10.1007/978-3-642-01850-3} {\emph {\bibinfo {title} {{Quantum chromodynamics
  on the lattice}}}},\ Vol.\ \bibinfo {volume} {788}\ (\bibinfo  {publisher}
  {Springer},\ \bibinfo {address} {Berlin},\ \bibinfo {year}
  {2010})\BibitemShut {NoStop}%
\bibitem [{\citenamefont {Wilcox}\ \emph {et~al.}(1992)\citenamefont {Wilcox},
  \citenamefont {Draper},\ and\ \citenamefont {Liu}}]{Wilcox:1991cq}%
  \BibitemOpen
  \bibfield  {author} {\bibinfo {author} {\bibfnamefont {Walter}\ \bibnamefont
  {Wilcox}}, \bibinfo {author} {\bibfnamefont {Terrence}\ \bibnamefont
  {Draper}}, \ and\ \bibinfo {author} {\bibfnamefont {Keh-Fei}\ \bibnamefont
  {Liu}},\ }\bibfield  {title} {\enquote {\bibinfo {title} {{Chiral limit of
  nucleon lattice electromagnetic form-factors}},}\ }\href {\doibase
  10.1103/PhysRevD.46.1109} {\bibfield  {journal} {\bibinfo  {journal} {Phys.
  Rev. D}\ }\textbf {\bibinfo {volume} {46}},\ \bibinfo {pages} {1109--1122}
  (\bibinfo {year} {1992})},\ \Eprint {http://arxiv.org/abs/hep-lat/9205015}
  {arXiv:hep-lat/9205015} \BibitemShut {NoStop}%
\bibitem [{\citenamefont {Alexandru}\ \emph {et~al.}(2005)\citenamefont
  {Alexandru}, \citenamefont {Faber}, \citenamefont {Horv{\'a}th},\ and\
  \citenamefont {Liu}}]{Alexandru_2005}%
  \BibitemOpen
  \bibfield  {author} {\bibinfo {author} {\bibfnamefont {Andrei}\ \bibnamefont
  {Alexandru}}, \bibinfo {author} {\bibfnamefont {Manfried}\ \bibnamefont
  {Faber}}, \bibinfo {author} {\bibfnamefont {Ivan}\ \bibnamefont
  {Horv{\'a}th}}, \ and\ \bibinfo {author} {\bibfnamefont {Keh-Fei}\
  \bibnamefont {Liu}},\ }\bibfield  {title} {\enquote {\bibinfo {title}
  {Lattice qcd at finite density via a new canonical approach},}\ }\href
  {\doibase 10.1103/physrevd.72.114513} {\bibfield  {journal} {\bibinfo
  {journal} {Physical Review D}\ }\textbf {\bibinfo {volume} {72}} (\bibinfo
  {year} {2005}),\ 10.1103/physrevd.72.114513}\BibitemShut {NoStop}%
\end{thebibliography}%

\appendix 
\clearpage
\begin{widetext}
\section{Operators and current conservation}
\label{sec:op}
To evaluate Eq.\eqref{eq:P1} in lattice QCD, we use standard annihilation ($\psi$) and creation ($\psi^\dagger$)  operators for a charged pion,
\beq
\psi_{\pi^+}(x)=\bar{d}(x) \gamma_5 u (x), \;\psi_{\pi^+}^\dagger(x)=-\bar{u}(x) \gamma_5 d (x).
\label{eq:op}
\eeq
We also consider rho meson two-point functions constructed from,
\beq
\psi_{\rho}(x)_i=\bar{d}(x) \gamma_i u (x),\quad i=1,2,3,
\label{eq:rho}
\eeq
and average over the spatial directions.
For Wilson fermions, the Dirac operator $M_q=\fsl{D}+m_q$ takes the standard form for a single quark flavor labeled by $q$,
\beq
M_q = \iden - \kappa_q \sum_{\mu} \Big[(1-\gamma_\mu) U_\mu + (1+\gamma_\mu) U_\mu ^\dagger\Big],
\label{eq:matW}
\eeq
where $\kappa_q=1/(2m_q+4)$ is the hopping parameter and $m_q$ the bare quark mass.

For current operators, we consider two options.
One is the lattice local (or point) current built from up and down quark fields,
\beq
j^{(PC)}_\mu\equiv Z_V \kappa\left(q_u \bar{u}\gamma_\mu u + q_d \bar{d}\gamma_\mu d \right).
\label{eq:PC}
\eeq
The factor $\kappa$ here is to account for the quark-field rescaling $\psi\to \sqrt{2\kappa} \psi$ in Wilson fermions. The factor 2 is canceled by the 1/2 factor in the definition of the vector current ${1\over 2}\bar{\psi}\gamma_\mu \psi$.
The charge factors are $q_u=2/3$ and $q_d=-1/3$ where the resulting $e^2=\alpha\approx 1/137$ in the four-point function has been absorbed in the definition of $\alpha^\pi_E$. 
The advantage of this operator is that it leads to simple correlation functions. 
The drawback is that the renormalization constant for the vector current ($Z_V$) has to be determined. 

We also consider conserved vector current on the lattice ($Z_V\equiv 1$) which can be derived by the Noether procedure. For the Wilson fermion action $S=\bar{\psi}_q M_q \psi_q$ built  from the matrix in Eq.\eqref{eq:matW}, 
the simplest way~\cite{karsten1981lattice} is to substitute the gauge fields by 
\beq
U_\mu(x) \to U_\mu(x) e^{i q_q v^q_\mu},
\eeq
and differentiate with respect to the external vector field $v^q_\mu$, then take $v^q_\mu\to 0$. The result is the point-split form
\beq
j^{(q,PS)}_\mu (x) =-i\,{\delta S \over \delta v^q_\mu}\bigg |_{v^q_\mu\to 0}
=- q_q \kappa_q \big[ 
\bar{\psi}_q(x) (1-\gamma_\mu) U_\mu(x) \psi_q(x+\hat{\mu}) 
-
\bar{\psi_q}(x+\hat{\mu}) (1+\gamma_\mu ) U_\mu^\dagger(x) \psi_q(x) 
\big].
\eeq
The phase factor $-i$ is explained in Ref.~\cite{Gattringer:2010zz}.
An alternative method~\cite{Wilcox:1991cq,Alexandru_2005} is through a local transformation on the quark fields, $\psi \to e^{-i\omega(x)}\psi$, and do variation ${\delta S \over \delta (\Delta_\mu\omega)}$ on the finite difference $\Delta_\mu\omega=\omega(x+\hat{\mu})-\omega(x)$. 
For two quark flavors (u and d), we have
\beqs
 j^{(PS)}_\mu (x) &=
 q_u {\kappa_u} \big[ 
-\bar{u}(x) (1-\gamma_\mu) U_\mu(x) u(x+\hat{\mu}) 
+
\bar{u}(x+\hat{\mu}) (1+\gamma_\mu ) U_\mu^\dagger(x) u(x) 
\big] 
\\ &
+  q_d {\kappa_d} \big[ 
-\bar{d}(x) (1-\gamma_\mu) U_\mu(x) d(x+\hat{\mu}) 
+ 
\bar{d}(x+\hat{\mu}) (1+\gamma_\mu ) U_\mu^\dagger(x) d(x) 
\big].
\label{eq:PS}
\eeqs
The conserved current for nhyp fermion has the same form, except the gauge links are nhyp-smeared.
Although conserved currents explicitly involve gauge fields and lead to more complicated correlation functions, they have the advantage of circumventing the renormalization issue.

Just like current conservation  guarantees the normalization condition in three-point functions, 
\beq
\sum_{\bm x_1}\langl \Omega| {\psi}(x)\, j^{(q,PS)}_4 (x_1)\,  \psi^\dagger(0) | \Omega\rangl =
q_q \langl \Omega |{\psi} (x) \psi^\dagger(0) | \Omega\rangl,
\label{eq:Q3pt}
\eeq
\comment{
for a hadron of interpolating field $\psi.
It holds for every time step $t_1$ between source and sink. 
This can be shown as a direct consequence of $\partial^\mu j^{(q)}_\mu(x) =0$ for a conserved current (for which the point-split $j^{(q,PS)}_\mu$ is an example). Taking the expectation value $\langl\cdots\rangl\equiv \langl\Omega|\psi^\dagger(\cdots)\psi|\Omega\rangl$ between hadrons, using finite differences on the lattice, and summing over spatial coordinates, gives the expression
\beq
\sum_{\bm x} \langl \sum_{\mu} \big[j^{(q)}_\mu(x+\hat{\mu})-j^{(q)}_\mu(x)\big] \rangl =0.
\eeq
Separating out the time components, 
\beq
\sum_{\bm x} \langl j^{(q)}_4(\bm x,t+1)-j^{(q)}_4(\bm x, t) \rangl =-\langl \sum_{\mu=1}^3 \bigg[\sum_{\bm x} j^{(q)}_\mu(\bm x+\hat{\mu},t)-\sum_{\bm x} j^{(q)}_\mu(\bm x,t)\bigg]\rangl.
\eeq
The two spatial sums are the same under periodic boundary conditions in space, leading to
\beq
\sum_{\bm x} \langl j^{(q)}_4(\bm x,t+1)\rangl=\sum_{\bm x} \langl j^{(q)}_4(\bm x, t) \rangl,
\eeq
which is a statement for conservation of the quark charge $q_q$. Since the argument does not explicitly involve gauge fields, the statement is true on every gauge configuration. 
}
a similar condition holds in four-point functions, 
\beq
\sum_{\bm x_2, \bm x_1}\langl \Omega | {\psi}(x) \,j^{(q_2,PS)}_4 (x_2)\, j^{(q_1,PS)}_4 (x_1)\,  \psi^\dagger(0) | \Omega\rangl =
q_1 q_2 \langl \Omega | {\psi}(x) \psi^\dagger(0) | \Omega \rangl.
\label{eq:Q4pt}
\eeq
In physical terms, the charge overlap at $\bm q=0$ on the left-hand-side is effectively reconstructing the two-point function. Each charge density is spread over all spatial sites on the lattice. By summing over $\bm x_1$ and $\bm x_2$ at zero momentum, we recover the total charge factor from each insertion, regardless of the time points of the insertions.
\comment{
The time independence of the condition with regard to current insertions can be argued from the bilinear divergence of conserved currents $\partial^\mu j^{(q_1)}_\mu \partial^\nu j^{(q_2)}_\nu =0$. Evaluating between hadrons and summing over spatial coordinates, we have on the lattice,
\beq
\sum_{\bm x_1,\bm x_2} \langl \sum_{\mu}  \big[j^{(q_1)}_\mu(x_1+\hat{\mu})-j^{(q_1)}_\mu(x_1)\big]
\sum_{\nu} \big[j^{(q_2)}_\nu(x_2+\hat{\nu})-j^{(q_2)}_\nu(x_2)\big] \rangl
=0.
\label{eq:Q1}
\eeq
If we define the spatial sums for time components,
\beq 
Q^{(q_1,q_2)}(t_1,t_2)\equiv \sum_{\bm x_1,\bm x_2} \langl j^{(q_1)}_4(\bm x_1,t_1) j^{(q_2)}_4(\bm x_2,t_2) \rangl, 
\eeq 
then Eq.\eqref{eq:Q1} leads to,
\beq
Q^{(q_1,q_2)}(t_1,t_2)-Q^{(q_1,q_2)}(t_1+1,t_2)-Q^{(q_1,q_2)}(t_1,t_2+1)+Q^{(q_1,q_2)}(t_1+1,t_2+1)=0,
\label{eq:Qt1t2}
\eeq
which is the lattice version of ${\partial^2Q^{(q_1,q_2)}(t_1,t_2)\over \partial t_1 \partial t_2}=0$.
This is a statement on the conservation of $Q^{(q_1,q_2)}(t_1,t_2)$ with respect to $t_1$ and $t_2$, and it is true on every gauge configuration. 
}
There is a subtle issue with four-point functions.  If the two currents couple to different quark lines ($q_1\neq q_2$), the conservation is for all combinations of $t_1$ and $t_2$ between source and sink, including $t_1=t_2$.
If they couple to the same quark line ($q_1= q_2$), the conservation is only true for $t_1\neq t_2$. The point $t_1=t_2$ introduces unwanted contact terms on the lattice and is avoided.
The issue is a lattice artifact; in the continuum, the contact interaction is regular and well-defined.
The conservation property in Eq.\eqref{eq:Q4pt} is used to validate the four-point diagrams in this work.

\section{Wick contractions}
\label{sec:wick}
Here we give the unnormalized  correlation functions in Eq.\eqref{eq:Qmn} by contracting out all quark-antiqurk pairs.

\subsection{Local current}
\label{sec:pc}
For point current (PC), using Eq.\eqref{eq:op} and Eq.\eqref{eq:PC}, the full correlation function has 20 diagrams,
\beqs
 \tilde{Q}_{\mu\nu}^{(PC)}(\bm q,t_3,t_2,t_1,t_0) &=\sum_{\bm x_2,\bm x_1} 
   e^{-i\bm q\cdot \bm x_2} e^{i\bm q\cdot \bm x_1} 
   \sum_{\bm x_3,\bm x_0} 
   \langle \Omega | \psi_{\pi^+}(\bm x_3,t_3)j^{(PC)}_{\mu}(\bm x_2,t_2)j^{(PC)}_{\nu}(\bm x_1,t_1)\psi_{\pi^+}^\dagger(\bm x_0,t_0)| \Omega\rangle \\
   & \equiv {Z_V^2\kappa^2\over 9} 
   \sum_{i=0}^{19} d_i(\bm q,t_3,t_2,t_1,t_0),
   \label{eq:dPC}
\eeqs
where
\beqs
d_{10}^{\text{A}}=-2\,\text{tr}\left[S_{u}(t_1, t_3)\gamma_5S_{d}(t_3, t_2)\gamma_{\mu}e^{-i\mathbf{q}}S_{d}(t_2, t_0)\gamma_5S_{u}(t_0, t_1)\gamma_{\nu}e^{i\mathbf{q}}\right]\\
d_{7}^{\text{A-bwd}}=-2\,\text{tr}\left[S_{u}(t_2, t_3)\gamma_5S_{d}(t_3, t_1)\gamma_{\nu}e^{i\mathbf{q}}S_{d}(t_1, t_0)\gamma_5S_{u}(t_0, t_2)\gamma_{\mu}e^{-i\mathbf{q}}\right]\\
d_{5}^{\text{B}}=4\,\text{tr}\left[S_{u}(t_2, t_3)\gamma_5S_{d}(t_3, t_0)\gamma_5S_{u}(t_0, t_1)\gamma_{\nu}e^{i\mathbf{q}}S_{u}(t_1, t_2)\gamma_{\mu}e^{-i\mathbf{q}}\right]\\
d_{15}^{\text{B-bwd}}=1\,\text{tr}\left[S_{u}(t_0, t_3)\gamma_5S_{d}(t_3, t_2)\gamma_{\mu}e^{-i\mathbf{q}}S_{d}(t_2, t_1)\gamma_{\nu}e^{i\mathbf{q}}S_{d}(t_1, t_0)\gamma_5\right]\\
d_{1}^{\text{C}}=4\,\text{tr}\left[S_{u}(t_1, t_3)\gamma_5S_{d}(t_3, t_0)\gamma_5S_{u}(t_0, t_2)\gamma_{\mu}e^{-i\mathbf{q}}S_{u}(t_2, t_1)\gamma_{\nu}e^{i\mathbf{q}}\right]\\
d_{17}^{\text{C-bwd}}=1\,\text{tr}\left[S_{u}(t_0, t_3)\gamma_5S_{d}(t_3, t_1)\gamma_{\nu}e^{i\mathbf{q}}S_{d}(t_1, t_2)\gamma_{\mu}e^{-i\mathbf{q}}S_{d}(t_2, t_0)\gamma_5\right]\\
d_{0}^{\text{D}}=-4\,\text{tr}\left[S_{u}(t_0, t_3)\gamma_5S_{d}(t_3, t_0)\gamma_5\right]\text{tr}\left[S_{u}(t_1, t_2)\gamma_{\mu}e^{-i\mathbf{q}}S_{u}(t_2, t_1)\gamma_{\nu}e^{i\mathbf{q}}\right]\\
d_{18}^{\text{D}}=-1\,\text{tr}\left[S_{u}(t_0, t_3)\gamma_5S_{d}(t_3, t_0)\gamma_5\right]\text{tr}\left[S_{d}(t_1, t_2)\gamma_{\mu}e^{-i\mathbf{q}}S_{d}(t_2, t_1)\gamma_{\nu}e^{i\mathbf{q}}\right]\\
d_{4}^{\text{El}}=-4\,\text{tr}\left[S_{u}(t_1, t_3)\gamma_5S_{d}(t_3, t_0)\gamma_5S_{u}(t_0, t_1)\gamma_{\nu}e^{i\mathbf{q}}\right]\text{tr}\left[S_{u}(t_2, t_2)\gamma_{\mu}e^{-i\mathbf{q}}\right]\\
d_{13}^{\text{El}}=2\,\text{tr}\left[S_{u}(t_1, t_3)\gamma_5S_{d}(t_3, t_0)\gamma_5S_{u}(t_0, t_1)\gamma_{\nu}e^{i\mathbf{q}}\right]\text{tr}\left[S_{d}(t_2, t_2)\gamma_{\mu}e^{-i\mathbf{q}}\right]\\
d_{6}^{\text{El-bwd}}=2\,\text{tr}\left[S_{u}(t_0, t_3)\gamma_5S_{d}(t_3, t_1)\gamma_{\nu}e^{i\mathbf{q}}S_{d}(t_1, t_0)\gamma_5\right]\text{tr}\left[S_{u}(t_2, t_2)\gamma_{\mu}e^{-i\mathbf{q}}\right]\\
d_{14}^{\text{El-bwd}}=-1\,\text{tr}\left[S_{u}(t_0, t_3)\gamma_5S_{d}(t_3, t_1)\gamma_{\nu}e^{i\mathbf{q}}S_{d}(t_1, t_0)\gamma_5\right]\text{tr}\left[S_{d}(t_2, t_2)\gamma_{\mu}e^{-i\mathbf{q}}\right]\\
d_{2}^{\text{Er}}=-4\,\text{tr}\left[S_{u}(t_2, t_3)\gamma_5S_{d}(t_3, t_0)\gamma_5S_{u}(t_0, t_2)\gamma_{\mu}e^{-i\mathbf{q}}\right]\text{tr}\left[S_{u}(t_1, t_1)\gamma_{\nu}e^{i\mathbf{q}}\right]\\
d_{8}^{\text{Er}}=2\,\text{tr}\left[S_{u}(t_2, t_3)\gamma_5S_{d}(t_3, t_0)\gamma_5S_{u}(t_0, t_2)\gamma_{\mu}e^{-i\mathbf{q}}\right]\text{tr}\left[S_{d}(t_1, t_1)\gamma_{\nu}e^{i\mathbf{q}}\right]\\
d_{11}^{\text{Er-bwd}}=2\,\text{tr}\left[S_{u}(t_0, t_3)\gamma_5S_{d}(t_3, t_2)\gamma_{\mu}e^{-i\mathbf{q}}S_{d}(t_2, t_0)\gamma_5\right]\text{tr}\left[S_{u}(t_1, t_1)\gamma_{\nu}e^{i\mathbf{q}}\right]\\
d_{16}^{\text{Er-bwd}}=-1\,\text{tr}\left[S_{u}(t_0, t_3)\gamma_5S_{d}(t_3, t_2)\gamma_{\mu}e^{-i\mathbf{q}}S_{d}(t_2, t_0)\gamma_5\right]\text{tr}\left[S_{d}(t_1, t_1)\gamma_{\nu}e^{i\mathbf{q}}\right]\\
d_{3}^{\text{F}}=4\,\text{tr}\left[S_{u}(t_0, t_3)\gamma_5S_{d}(t_3, t_0)\gamma_5\right]\text{tr}\left[S_{u}(t_2, t_2)\gamma_{\mu}e^{-i\mathbf{q}}\right]\text{tr}\left[S_{u}(t_1, t_1)\gamma_{\nu}e^{i\mathbf{q}}\right]\\
d_{9}^{\text{F}}=-2\,\text{tr}\left[S_{u}(t_0, t_3)\gamma_5S_{d}(t_3, t_0)\gamma_5\right]\text{tr}\left[S_{u}(t_2, t_2)\gamma_{\mu}e^{-i\mathbf{q}}\right]\text{tr}\left[S_{d}(t_1, t_1)\gamma_{\nu}e^{i\mathbf{q}}\right]\\
d_{12}^{\text{F}}=-2\,\text{tr}\left[S_{u}(t_0, t_3)\gamma_5S_{d}(t_3, t_0)\gamma_5\right]\text{tr}\left[S_{d}(t_2, t_2)\gamma_{\mu}e^{-i\mathbf{q}}\right]\text{tr}\left[S_{u}(t_1, t_1)\gamma_{\nu}e^{i\mathbf{q}}\right]\\
d_{19}^{\text{F}}=1\,\text{tr}\left[S_{u}(t_0, t_3)\gamma_5S_{d}(t_3, t_0)\gamma_5\right]\text{tr}\left[S_{d}(t_2, t_2)\gamma_{\mu}e^{-i\mathbf{q}}\right]\text{tr}\left[S_{d}(t_1, t_1)\gamma_{\nu}e^{i\mathbf{q}}\right]
\eeqs
\normalsize
We use a matrix notation that highlights time dependence. The trace is over spin and color. 
The momentum factor is defined by a diagonal matrix,
\beq
[e^{\pm i{\bf q}}]_{s,c,{\bf x};s',c',{\bf x}'}\equiv 
\delta_{ss'}\delta_{cc'} \delta_{{\bf x},{\bf x'}}
e^{\pm i{\bf q}\cdot{\bf x}}.
\label{eq:mom}
\eeq
 The spatial sums over $(\bm x_2, \bm x_1, \bm x_3, \bm x_0)$ are implicit in the matrix multiplications.
 We use $S(t_2,t_1)$ to denote a quark propagator from $t_1$ to $t_2$ (from right to left), obtained from the inverse of quark matrix $M$ with a source $M x=b$, see Eq.\eqref{eq:PSP}. 
The terms are grouped into six distinct topological diagrams depicted in Fig~\ref{fig:4pt}, labeled by superscripts on $d_i$.
If isospin limit ($\kappa_u=\kappa_d=\kappa$) is taken, we get 12 diagrams (first six connected, the rest disconnected), 
\beqs
d_{4}^{\text{A}}=-2\,\text{tr}\left[S(t_1, t_3)\gamma_5S(t_3, t_2)\gamma_{\mu}e^{-i\mathbf{q}}S(t_2, t_0)\gamma_5S(t_0, t_1)\gamma_{\nu}e^{i\mathbf{q}}\right]\\
d_{2}^{\text{A-bwd}}=-2\,\text{tr}\left[S(t_2, t_3)\gamma_5S(t_3, t_1)\gamma_{\nu}e^{i\mathbf{q}}S(t_1, t_0)\gamma_5S(t_0, t_2)\gamma_{\mu}e^{-i\mathbf{q}}\right]\\
d_{1}^{\text{B}}=4\,\text{tr}\left[S(t_2, t_3)\gamma_5S(t_3, t_0)\gamma_5S(t_0, t_1)\gamma_{\nu}e^{i\mathbf{q}}S(t_1, t_2)\gamma_{\mu}e^{-i\mathbf{q}}\right]\\
d_{7}^{\text{B-bwd}}=1\,\text{tr}\left[S(t_0, t_3)\gamma_5S(t_3, t_2)\gamma_{\mu}e^{-i\mathbf{q}}S(t_2, t_1)\gamma_{\nu}e^{i\mathbf{q}}S(t_1, t_0)\gamma_5\right]\\
d_{0}^{\text{C}}=4\,\text{tr}\left[S(t_1, t_3)\gamma_5S(t_3, t_0)\gamma_5S(t_0, t_2)\gamma_{\mu}e^{-i\mathbf{q}}S(t_2, t_1)\gamma_{\nu}e^{i\mathbf{q}}\right]\\
d_{9}^{\text{C-bwd}}=1\,\text{tr}\left[S(t_0, t_3)\gamma_5S(t_3, t_1)\gamma_{\nu}e^{i\mathbf{q}}S(t_1, t_2)\gamma_{\mu}e^{-i\mathbf{q}}S(t_2, t_0)\gamma_5\right]\\
d_{10}^{\text{D}}=-5\,\text{tr}\left[S(t_0, t_3)\gamma_5S(t_3, t_0)\gamma_5\right]\text{tr}\left[S(t_1, t_2)\gamma_{\mu}e^{-i\mathbf{q}}S(t_2, t_1)\gamma_{\nu}e^{i\mathbf{q}}\right]\\
d_{5}^{\text{El}}=-2\,\text{tr}\left[S(t_1, t_3)\gamma_5S(t_3, t_0)\gamma_5S(t_0, t_1)\gamma_{\nu}e^{i\mathbf{q}}\right]\text{tr}\left[S(t_2, t_2)\gamma_{\mu}e^{-i\mathbf{q}}\right]\\
d_{6}^{\text{El-bwd}}=1\,\text{tr}\left[S(t_0, t_3)\gamma_5S(t_3, t_1)\gamma_{\nu}e^{i\mathbf{q}}S(t_1, t_0)\gamma_5\right]\text{tr}\left[S(t_2, t_2)\gamma_{\mu}e^{-i\mathbf{q}}\right]\\
d_{3}^{\text{Er}}=-2\,\text{tr}\left[S(t_2, t_3)\gamma_5S(t_3, t_0)\gamma_5S(t_0, t_2)\gamma_{\mu}e^{-i\mathbf{q}}\right]\text{tr}\left[S(t_1, t_1)\gamma_{\nu}e^{i\mathbf{q}}\right]\\
d_{8}^{\text{Er-bwd}}=1\,\text{tr}\left[S(t_0, t_3)\gamma_5S(t_3, t_2)\gamma_{\mu}e^{-i\mathbf{q}}S(t_2, t_0)\gamma_5\right]\text{tr}\left[S(t_1, t_1)\gamma_{\nu}e^{i\mathbf{q}}\right]\\
d_{11}^{\text{F}}=1\,\text{tr}\left[S(t_0, t_3)\gamma_5S(t_3, t_0)\gamma_5\right]\text{tr}\left[S(t_2, t_2)\gamma_{\mu}e^{-i\mathbf{q}}\right]\text{tr}\left[S(t_1, t_1)\gamma_{\nu}e^{i\mathbf{q}}\right]
\label{eq:dPC12}
\eeqs
\normalsize

\comment{
For gamma matrices, we use the following convention (MILC),
\beqs
&\{\gamma_1,\gamma_2,\gamma_3,\gamma_4,\gamma_5\}= \\
&\left\{\left(
\begin{array}{cccc}
 0 & 0 & 0 & i \\
 0 & 0 & i & 0 \\
 0 & -i & 0 & 0 \\
 -i & 0 & 0 & 0 \\
\end{array}
\right),\left(
\begin{array}{cccc}
 0 & 0 & 0 & 1 \\
 0 & 0 & -1 & 0 \\
 0 & -1 & 0 & 0 \\
 1 & 0 & 0 & 0 \\
\end{array}
\right),\left(
\begin{array}{cccc}
 0 & 0 & i & 0 \\
 0 & 0 & 0 & -i \\
 -i & 0 & 0 & 0 \\
 0 & i & 0 & 0 \\
\end{array}
\right),\left(
\begin{array}{cccc}
 0 & 0 & -1 & 0 \\
 0 & 0 & 0 & -1 \\
 -1 & 0 & 0 & 0 \\
 0 & -1 & 0 & 0 \\
\end{array}
\right),\left(
\begin{array}{cccc}
 1 & 0 & 0 & 0 \\
 0 & 1 & 0 & 0 \\
 0 & 0 & -1 & 0 \\
 0 & 0 & 0 & -1 \\
\end{array}
\right)\right\}.
\eeqs
}

\subsection{Conserved current}
\label{sec:ps}
For point-split current (PS), using Eq.\eqref{eq:op} and Eq.\eqref{eq:PS},
Wick contraction yields 80 diagrams (not shown here) if $u$ and $d$ are distinct.
%
If isospin limit is taken, there are  48 diagrams which we express as,
\beqs
   \tilde{Q}_{\mu\nu}^{(PS)}(\bm q,t_3,t_2,t_1,t_0) &=\sum_{\bm x_2,\bm x_1} 
   e^{-i\bm q\cdot \bm x_2} e^{i\bm q\cdot \bm x_1} 
   \sum_{\bm x_3,\bm x_0} 
   \langle \Omega| \psi_{\pi^+}(\bm x_3,t_3)j^{(PS)}_{\mu}(\bm x_2,t_2)j^{(PS)}_{\nu}(\bm x_1,t_1)\psi_{\pi^+}^\dagger(\bm x_0,t_0)|\Omega\rangle \\
   & \equiv  {\kappa^2\over 9} 
   \sum_{i=0}^{47} d_i(\bm q,t_3,t_2,t_1,t_0).
   \label{eq:dPS}
\eeqs

 \fontsize{9}{9}
The 24 connected diagrams are given by, 
\beqs
d_{16}^{\text{A}}=-2\,\text{tr}\left[S(t_1+\hat{\nu_4}, t_3)(\gamma_5)S(t_3, t_2)(1-\gamma_{\mu})e^{-i\mathbf{q}}U_{\mu}(t_2,t_2+\hat{\mu_4})S(t_2+\hat{\mu_4}, t_0)(\gamma_5)S(t_0, t_1)(1-\gamma_{\nu})e^{i\mathbf{q}}U_{\nu}(t_1,t_1+\hat{\nu_4})\right]\\
d_{18}^{\text{A}}=2\,\text{tr}\left[S(t_1+\hat{\nu_4}, t_3)(\gamma_5)S(t_3, t_2+\hat{\mu_4})(1+\gamma_{\mu})U_{\mu}^{\dagger}(t_2+\hat{\mu_4},t_2)e^{-i\mathbf{q}}S(t_2, t_0)(\gamma_5)S(t_0, t_1)(1-\gamma_{\nu})e^{i\mathbf{q}}U_{\nu}(t_1,t_1+\hat{\nu_4})\right]\\
d_{20}^{\text{A}}=2\,\text{tr}\left[S(t_1, t_3)(\gamma_5)S(t_3, t_2)(1-\gamma_{\mu})e^{-i\mathbf{q}}U_{\mu}(t_2,t_2+\hat{\mu_4})S(t_2+\hat{\mu_4}, t_0)(\gamma_5)S(t_0, t_1+\hat{\nu_4})(1+\gamma_{\nu})U_{\nu}^{\dagger}(t_1+\hat{\nu_4},t_1)e^{i\mathbf{q}}\right]\\
d_{22}^{\text{A}}=-2\,\text{tr}\left[S(t_1, t_3)(\gamma_5)S(t_3, t_2+\hat{\mu_4})(1+\gamma_{\mu})U_{\mu}^{\dagger}(t_2+\hat{\mu_4},t_2)e^{-i\mathbf{q}}S(t_2, t_0)(\gamma_5)S(t_0, t_1+\hat{\nu_4})(1+\gamma_{\nu})U_{\nu}^{\dagger}(t_1+\hat{\nu_4},t_1)e^{i\mathbf{q}}\right]\\
d_{8}^{\text{A-bwd}}=-2\,\text{tr}\left[S(t_2+\hat{\mu_4}, t_3)(\gamma_5)S(t_3, t_1)(1-\gamma_{\nu})e^{i\mathbf{q}}U_{\nu}(t_1,t_1+\hat{\nu_4})S(t_1+\hat{\nu_4}, t_0)(\gamma_5)S(t_0, t_2)(1-\gamma_{\mu})e^{-i\mathbf{q}}U_{\mu}(t_2,t_2+\hat{\mu_4})\right]\\
d_{10}^{\text{A-bwd}}=2\,\text{tr}\left[S(t_2, t_3)(\gamma_5)S(t_3, t_1)(1-\gamma_{\nu})e^{i\mathbf{q}}U_{\nu}(t_1,t_1+\hat{\nu_4})S(t_1+\hat{\nu_4}, t_0)(\gamma_5)S(t_0, t_2+\hat{\mu_4})(1+\gamma_{\mu})U_{\mu}^{\dagger}(t_2+\hat{\mu_4},t_2)e^{-i\mathbf{q}}\right]\\
d_{12}^{\text{A-bwd}}=2\,\text{tr}\left[S(t_2+\hat{\mu_4}, t_3)(\gamma_5)S(t_3, t_1+\hat{\nu_4})(1+\gamma_{\nu})U_{\nu}^{\dagger}(t_1+\hat{\nu_4},t_1)e^{i\mathbf{q}}S(t_1, t_0)(\gamma_5)S(t_0, t_2)(1-\gamma_{\mu})e^{-i\mathbf{q}}U_{\mu}(t_2,t_2+\hat{\mu_4})\right]\\
d_{14}^{\text{A-bwd}}=-2\,\text{tr}\left[S(t_2, t_3)(\gamma_5)S(t_3, t_1+\hat{\nu_4})(1+\gamma_{\nu})U_{\nu}^{\dagger}(t_1+\hat{\nu_4},t_1)e^{i\mathbf{q}}S(t_1, t_0)(\gamma_5)S(t_0, t_2+\hat{\mu_4})(1+\gamma_{\mu})U_{\mu}^{\dagger}(t_2+\hat{\mu_4},t_2)e^{-i\mathbf{q}}\right]\\
d_{1}^{\text{B}}=4\,\text{tr}\left[S(t_2+\hat{\mu_4}, t_3)(\gamma_5)S(t_3, t_0)(\gamma_5)S(t_0, t_1)(1-\gamma_{\nu})e^{i\mathbf{q}}U_{\nu}(t_1,t_1+\hat{\nu_4})S(t_1+\hat{\nu_4}, t_2)(1-\gamma_{\mu})e^{-i\mathbf{q}}U_{\mu}(t_2,t_2+\hat{\mu_4})\right]\\
d_{3}^{\text{B}}=-4\,\text{tr}\left[S(t_2, t_3)(\gamma_5)S(t_3, t_0)(\gamma_5)S(t_0, t_1)(1-\gamma_{\nu})e^{i\mathbf{q}}U_{\nu}(t_1,t_1+\hat{\nu_4})S(t_1+\hat{\nu_4}, t_2+\hat{\mu_4})(1+\gamma_{\mu})U_{\mu}^{\dagger}(t_2+\hat{\mu_4},t_2)e^{-i\mathbf{q}}\right]\\
d_{5}^{\text{B}}=-4\,\text{tr}\left[S(t_2+\hat{\mu_4}, t_3)(\gamma_5)S(t_3, t_0)(\gamma_5)S(t_0, t_1+\hat{\nu_4})(1+\gamma_{\nu})U_{\nu}^{\dagger}(t_1+\hat{\nu_4},t_1)e^{i\mathbf{q}}S(t_1, t_2)(1-\gamma_{\mu})e^{-i\mathbf{q}}U_{\mu}(t_2,t_2+\hat{\mu_4})\right]\\
d_{7}^{\text{B}}=4\,\text{tr}\left[S(t_2, t_3)(\gamma_5)S(t_3, t_0)(\gamma_5)S(t_0, t_1+\hat{\nu_4})(1+\gamma_{\nu})U_{\nu}^{\dagger}(t_1+\hat{\nu_4},t_1)e^{i\mathbf{q}}S(t_1, t_2+\hat{\mu_4})(1+\gamma_{\mu})U_{\mu}^{\dagger}(t_2+\hat{\mu_4},t_2)e^{-i\mathbf{q}}\right]\\
d_{25}^{\text{B-bwd}}=1\,\text{tr}\left[S(t_0, t_3)(\gamma_5)S(t_3, t_2)(1-\gamma_{\mu})e^{-i\mathbf{q}}U_{\mu}(t_2,t_2+\hat{\mu_4})S(t_2+\hat{\mu_4}, t_1)(1-\gamma_{\nu})e^{i\mathbf{q}}U_{\nu}(t_1,t_1+\hat{\nu_4})S(t_1+\hat{\nu_4}, t_0)(\gamma_5)\right]\\
d_{31}^{\text{B-bwd}}=-1\,\text{tr}\left[S(t_0, t_3)(\gamma_5)S(t_3, t_2+\hat{\mu_4})(1+\gamma_{\mu})U_{\mu}^{\dagger}(t_2+\hat{\mu_4},t_2)e^{-i\mathbf{q}}S(t_2, t_1)(1-\gamma_{\nu})e^{i\mathbf{q}}U_{\nu}(t_1,t_1+\hat{\nu_4})S(t_1+\hat{\nu_4}, t_0)(\gamma_5)\right]\\
d_{37}^{\text{B-bwd}}=-1\,\text{tr}\left[S(t_0, t_3)(\gamma_5)S(t_3, t_2)(1-\gamma_{\mu})e^{-i\mathbf{q}}U_{\mu}(t_2,t_2+\hat{\mu_4})S(t_2+\hat{\mu_4}, t_1+\hat{\nu_4})(1+\gamma_{\nu})U_{\nu}^{\dagger}(t_1+\hat{\nu_4},t_1)e^{i\mathbf{q}}S(t_1, t_0)(\gamma_5)\right]\\
d_{43}^{\text{B-bwd}}=1\,\text{tr}\left[S(t_0, t_3)(\gamma_5)S(t_3, t_2+\hat{\mu_4})(1+\gamma_{\mu})U_{\mu}^{\dagger}(t_2+\hat{\mu_4},t_2)e^{-i\mathbf{q}}S(t_2, t_1+\hat{\nu_4})(1+\gamma_{\nu})U_{\nu}^{\dagger}(t_1+\hat{\nu_4},t_1)e^{i\mathbf{q}}S(t_1, t_0)(\gamma_5)\right]\\
d_{0}^{\text{C}}=4\,\text{tr}\left[S(t_1+\hat{\nu_4}, t_3)(\gamma_5)S(t_3, t_0)(\gamma_5)S(t_0, t_2)(1-\gamma_{\mu})e^{-i\mathbf{q}}U_{\mu}(t_2,t_2+\hat{\mu_4})S(t_2+\hat{\mu_4}, t_1)(1-\gamma_{\nu})e^{i\mathbf{q}}U_{\nu}(t_1,t_1+\hat{\nu_4})\right]\\
d_{2}^{\text{C}}=-4\,\text{tr}\left[S(t_1+\hat{\nu_4}, t_3)(\gamma_5)S(t_3, t_0)(\gamma_5)S(t_0, t_2+\hat{\mu_4})(1+\gamma_{\mu})U_{\mu}^{\dagger}(t_2+\hat{\mu_4},t_2)e^{-i\mathbf{q}}S(t_2, t_1)(1-\gamma_{\nu})e^{i\mathbf{q}}U_{\nu}(t_1,t_1+\hat{\nu_4})\right]\\
d_{4}^{\text{C}}=-4\,\text{tr}\left[S(t_1, t_3)(\gamma_5)S(t_3, t_0)(\gamma_5)S(t_0, t_2)(1-\gamma_{\mu})e^{-i\mathbf{q}}U_{\mu}(t_2,t_2+\hat{\mu_4})S(t_2+\hat{\mu_4}, t_1+\hat{\nu_4})(1+\gamma_{\nu})U_{\nu}^{\dagger}(t_1+\hat{\nu_4},t_1)e^{i\mathbf{q}}\right]\\
d_{6}^{\text{C}}=4\,\text{tr}\left[S(t_1, t_3)(\gamma_5)S(t_3, t_0)(\gamma_5)S(t_0, t_2+\hat{\mu_4})(1+\gamma_{\mu})U_{\mu}^{\dagger}(t_2+\hat{\mu_4},t_2)e^{-i\mathbf{q}}S(t_2, t_1+\hat{\nu_4})(1+\gamma_{\nu})U_{\nu}^{\dagger}(t_1+\hat{\nu_4},t_1)e^{i\mathbf{q}}\right]\\
d_{27}^{\text{C-bwd}}=1\,\text{tr}\left[S(t_0, t_3)(\gamma_5)S(t_3, t_1)(1-\gamma_{\nu})e^{i\mathbf{q}}U_{\nu}(t_1,t_1+\hat{\nu_4})S(t_1+\hat{\nu_4}, t_2)(1-\gamma_{\mu})e^{-i\mathbf{q}}U_{\mu}(t_2,t_2+\hat{\mu_4})S(t_2+\hat{\mu_4}, t_0)(\gamma_5)\right]\\
d_{33}^{\text{C-bwd}}=-1\,\text{tr}\left[S(t_0, t_3)(\gamma_5)S(t_3, t_1)(1-\gamma_{\nu})e^{i\mathbf{q}}U_{\nu}(t_1,t_1+\hat{\nu_4})S(t_1+\hat{\nu_4}, t_2+\hat{\mu_4})(1+\gamma_{\mu})U_{\mu}^{\dagger}(t_2+\hat{\mu_4},t_2)e^{-i\mathbf{q}}S(t_2, t_0)(\gamma_5)\right]\\
d_{39}^{\text{C-bwd}}=-1\,\text{tr}\left[S(t_0, t_3)(\gamma_5)S(t_3, t_1+\hat{\nu_4})(1+\gamma_{\nu})U_{\nu}^{\dagger}(t_1+\hat{\nu_4},t_1)e^{i\mathbf{q}}S(t_1, t_2)(1-\gamma_{\mu})e^{-i\mathbf{q}}U_{\mu}(t_2,t_2+\hat{\mu_4})S(t_2+\hat{\mu_4}, t_0)(\gamma_5)\right]\\
d_{45}^{\text{C-bwd}}=1\,\text{tr}\left[S(t_0, t_3)(\gamma_5)S(t_3, t_1+\hat{\nu_4})(1+\gamma_{\nu})U_{\nu}^{\dagger}(t_1+\hat{\nu_4},t_1)e^{i\mathbf{q}}S(t_1, t_2+\hat{\mu_4})(1+\gamma_{\mu})U_{\mu}^{\dagger}(t_2+\hat{\mu_4},t_2)e^{-i\mathbf{q}}S(t_2, t_0)(\gamma_5)\right]\\
\label{eq:dPS1}
\eeqs
The 24 disconnected diagrams are given by,
\beqs
d_{28}^{\text{D}}=-5\,\text{tr}\left[S(t_0, t_3)(\gamma_5)S(t_3, t_0)(\gamma_5)\right]\text{tr}\left[S(t_1+\hat{\nu_4}, t_2)(1-\gamma_{\mu})e^{-i\mathbf{q}}U_{\mu}(t_2,t_2+\hat{\mu_4})S(t_2+\hat{\mu_4}, t_1)(1-\gamma_{\nu})e^{i\mathbf{q}}U_{\nu}(t_1,t_1+\hat{\nu_4})\right]\\
d_{34}^{\text{D}}=5\,\text{tr}\left[S(t_0, t_3)(\gamma_5)S(t_3, t_0)(\gamma_5)\right]\text{tr}\left[S(t_1+\hat{\nu_4}, t_2+\hat{\mu_4})(1+\gamma_{\mu})U_{\mu}^{\dagger}(t_2+\hat{\mu_4},t_2)e^{-i\mathbf{q}}S(t_2, t_1)(1-\gamma_{\nu})e^{i\mathbf{q}}U_{\nu}(t_1,t_1+\hat{\nu_4})\right]\\
d_{40}^{\text{D}}=5\,\text{tr}\left[S(t_0, t_3)(\gamma_5)S(t_3, t_0)(\gamma_5)\right]\text{tr}\left[S(t_1, t_2)(1-\gamma_{\mu})e^{-i\mathbf{q}}U_{\mu}(t_2,t_2+\hat{\mu_4})S(t_2+\hat{\mu_4}, t_1+\hat{\nu_4})(1+\gamma_{\nu})U_{\nu}^{\dagger}(t_1+\hat{\nu_4},t_1)e^{i\mathbf{q}}\right]\\
d_{46}^{\text{D}}=-5\,\text{tr}\left[S(t_0, t_3)(\gamma_5)S(t_3, t_0)(\gamma_5)\right]\text{tr}\left[S(t_1, t_2+\hat{\mu_4})(1+\gamma_{\mu})U_{\mu}^{\dagger}(t_2+\hat{\mu_4},t_2)e^{-i\mathbf{q}}S(t_2, t_1+\hat{\nu_4})(1+\gamma_{\nu})U_{\nu}^{\dagger}(t_1+\hat{\nu_4},t_1)e^{i\mathbf{q}}\right]\\
d_{17}^{\text{El}}=-2\,\text{tr}\left[S(t_1+\hat{\nu_4}, t_3)(\gamma_5)S(t_3, t_0)(\gamma_5)S(t_0, t_1)(1-\gamma_{\nu})e^{i\mathbf{q}}U_{\nu}(t_1,t_1+\hat{\nu_4})\right]\text{tr}\left[S(t_2+\hat{\mu_4}, t_2)(1-\gamma_{\mu})e^{-i\mathbf{q}}U_{\mu}(t_2,t_2+\hat{\mu_4})\right]\\
d_{19}^{\text{El}}=2\,\text{tr}\left[S(t_1+\hat{\nu_4}, t_3)(\gamma_5)S(t_3, t_0)(\gamma_5)S(t_0, t_1)(1-\gamma_{\nu})e^{i\mathbf{q}}U_{\nu}(t_1,t_1+\hat{\nu_4})\right]\text{tr}\left[S(t_2, t_2+\hat{\mu_4})(1+\gamma_{\mu})U_{\mu}^{\dagger}(t_2+\hat{\mu_4},t_2)e^{-i\mathbf{q}}\right]\\
d_{21}^{\text{El}}=2\,\text{tr}\left[S(t_1, t_3)(\gamma_5)S(t_3, t_0)(\gamma_5)S(t_0, t_1+\hat{\nu_4})(1+\gamma_{\nu})U_{\nu}^{\dagger}(t_1+\hat{\nu_4},t_1)e^{i\mathbf{q}}\right]\text{tr}\left[S(t_2+\hat{\mu_4}, t_2)(1-\gamma_{\mu})e^{-i\mathbf{q}}U_{\mu}(t_2,t_2+\hat{\mu_4})\right]\\
d_{23}^{\text{El}}=-2\,\text{tr}\left[S(t_1, t_3)(\gamma_5)S(t_3, t_0)(\gamma_5)S(t_0, t_1+\hat{\nu_4})(1+\gamma_{\nu})U_{\nu}^{\dagger}(t_1+\hat{\nu_4},t_1)e^{i\mathbf{q}}\right]\text{tr}\left[S(t_2, t_2+\hat{\mu_4})(1+\gamma_{\mu})U_{\mu}^{\dagger}(t_2+\hat{\mu_4},t_2)e^{-i\mathbf{q}}\right]\\
d_{24}^{\text{El-bwd}}=1\,\text{tr}\left[S(t_0, t_3)(\gamma_5)S(t_3, t_1)(1-\gamma_{\nu})e^{i\mathbf{q}}U_{\nu}(t_1,t_1+\hat{\nu_4})S(t_1+\hat{\nu_4}, t_0)(\gamma_5)\right]\text{tr}\left[S(t_2+\hat{\mu_4}, t_2)(1-\gamma_{\mu})e^{-i\mathbf{q}}U_{\mu}(t_2,t_2+\hat{\mu_4})\right]\\
d_{30}^{\text{El-bwd}}=-1\,\text{tr}\left[S(t_0, t_3)(\gamma_5)S(t_3, t_1)(1-\gamma_{\nu})e^{i\mathbf{q}}U_{\nu}(t_1,t_1+\hat{\nu_4})S(t_1+\hat{\nu_4}, t_0)(\gamma_5)\right]\text{tr}\left[S(t_2, t_2+\hat{\mu_4})(1+\gamma_{\mu})U_{\mu}^{\dagger}(t_2+\hat{\mu_4},t_2)e^{-i\mathbf{q}}\right]\\
d_{36}^{\text{El-bwd}}=-1\,\text{tr}\left[S(t_0, t_3)(\gamma_5)S(t_3, t_1+\hat{\nu_4})(1+\gamma_{\nu})U_{\nu}^{\dagger}(t_1+\hat{\nu_4},t_1)e^{i\mathbf{q}}S(t_1, t_0)(\gamma_5)\right]\text{tr}\left[S(t_2+\hat{\mu_4}, t_2)(1-\gamma_{\mu})e^{-i\mathbf{q}}U_{\mu}(t_2,t_2+\hat{\mu_4})\right]\\
d_{42}^{\text{El-bwd}}=1\,\text{tr}\left[S(t_0, t_3)(\gamma_5)S(t_3, t_1+\hat{\nu_4})(1+\gamma_{\nu})U_{\nu}^{\dagger}(t_1+\hat{\nu_4},t_1)e^{i\mathbf{q}}S(t_1, t_0)(\gamma_5)\right]\text{tr}\left[S(t_2, t_2+\hat{\mu_4})(1+\gamma_{\mu})U_{\mu}^{\dagger}(t_2+\hat{\mu_4},t_2)e^{-i\mathbf{q}}\right]\\
d_{9}^{\text{Er}}=-2\,\text{tr}\left[S(t_2+\hat{\mu_4}, t_3)(\gamma_5)S(t_3, t_0)(\gamma_5)S(t_0, t_2)(1-\gamma_{\mu})e^{-i\mathbf{q}}U_{\mu}(t_2,t_2+\hat{\mu_4})\right]\text{tr}\left[S(t_1+\hat{\nu_4}, t_1)(1-\gamma_{\nu})e^{i\mathbf{q}}U_{\nu}(t_1,t_1+\hat{\nu_4})\right]\\
d_{11}^{\text{Er}}=2\,\text{tr}\left[S(t_2, t_3)(\gamma_5)S(t_3, t_0)(\gamma_5)S(t_0, t_2+\hat{\mu_4})(1+\gamma_{\mu})U_{\mu}^{\dagger}(t_2+\hat{\mu_4},t_2)e^{-i\mathbf{q}}\right]\text{tr}\left[S(t_1+\hat{\nu_4}, t_1)(1-\gamma_{\nu})e^{i\mathbf{q}}U_{\nu}(t_1,t_1+\hat{\nu_4})\right]\\
d_{13}^{\text{Er}}=2\,\text{tr}\left[S(t_2+\hat{\mu_4}, t_3)(\gamma_5)S(t_3, t_0)(\gamma_5)S(t_0, t_2)(1-\gamma_{\mu})e^{-i\mathbf{q}}U_{\mu}(t_2,t_2+\hat{\mu_4})\right]\text{tr}\left[S(t_1, t_1+\hat{\nu_4})(1+\gamma_{\nu})U_{\nu}^{\dagger}(t_1+\hat{\nu_4},t_1)e^{i\mathbf{q}}\right]\\
d_{15}^{\text{Er}}=-2\,\text{tr}\left[S(t_2, t_3)(\gamma_5)S(t_3, t_0)(\gamma_5)S(t_0, t_2+\hat{\mu_4})(1+\gamma_{\mu})U_{\mu}^{\dagger}(t_2+\hat{\mu_4},t_2)e^{-i\mathbf{q}}\right]\text{tr}\left[S(t_1, t_1+\hat{\nu_4})(1+\gamma_{\nu})U_{\nu}^{\dagger}(t_1+\hat{\nu_4},t_1)e^{i\mathbf{q}}\right]\\
d_{26}^{\text{Er-bwd}}=1\,\text{tr}\left[S(t_0, t_3)(\gamma_5)S(t_3, t_2)(1-\gamma_{\mu})e^{-i\mathbf{q}}U_{\mu}(t_2,t_2+\hat{\mu_4})S(t_2+\hat{\mu_4}, t_0)(\gamma_5)\right]\text{tr}\left[S(t_1+\hat{\nu_4}, t_1)(1-\gamma_{\nu})e^{i\mathbf{q}}U_{\nu}(t_1,t_1+\hat{\nu_4})\right]\\
d_{32}^{\text{Er-bwd}}=-1\,\text{tr}\left[S(t_0, t_3)(\gamma_5)S(t_3, t_2+\hat{\mu_4})(1+\gamma_{\mu})U_{\mu}^{\dagger}(t_2+\hat{\mu_4},t_2)e^{-i\mathbf{q}}S(t_2, t_0)(\gamma_5)\right]\text{tr}\left[S(t_1+\hat{\nu_4}, t_1)(1-\gamma_{\nu})e^{i\mathbf{q}}U_{\nu}(t_1,t_1+\hat{\nu_4})\right]\\
d_{38}^{\text{Er-bwd}}=-1\,\text{tr}\left[S(t_0, t_3)(\gamma_5)S(t_3, t_2)(1-\gamma_{\mu})e^{-i\mathbf{q}}U_{\mu}(t_2,t_2+\hat{\mu_4})S(t_2+\hat{\mu_4}, t_0)(\gamma_5)\right]\text{tr}\left[S(t_1, t_1+\hat{\nu_4})(1+\gamma_{\nu})U_{\nu}^{\dagger}(t_1+\hat{\nu_4},t_1)e^{i\mathbf{q}}\right]\\
d_{44}^{\text{Er-bwd}}=1\,\text{tr}\left[S(t_0, t_3)(\gamma_5)S(t_3, t_2+\hat{\mu_4})(1+\gamma_{\mu})U_{\mu}^{\dagger}(t_2+\hat{\mu_4},t_2)e^{-i\mathbf{q}}S(t_2, t_0)(\gamma_5)\right]\text{tr}\left[S(t_1, t_1+\hat{\nu_4})(1+\gamma_{\nu})U_{\nu}^{\dagger}(t_1+\hat{\nu_4},t_1)e^{i\mathbf{q}}\right]\\
d_{29}^{\text{F}}=1\,\text{tr}\left[S(t_0, t_3)(\gamma_5)S(t_3, t_0)(\gamma_5)\right]\text{tr}\left[S(t_2+\hat{\mu_4}, t_2)(1-\gamma_{\mu})e^{-i\mathbf{q}}U_{\mu}(t_2,t_2+\hat{\mu_4})\right]\text{tr}\left[S(t_1+\hat{\nu_4}, t_1)(1-\gamma_{\nu})e^{i\mathbf{q}}U_{\nu}(t_1,t_1+\hat{\nu_4})\right]\\
d_{35}^{\text{F}}=-1\,\text{tr}\left[S(t_0, t_3)(\gamma_5)S(t_3, t_0)(\gamma_5)\right]\text{tr}\left[S(t_2, t_2+\hat{\mu_4})(1+\gamma_{\mu})U_{\mu}^{\dagger}(t_2+\hat{\mu_4},t_2)e^{-i\mathbf{q}}\right]\text{tr}\left[S(t_1+\hat{\nu_4}, t_1)(1-\gamma_{\nu})e^{i\mathbf{q}}U_{\nu}(t_1,t_1+\hat{\nu_4})\right]\\
d_{41}^{\text{F}}=-1\,\text{tr}\left[S(t_0, t_3)(\gamma_5)S(t_3, t_0)(\gamma_5)\right]\text{tr}\left[S(t_2+\hat{\mu_4}, t_2)(1-\gamma_{\mu})e^{-i\mathbf{q}}U_{\mu}(t_2,t_2+\hat{\mu_4})\right]\text{tr}\left[S(t_1, t_1+\hat{\nu_4})(1+\gamma_{\nu})U_{\nu}^{\dagger}(t_1+\hat{\nu_4},t_1)e^{i\mathbf{q}}\right]\\
d_{47}^{\text{F}}=1\,\text{tr}\left[S(t_0, t_3)(\gamma_5)S(t_3, t_0)(\gamma_5)\right]\text{tr}\left[S(t_2, t_2+\hat{\mu_4})(1+\gamma_{\mu})U_{\mu}^{\dagger}(t_2+\hat{\mu_4},t_2)e^{-i\mathbf{q}}\right]\text{tr}\left[S(t_1, t_1+\hat{\nu_4})(1+\gamma_{\nu})U_{\nu}^{\dagger}(t_1+\hat{\nu_4},t_1)e^{i\mathbf{q}}\right]
\label{eq:dPS2}
\eeqs
 \normalsize
The shifted quark propagators have the following meaning depending on whether the current is split in temporal or spatial directions, for example,
\beqs
S(t_3, t_2+\hat{\mu}_4) \equiv 
 \begin{cases} S(t_3,t_2+1) =  P(t_3)M^{-1}P(t_2+1)^T, & \text{if } \mu= 4\\ 
 S(t_3,t_2) = P(t_3)M^{-1}P(t_2)^T, &\text{if } \mu\neq 4,
 \end{cases}
 \eeqs
 where the projector $P(t)$ is defined in Eq.\eqref{eq:Pt}. 
The associated gauge links have the meaning,
 \beqs
 U_{\mu}(t_2,t_2+\hat{\mu}_4) &\equiv
 \begin{cases} U_4(t_2,t_2+1),  & \text{if } \mu= 4\\ 
 U_\mu(t_2,t_2), &\text{if } \mu \neq 4,
 \end{cases} \\ 
  U_{\mu}^{\dagger}(t_2+\hat{\mu}_4,t_2) &\equiv
 \begin{cases} U_4^{\dagger}(t_2+1,t_2), & \text{if } \mu= 4\\ 
 U_\mu^{\dagger}(t_2,t_2), &\text{if } \mu \neq 4,
 \end{cases}
 \eeqs
where the gauge links are defined in Eq.\eqref{eq:U}. 
So the split in time is explicitly carried in both the propagators and gauge links, whereas the split in space is only implicitly carried in the gauge links.
Note the placement of $e^{\pm i\bm q}$ in relation to $U$ and $U^\dagger$. They do not commute when the currents are split in spatial directions.

\section{Wall source implementation}
\label{sec:wall}

We introduce a rigorous matrix notation to elucidate the implementation of wall sources.
We define wall sources as a vector in spatial coordinates, diagonal in spin and color,
\beq
[{\cal W}]_{s,c,{\bf x}; s',c'} \equiv \delta_{ss'}\delta_{cc'}.
\eeq
That is, all spatial entries of the real part are set to 1, imaginary part to zero. It can be placed at any time slice.
Under a gauge transformation $G$, the gauge average is
\beq
\langl G(t) {\cal W} {\cal W}^T G(t)^\dagger \rangl_G = \iden_{\bm x,s,c},\quad
\text{where }
\left[G(t){\cal W}\right]_{\bf x} = G(t,\bf x) \iden_{s,c}.
\eeq
More explicitly,
\beq
\left[\langl G(t) {\cal W} {\cal W}^T G(t)^\dagger\rangl_G\right]_{\bm x,\bm y} = \frac1{|G|}\int DG\, G(t,{\bm x}) \iden_\text{spin} G(t,\bm y)^\dagger \iden_\text{spin} = \delta_{\bm x, \bm y} \iden_s \iden_c.
\eeq
We insert the wall source in between a pair of quark propagators in the path integral by the following steps, only highlighting the time dependence in $S$ to keep the notation simple,
\fontsize{11}{11}
\beqs
& \int DU P(U) \Tr_{\bm x,s,c} \Big[\ldots S[U](t',t) S[U](t, t'') \ldots \Big] \\ 
&=  
\int DU P(U)\Tr_{\bm x,s,c} \Big[\ldots S[U](t',t)\, \iden_{\bm x,s,c} \,S[U](t, t'') \ldots
\Big]  \\
&= \frac1{|G|} \int DG \int DU P(U)
\Tr_{\bm x,s,c} \Big[\ldots S[U](t',t) G(t) {\cal W} {\cal W}^T G(t)^\dagger S[U](t, t'') \ldots
\Big] \\
&= \frac1{|G|} \int DG \int DU P(U_G)
\Tr_{\bm x,s,c} \Big[\ldots S[U_G](t',t) G(t) {\cal W} {\cal W}^T G(t)^\dagger S[U_G](t, t'') \ldots
\Big] \\
&= \frac1{|G|} \int DG \int DU P(U)
\Tr_{\bm x,s,c} \Big[\ldots S[U](t',t)  {\cal W} {\cal W}^T  S[U](t, t'') \ldots
\Big] \\
&= \int DU P(U)
\Tr_{\bm x,s,c} \Big[\ldots S[U](t',t)  {\cal W} {\cal W}^T  S[U](t, t'') \ldots
\Big] \\
&= \int DU P(U)
\Tr_{s,c} \Big[{\cal W}^T  S[U](t, t'') \ldots S[U](t',t)  {\cal W} 
\Big].
\label{eq:path}
\eeqs
In the last step, we use the cyclic property of trace $\Tr AB= \Tr BA$. 
We also used the property that under a gauge transformation
$U_\mu\to (U_G)_\mu \equiv GU_\mu G^\dagger$, the propagator transforms as,
\beq
S[U_G](t,t') = G(t)S[U](t,t')G(t')^\dagger.
\eeq
More explicitly, the gauge links are,
\beq
(U_\mu)_{\bm x,t; \bm x',t'} = \delta_{(\bm x,t),(\bm x',t') - \mu} U_\mu(\bm x, t) \iden_s,
\label{eq:U}
\eeq
and its gauge transformation is 
\beq
(G U_\mu G^\dagger)_{x,y} = G(x) [U_\mu]_{x,y} G(y)^\dagger =
G(x) \delta_{x,y-\mu} U_\mu(x) G(y)^\dagger =
\delta_{x,y-\mu} G(x) U_\mu(x) G(x+\mu)^\dagger.
\eeq
Note that we will use 
\beq
U_\mu(t,t')=P(t) U_\mu P(t')^T\qquad
\text{and}\qquad 
U_\mu^\dagger(t,t') = P(t) U_\mu^\dagger P(t')^T.
\eeq
    Here $P(t)$ is defined as projection to a time slice (not to be confused with the weighting factor $P(U)$ in the path integral in Eq.\eqref{eq:path}),
\beq
[P(t_p)]_{s,c,{\bf x};s',c',t',{\bf x'}}  
\equiv
\delta_{t_p,t'}\delta_{ss'}\delta_{cc'}\delta_{{\bf x},{\bf x'}},
\label{eq:Pt}
\eeq
which is diagonal in spin, color, and space.
When we take the dagger of $U_\mu(t,t')$, we need to switch the time arguments since
\beq
[U_\mu(t,t')]^\dagger
= [P(t) U_\mu P(t')^T]^\dagger
= P(t') U_\mu^\dagger P(t)^T = 
U_\mu^\dagger(t',t)\,.
\eeq

Operationally, a quark propagator can be written in terms of the inverse of the quark matrix as,
\beq
S(t,t') \equiv P(t) M_q^{-1} P(t')^T.
\label{eq:PSP}
\eeq
For Wilson-type fermions, $M_q$ satisfies the $\gamma_5$-hermiticity relation 
\beq 
M_q^\dagger=\gamma_5 M_q\gamma_5, \quad\left(M^{-1}_q\right)^\dagger=\gamma_5 M^{-1}_q\gamma_5.
\eeq
Examples on how to use the notation to calculate two-point and four-point correlation functions are discussed in Sec.~\ref{sec:cfun}. 

\section{Form factor from four-point functions}
\label{sec:ff4}

%
\begin{table}[b!]
\caption{Pion form factor $F_\pi(\bm q^2)$ from four-point functions.  An example of the data to be fitted is given in Fig.~\ref{fig:Q44ab}. The fit form is in Eq.\eqref{eq:Q44elas} with $F_\pi$ and $E_\pi$ treated as free parameters and $m_\pi$ taken from the measured value. For comparison,  the $E_\pi$ from the continuum dispersion relation is provided with the same $m_\pi$ values. The four columns correspond to $\bm q=\{0,0,1\}, \{0,1,1\}, \{1,1,1\},  \{0,0,2\}$ from left to right.}
\label{tab:ff}
\begin{tabular}{c}
$      
\renewcommand{\arraystretch}{1.2}
\begin{array}{l|cccc}
\hline
  & \text{} & m_{\pi }\text{=1100 MeV} & \text{} & \text{} \\
\hline
 F_{\pi } & 0.8209\pm 0.0023 & 0.7213\pm 0.0023 & 0.650\pm 0.004 & 0.604\pm 0.005 \\
 E_{\pi }\text{ fit} & 1.2556\pm 0.0016 & 1.4021\pm 0.0027 & 1.530\pm 0.004 & 1.644\pm 0.006 \\
 E_{\pi }\text{ continuum} & 1.2597\pm 0.0010 & 1.3976\pm 0.0009 & 1.5230\pm 0.0009 & 1.6389\pm 0.0008 \\
 \text{Fit range} & \text{$\{$7,9$\}$} & \text{$\{$6,8$\}$} & \text{$\{$7,10$\}$} & \text{$\{$7,12$\}$} \\
 \chi ^2\text{/dof} & 2.00 & 1.40 & 2.70 & 1.90 \\
 \hline
  & \text{} & m_{\pi }\text{=800 MeV} & \text{} & \text{} \\
  \hline
 F_{\pi } & 0.7677\pm 0.0027 & 0.646\pm 0.006 & 0.568\pm 0.010 & 0.552\pm 0.011 \\
 E_{\pi }\text{ fit} & 0.9967\pm 0.0020 & 1.163\pm 0.005 & 1.308\pm 0.010 & 1.463\pm 0.013 \\
 E_{\pi }\text{ continuum} & 0.9992\pm 0.0009 & 1.1682\pm 0.0007 & 1.3157\pm 0.0007 & 1.4483\pm 0.0006 \\
 \text{Fit range} & \text{$\{$9,13$\}$} & \text{$\{$10,17$\}$} & \text{$\{$10,17$\}$} & \text{$\{$9,14$\}$} \\
 \chi ^2\text{/dof} & 1.40 & 1.40 & 1.10 & 0.72 \\
\hline
  & \text{} & m_{\pi }\text{=600 MeV} & \text{} & \text{} \\
\hline
 F_{\pi } & 0.7412\pm 0.0015 & 0.6360\pm 0.0025 & 0.583\pm 0.004 & 0.525\pm 0.012 \\
 E_{\pi }\text{ fit} & 0.8508\pm 0.0017 & 1.050\pm 0.004 & 1.231\pm 0.007 & 1.354\pm 0.016 \\
 E_{\pi }\text{ continuum} & 0.8500\pm 0.0010 & 1.0435\pm 0.0008 & 1.2063\pm 0.0007 & 1.3497\pm 0.0006 \\
 \text{Fit range} & \text{$\{$4,13$\}$} & \text{$\{$6,15$\}$} & \text{$\{$6,9$\}$} & \text{$\{$8,13$\}$} \\
 \chi ^2\text{/dof} & 0.52 & 1.30 & 1.50 & 0.39 \\
\hline
  & \text{} & m_{\pi }\text{=360 MeV} & \text{} & \text{} \\
\hline
 F_{\pi } & 0.720\pm 0.004 & 0.616\pm 0.005 & 0.554\pm 0.006 & 0.530\pm 0.008 \\
 E_{\pi }\text{ fit} & 0.695\pm 0.005 & 0.911\pm 0.009 & 1.076\pm 0.013 & 1.258\pm 0.018 \\
 E_{\pi }\text{ continuum} & 0.7082\pm 0.0011 & 0.9316\pm 0.0009 & 1.1110\pm 0.0007 & 1.2652\pm 0.0006 \\
 \text{Fit range} & \text{$\{$6,11$\}$} & \text{$\{$6,11$\}$} & \text{$\{$6,11$\}$} & \text{$\{$6,11$\}$} \\
 \chi ^2\text{/dof} & 0.68 & 0.34 & 0.68 & 1.00 \\
 \hline
\end{array}$
\end{tabular}
\end{table}

\end{widetext}
\end{document}